# RADIATIVE TRANSFER IN SPHERES
# I. ANALYTICAL SOLUTIONS

Charles H. Aboughantous
Principal Investigator

Department of Physics and Astronomy, Louisiana State University
Baton Rouge, Louisiana 70803

August 2001

Sponsored by

Naval Research Laboratories, Plasma Dynamics Branch
Plasma Physics Division
Washington DC, 20375

NRL PA 010

# ABSTRACT

A nonsingular analytical solution for the transfer equation in a pure absorber is obtained in central symmetry and in a monochromatic radiation field. The native regular singularity of the equation is removed by applying a linear transformation to the frame of reference. Two different approaches are used to carry out the solution. In the first approach the angular derivative is interpreted in an original way that made it possible to discard this derivative from the equation for all black body media without upsetting the conservation of energy. In this approach the analytic solution is expressible in terms of exponential integrals without approximations but for practical considerations the solution is presented in the form of Gauss-Legendre quadrature for quantitative evaluation of the solutions. In the second approach the angular derivative is approximated by a new set of discrete ordinates that guarantees the closer of the set of equations and the conservation of energy. The solutions from the two approaches are numerically identical in homogeneous and non-homogeneous cold media as well as in local thermodynamic equilibrium and for all optical depths that can be managed with 16 significant digits arithmetics without numerical instabilities. The solution in either approach is made possible by splitting the original transfer equation into two equations, one for the centripetal energy flow and another one for the centrifugal energy flow. The end-points form of the solution proved to be very effective in numerical computations without spatial truncation errors.

# TABLE OF CONTENTS

# INTRODUCTION

The radiative transfer problem has been the attraction to scientists since at least early the twentieth century. A great deal of efforts was devoted to obtain reliable and dependable solutions for the governing equation but only qualified successes were recorded. Approximations were introduced and have proven to be effective in slab geometry [1] [2], but none is equally satisfactory in spherical geometry. The spherical geometry solution for the transfer equation is currently sought in addressing a number of problems. Among others are the astrophysical and the atmospheric radiative transfer and the diagnostic problem of the imploded pellet of the laser induced fusion technology. Our efforts in this undertaking are steered toward this latter aspect of the radiative transfer problem with a specific address to the analytic solution of the time independent equation.

The most widely used approach for solving the transfer equation in curved geometry is a numerical scheme based on approximating the angular derivative by a set of *angular parameters*. This is the traditional discrete ordinates model in curved geometry. The set was derived from the abscissas and weights of Gauss-Legendre (GL) quadrature but the actual integration was made with a quadrature developed specifically for that purpose [3]. Although the model is conditioned to satisfy the conservation of energy, i.e., the flux obeys the inverse square law, it fails to satisfy the *closure* condition: there are $N$ linear equations with $N$+1 unknowns. The additional unknown, labeled the *starter solution*, is borrowed from the solution of the problem in slab geometry for the direction cosine $\mu = -1$.

It appeared over the years that this approach produces a reasonable approximation for most common problems, but it is not up to the needs of the users of the transfer equations in optically thick media. In part because it requires too fine a spatial grid, but also it produces an asymmetric solution in otherwise perfectly defined symmetric problem [4]. The asymmetry is described as a dip in the intensity profile at the center of the sphere. There is no proof, though, whether this dip is a numerical aberration or a characteristic property of the solution of the transport equation. Notwithstanding the lack of such a proof, ad hoc recipes were introduced to cure the dip and make the intensity profile a monotonic function in the sphere. Among those is the *starter direction cosine* that depends on the radius of the sphere, save it is allowed to take values of magnitudes larger than 1.0 [5].

We revisited the structure of the discrete ordinates set with a new approach and built a new set based on GL quadrature. The new set comprises only positive direction cosines for all specific intensities. The new set of discrete ordinates enabled transcribing the transfer equation into a *complete* set of equations. By *complete* we mean the set of equations is closed ($N$ equations in $N$ unknowns) and conservative (the solution satisfies the conservation relation). The completeness of the set is accomplished by splitting the transfer equation into two equations, one equation for the centripetal energy flow and another one for the centrifugal energy flow.

The splitting of the transfer equation into two equations enabled the construction of a new diffusion model by dropping the angular derivative from the transfer equations. This diffusion model is valid for transfer problems in central symmetry whenever the radiation field is a competition of both the centripetal and the centrifugal flow of energy in homogeneous and heterogeneous media. The solution of these diffusion equations is delightfully simple and accurate.

A non-physical feature of the transfer equation in spherical geometry is that it is singular at the center of the sphere. This feature is intrinsic to the transfer equation in its native form as an abstract mathematical equation in spherical geometry. We cured this problem by an appropriate transformation of the frame of reference.

The solutions for the discrete ordinates equations and the diffusion equations are presented in two forms: continuous in $r$ and end-points form, and tested quantitatively. The end-points solution is particularly attractive in numerical computations in optically thick media.

# PART I

# THE TRANSFER EQUATION

## 1.1 The traditional transfer equation

Consider a monochromatic spherical source of radius $\varepsilon$, which we will be referring to henceforth as the *core*, surrounded by a mantle of a homogeneous isothermal pure absorber of radius $R$ (fig. 1.1). The radiance of the source is isotropic and uniformly distributed on the surface. The ensemble is immersed in a uniform and isotropic radiation field. Under these LTE conditions the mantle is an isotropic and uniform emitter of emission density $\kappa B$ where $B$ is the black body constant and $\kappa$ is a plankian opacity.

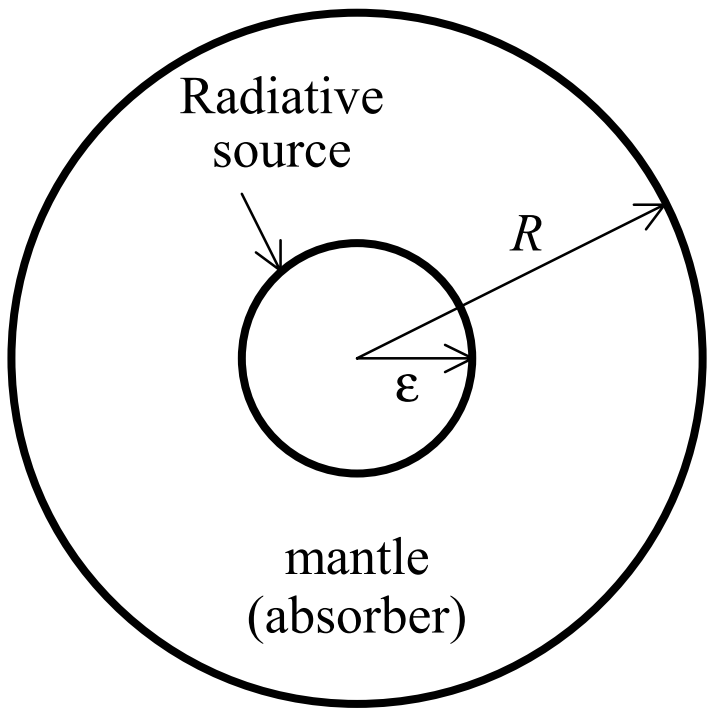

**Figure 1.1**

It is readily apparent that the problem at hand is perfectly symmetric about the center of the core: the symmetry is in the geometry as well as in the emission, transmission and the boundary conditions. Hence, the transfer equation for this problem may be written as:

$$\mu \partial_r \psi_r + \frac{\nu^2}{r} \partial_\mu \psi_r + \kappa \psi_r = \kappa B \tag{1.1}$$

where $\partial_r$ is the tensor notation for the derivative with respect to $r$, $\psi$ is the specific intensity[✩] and $\mu$ the direction cosine on $[-1, +1]$. The subscript in $\psi_r$ is for the spatial variable; the angular variable is omitted, it is implied by the mere use of the letter $\psi$ unless otherwise indicated. We used the symbol $\nu = \sqrt{1-\mu^2} = \sin\theta \Rightarrow \partial_\mu \nu^2 = -2\mu$ for notational convenience. First we note that eq. (1.1) is conservative. If it is integrated on $\mu \in [-1, +1]$ we obtain the conservation relation:

$$\frac{1}{r^2}\partial_r\left(r^2 F_r\right) + \kappa\varphi_r = \kappa B \tag{1.2}$$

where $F$ is the flux and $\varphi$ is the total intensity. The integration on the angular domain is rather straightforward if we transform eq. (1.1) by adding and subtracting $2\mu\psi_r/r$ to the left side to obtain:

$$\frac{\mu}{r^2}\partial_r\left(r^2\psi_r\right) + \frac{1}{r}\partial_\mu\left[\nu^2\psi_r\right] + \kappa\psi_r = \kappa B \tag{1.3}$$

Upon integration of eq. (1.3) over all directions, the conservation relation (1.2) is obtained.

The equation (1.3) is referred to as the conservation form of the transfer equation although it is by no means any more conservative than eq. (1.1). It appears that this characterization originates from the fact that when the derivatives of eq. (1.1) are discretized for numerical computations, the solution does not satisfy the conservation relation in the manner eq. (1.3) does. This is

✩ We use the symbol $\psi$ for the intensity to avoid confusion of the traditional symbol $I$ with the identity matrix.

the result of a human error in the discretization of the derivatives. The equation (1.1) is sound and compliant to discretization provided it is done correctly.

We wish to emphasize that the conservation relation (1.2) is a necessary condition for the solution to be physically acceptable, but it is not a sufficient condition for that purpose; the same applies to relation (1.2). Indeed, this conservation relation can always be satisfied in a central symmetry problem $\forall\psi_0 \leq \infty$. An exact solution for the transfer equation conditioned by the conservation relation alone may not necessarily be a correct solution. This situation occurs when the solution is mathematically exact but physically unsound.

## 1.2 The modified transfer equation

We recognize that eq. (1.1) is plagued by a regular singularity at $r = 0$, which is the case with a solid sphere with origin at the center. Mathematical treaties on differential equations preach that the solution of this class of equations is characteristically singular at the origin. The physical interpretation of this mathematical language is that the center of the sphere is a point of accumulation of energy, or equivalently an adiabatic boundary, which is not true. Indeed, a differential pencil of radiation that enters the sphere at one point heading toward the center will emerge only attenuated from the opposite point on the surface of the sphere, conceivably a pure absorber. This could have not been the case if the center was adiabatic.

The mythical singularity at the origin is understandable in the mathematical abstract of equations with regular singularity. The mathematical treaties assume the domain of definition of these equations is $(0, R]$. By this representation the frame of reference is placed at the center of the sphere which itself is an asymptotic boundary: the radiation that enters the sphere at the surface never reaches the center. This is not a natural physical process. It is a human error resulting from chiseling the physics to suit the mathematical formalism while it should be the other way around. Clearly, the central frame of reference, or the *c-frame*, is not adequate for our purpose.

In our physical problem, the appropriate lower limit of the domain of definition must be at the point where the radiation begins its journey propagating into the sphere. In the case of a solid sphere the domain of $r$ must be $[-R, +R]$. Unfortunately, the central symmetry geometry with radial origin at the center of the sphere inhibits negative values for the radial variable. The $[-R, +R]$ domain is not adequate for our purpose if we insist to preserve the central symmetry requirement $r \geq 0$. An equivalent domain may be obtained by applying a liner transformation to the frame of reference. The simplest would be shifting the $c$-frame by $R$ to the left (fig. 1.2). We define the domain of $r$ in the *shifted frame of reference*, or the *s-frame*, to be $[R, 2R]$. It follows that the independent variable in the $s$-frame becomes $r_s \in [R+r, 2R]$ and $r \in [0, R]$. Accordingly, the intensity also must be shifted. Using our subscript notation:

$$\psi_r \rightarrow \psi_{R;r}\ ; \quad \psi_0 \rightarrow \psi_{R;0}\ ; \quad \psi_R \rightarrow \psi_{R;R} \tag{1.4}$$

where the subscript $R;r$ mean $R$ is a fixed parameter in the $s$-frame and $r$ is the independent variable on $[0, R]$. The representation $\psi_{R;0}$ means that the intensity at the center of the sphere is a function of the radius $R$ of the sphere, a physically sound quantity. By contrast, the intensity at the center in the $c$-frame is singular as a result of the implication of a non-physical asymptotic boundary at that point. From this point forward the fixed parameter $R$ displayed as the first subscript in expressions (1.4) will be omitted and that should not introduce any confusion. The indexing in multi-media spheres will be defined in due course.

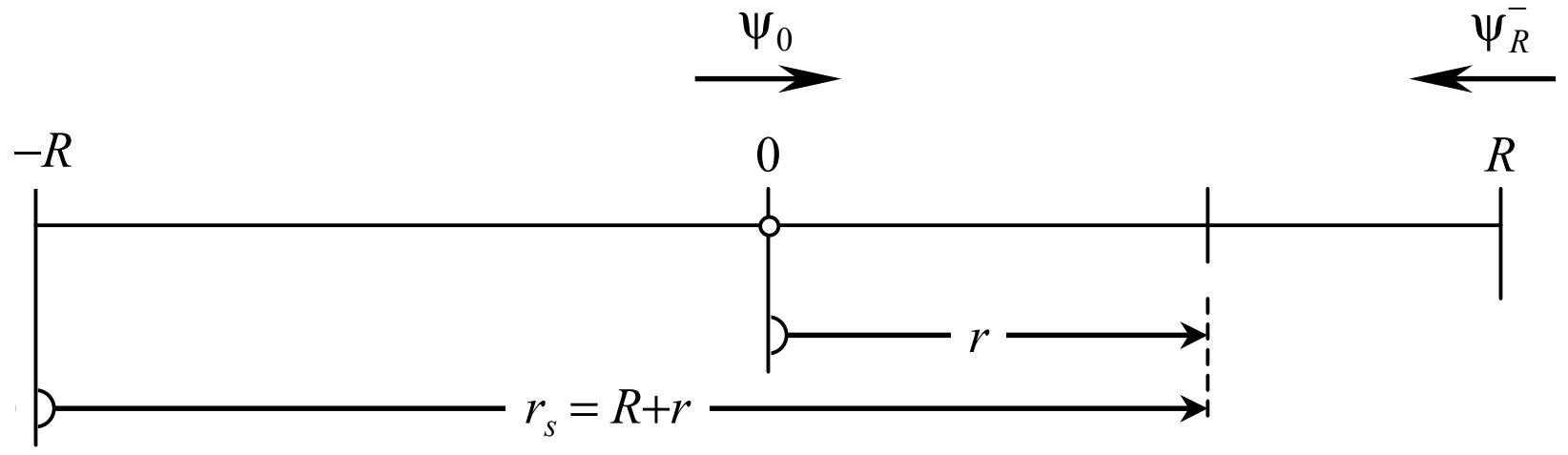

**Figure 1.2**. A schematic showing the independent variable $r$ in the $s$-frame and the boundary conditions for the centripetal and the centrifugal intensities.

The implementation of the shifted variable requires a revision to our definition of the problem: the equation (1.1) has to be expressed in the $s$-frame. Since the differential $\mathrm{d}r = \mathrm{d}(R+r)$, all we need to shift eq. (1.1) is to replace $r$ by $R+r$ so that our equation in the $s$-frame becomes:

$$\mu \partial_r \psi + \frac{\nu^2}{R+r} \partial_\mu \psi + \kappa\psi = \kappa B \qquad r \in [0, R] \qquad (1.5)$$

The advantage of working with eq. (1.5) is that it is not singular at the center of the sphere, i.e., at $r = 0$, and it is meaningful in the sense of identity (1.4). Albeit this advantage, eq. (1.5) is, like the non-shifted eq. (1.1), quite tenacious to yielding to the techniques of discrete ordinates❖ (see Part II). Alternatively, we convert eq. (1.5) to the perpetrated conservation form of the transfer equation by adding and subtracting $2\mu\psi_r/(R+r)$ to the left side. The transformed equation on the domain $[0, R]$ becomes:

$$\frac{\mu}{(R+r)^2} \partial_r\left[(R+r)^2\psi_r\right] + \frac{1}{R+r} \partial_\mu\left(\nu^2\psi_r\right) + \kappa\psi_r = \kappa B \qquad (1.6)$$

Similarly we obtain the differential conservation relation in $s$-frame:

$$\frac{1}{(R+r)^2} \partial_r\left[(R+r)^2 F_r\right] + \kappa\varphi_r = \kappa B \qquad (1.7)$$

The solution for eq. (1.6) will be sought subject to the conservation condition (1.7) and the condition of an infinite medium: for $R \to \infty \;\Rightarrow\; \psi_{\infty;r} \equiv \varphi_{\infty;r} = B$, where $\varphi$ is the total intensity. While the conservation and the infinite medium conditions are necessary for the consistency of the mathematical model, they are not sufficient to imply the correct solution. Additional constraints should be considered for that purpose (sec. 1.4).

## 1.3 The reduced mathematical problem

In our search for the solution of eq. (1.1) we find it helpful to benefit from the symmetry of the

---

❖ This problem is associated with the angular derivative and is unrelated to the shift of the spatial frame of reference.

problem. We split the equation into two equations on the angular domain, one for the *centrifugal intensity* $\psi^+$ and a second one for the *centripetal intensity* $\psi^-$, and both of the equations are defined on $\mu > 0$. To accomplish this we change the sign of the divergence operator in the centripetal equation. Hence, we have:

centrifugal equation: $$\mu\partial_r\psi^+ + \frac{\nu^2}{r}\partial_\mu\psi^+ + \kappa\psi^+ = \kappa B \qquad \mu > 0 \tag{1.8a}$$

centripetal equation: $$-\mu\partial_r\psi^- - \frac{\nu^2}{r}\partial_\mu\psi^- + \kappa\psi^- = \kappa B \qquad \mu > 0 \tag{1.8b}$$

The splitting of eq. (1.1) into one centripetal and one centrifugal equation applies to the conservation form of the transfer equation (1.3) as well, and eventually to eq. (1.6). Here, we will have to solve the two equations on $\mu > 0$:

$$\frac{\mu}{(R+r)^2}\partial_r\left[(R+r)^2\psi_r^+\right] + \frac{1}{R+r}\partial_\mu\left(\nu^2\psi_r^+\right) + \kappa\psi_r^+ = \kappa B \qquad r\in[0, R] \tag{1.9a}$$

$$\frac{-\mu}{(R+r)^2}\partial_r\left[(R+r)^2\psi_r^-\right] - \frac{1}{R+r}\partial_\mu\left(\nu^2\psi_r^-\right) + \kappa\psi_r^- = \kappa B \qquad r\in[0, R] \tag{1.9b}$$

The splitting of the transfer equation into two equations describing quantities in opposite directions but on the same positive interval of direction cosines is not new. It was recently introduced and implemented in the $S_{2N}$ discrete ordinates problem in slab geometry [2]. Credits should also be given to Feautrier [6] who apparently was the first to introduce this concept much earlier in the $S_2$ discrete ordinates equations in slab geometry.

The splitting of the transfer equation in spherical geometry carries with it a welcome bonus. If we integrate eq. (1.9a) on $\mu \in [0, 1]$, the angular derivative at a radial position $r$ yields only one non-vanishing term $+\psi_r^+(\mu = 0)$. Similarly, the angular derivative of eq. (1.9b) yields only one non-vanishing term $-\psi_r^-(\mu = 0)$. The sum of these two quantities vanishes at all points on the radial axis in central symmetry, a well-known property. What is not known about this property is that we can live without the angular derivative in each of the equations (1.9). This is always true whenever the problem at hand is about energy transfer in a sphere with central symmetry where both of the equations are needed for a complete description of the transfer. This is the underlying hypothesis of the *incomplete diffusion theory* of Part III. If, however, the problem is about radiating energy from a spherical surface, or a point source for that matter, in a non-scattering medium, only eq. (1.9a) is needed and the angular derivative must be there in full force. This problem is discussed in Part IV.

Finally, a less cumbersome transcription of the equations may be obtained by defining the function $\vartheta = R + r$. Then we have:

Centrifugal equation: $$\frac{\mu}{\vartheta^2}\partial_r\left(\vartheta^2\psi_r^+\right) + \frac{1}{\vartheta}\partial_\mu\left(\nu^2\psi_r^+\right) + \kappa\psi_r^+ = \kappa B \tag{1.10a}$$

Centrifugal equation: $$-\frac{\mu}{\vartheta^2}\partial_r\left(\vartheta^2\psi_r^-\right) - \frac{1}{\vartheta}\partial_\mu\left(\nu^2\psi_r^-\right) + \kappa\psi_r^- = \kappa B \tag{1.10b}$$

Differential conservation relation: $\frac{1}{\vartheta^2}\partial_r\left(\vartheta^2 F_r\right) + \kappa\varphi_r = \kappa B$ (1.11)

The two sets of equation (1.9) and (1.10) are identical except for the notation. Therefore, in all what follows a reference to either one of them implies a reference to the other as well.

The differential conservation relation (1.11) is sufficient to verify that the solution of the transfer equation is correct, but it does not provide information about how much energy is deposited within the volume of the sphere. For that part we have to integrate eq. (1.11) over the volume of the sphere in the *s*-frame. To accomplish this we have to define the differential volume in such a way that preserves the invariance of the volume of the sphere under linear transformation of the frame of reference. The hint about the new differential volume is provided by the flux term of eq. (1.11): the volume integral of the derivative converts to a surface integral if the differential volume satisfies the correspondence relation:

$$\left.\begin{array}{c} c\text{-}frame \\ 4\pi r^2\,\mathrm{d}r \end{array}\right\} \rightarrow \left\{\begin{array}{c} s\text{-}frame \\ \alpha 4\pi\vartheta^2\,\mathrm{d}r \end{array}\right. \tag{1.12}$$

where $\alpha$ is a transformation constant that preserves the invariance of the volume of the sphere under the prescribed linear transformation. It follows that the integral of eq. (1.11) becomes:

$$\alpha 4\pi\int_{R_1}^{R_2}\mathrm{d}\left(\vartheta^2 F_r\right) + \alpha 4\pi\kappa\int_{R_1}^{R_2}\varphi_r\vartheta^2\mathrm{d}r = \alpha 4\pi\kappa B\int_{R_1}^{R_2}\vartheta^2\mathrm{d}r \tag{1.13}$$

The lower radius $R_1$ is intended to be the inner radius of a mantle. It is just 0 for a solid sphere. Clearly, the quantity $4\alpha\pi$ is not needed for the mere purpose of verifying the energy balance.

## 1.4 Determination of the transformation factor

It is apparent from eq. (1.13) that the value of $\alpha$ depends on the geometric configuration of the problem under consideration. In what follows we determine the value of $\alpha$ in most common configurations. The method can be extended to other configurations as needed.

### (*a*) *A cold medium embedding a spherical source*

The medium in this configuration is a pure absorber without re-emission; it can be a mantle of finite outer radius $R$ or an infinite medium (fig. 1.1). The radiance of the core sets a centrifugal radiation field in the medium that is completely described by the homogeneous part of eq. (1.3). The *c*-frame is adequate for this problem and eq. (1.3) is the proper one to be considered for this problem. The radiation field for this problem is defined on $r \geq \varepsilon > 0$. By this requirement the equation is never singular: if $\varepsilon$ is set to zero the radiance $\psi_\varepsilon^+$ vanishes taking with it the radiation field altogether (see sec. 3.3b).

The conservation relation (1.2) applies to this problem by setting the black body constant to zero. To obtain the volume integral of this conservation relation all is needed is to operate $4\pi r^2\mathrm{d}r$ on the differential conservation relation (1.2) and integrate on $[\varepsilon, R]$ however large $R$ may be. The factor $\alpha$ has no place in this problem.

**(*b*) *A cold solid sphere in an isotropic radiation field***

This is a homogeneous sphere of finite radius $R$, a pure absorber without re-emission, immersed in a uniform isotropic radiation field (fig. 1.3). The natural boundary condition for this sphere is the centripetal intensity $\psi_R^-$. The equations (1.8) are singular at the center of this sphere. Therefore, the $s$-frame is adequate for this problem. The integral conservation relation (1.13) is completely defined by setting the black body constant to zero. Then we have:

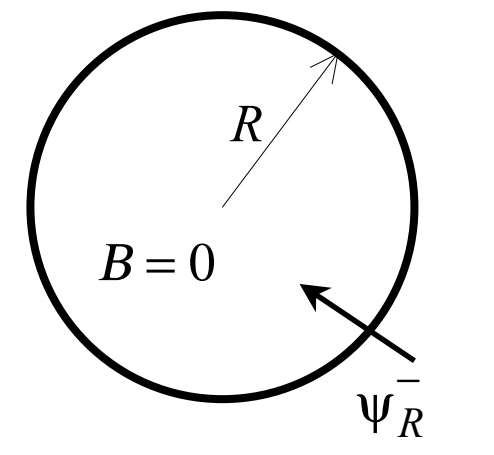

**Figure 1.3**

$$4\pi\alpha^c\left(\vartheta_R^2 F_R - \vartheta_0^2 F_0\right) + 4\pi\alpha^c \int_0^R \varphi_r\, \vartheta^2\, \mathrm{d}r = 0 \quad \Rightarrow \quad \alpha^c = 1/4$$

The brackets term of this equation is the total energy input into the sphere, it is independent of the frame of reference. The factor $\alpha$ can be evaluated by equating the bracketed term with $4\pi R^2 F_R$. By setting the flux at the center of the sphere to zero we obtain $\alpha^c = 1/4$; the superscript $c$ stands for *cold* ⌘, a reference to an absorptive medium without emission.

**(*c*) *A black body sphere surrounded by a cold medium***

The sphere is a homogeneous and isothermal medium with a uniformly distributed source $\kappa B$, and surrounded by vacuum (fig. 1.4). The radiation field within the sphere is described by eqs. (1.10). In order to determine the factor $\alpha$, we demand that the total emitted black body energy be invariably the same in all frame of references. Hence, we must have:

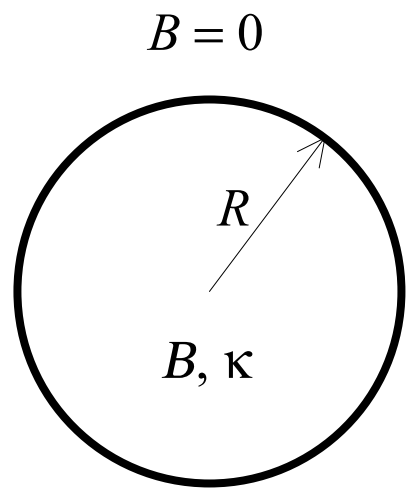

**Figure 1.4**

$$\frac{4\pi R^3}{3}\kappa B = \alpha^h 4\pi\kappa B \int_0^R (R+r)^2\, \mathrm{d}r \quad \Rightarrow \quad \alpha^h = \frac{1}{7}$$

The superscript $h$ stands for *hot* ⌘, a reference to a medium with black body emission.

**(*d*) *A black body sphere immersed in a radiation field***

This problem is the composition of the two previous configurations (*b*) and (*c*). Tow independent processes set the radiation field in the sphere: the input from an external source reaching the sphere at its boundary as an input boundary value $\psi_R^-$, and the energy generated by a black body emission $\kappa B$. Therefore, the intensity in the sphere is the sum of two intensities $\varphi = \varphi^c + \varphi^h$. The same for the fluxes at the surface: $F = F^c + F^h$. Since the two processes are independent, the conservation relations apply to each process separately. The transformation factor $\alpha^c$ of sec. (*b*) applies to the conservation of the cold process and the factor $\alpha^h$ of part (*c*) applies to the hot process.

⌘ A superscript $c$, or $h$, is used all throughout to indicate a reference to a cold, or a hot medium. A superscript $c$ does not refer to the central frame of reference.

### (*e*) *A black body mantle exposed to radiation at both sides*

The mantle is defined by its inner radius $R_1$ and outer radius $R_2$. It is an isothermal homogeneous medium characterized by its black body emission $\kappa B$ and exposed to isotropic radiation fields on both of its surfaces (fig. 1.5). The incident intensities $\psi^+_{R_1}$ and $\psi^-_{R_2}$ set a cold radiation field in the mantel, and the black body emission sets a hot radiation field. Since these two radiation fields are independent processes, the conservation relations should apply to each one of them separately. Hence, two values of the transformation factor $\alpha$ needs to be known. The cold factor $\alpha^c$ is the same as in part (*b*). The hot factor is determined by requiring:

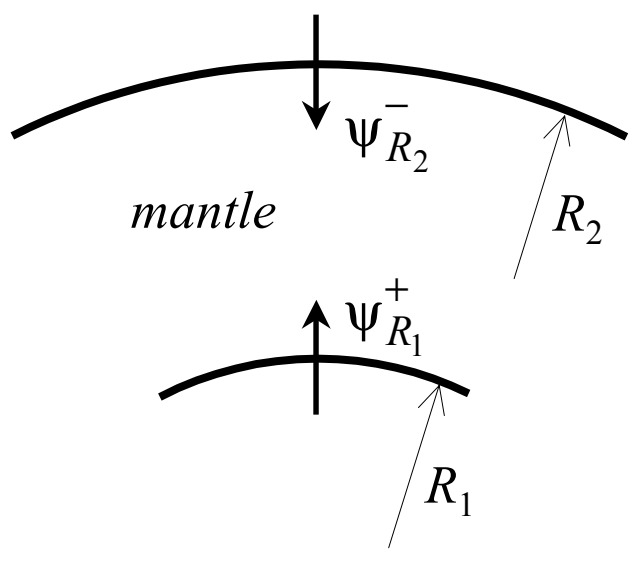

**Figure 1.5**

$$\frac{4\pi\left(R_2^3 - R_1^3\right)}{3}\kappa B = 4\pi\alpha^h \kappa B \int_{R_1}^{R_2} (R+r)^2 \, \mathrm{d}r \quad \Rightarrow \quad \alpha^h = \frac{\left(R_2^3 - R_1^3\right)}{8R_2^3 - (R_2 + R_1)^3}$$

In the limiting case as $R_1 \to 0$, the mantle reduces to a solid sphere and $\alpha^h \to 1/7$, the same factor as in part (*c*).

PART II

# THE DISCRETE ORDINATES SET

## 2.1 Motive

Currently, the discrete angular domain in curved geometry is implemented by approximating the angular derivative with a set of *angular parameters*. The resulting set of discrete ordinates equations is not closed: there is one unknown more than there are equations. Consequently, the solution of such a set of equations is initiated with a starter solution not belonging to the set of discrete ordinates under consideration. The result is a solution for the total intensity that is not monotonic[❅] within a homogeneous isothermal sphere. That solution is reported graphically with a dip at the center of a homogeneous pure scatterer sphere with a uniformly distributed source and the dip is described non-physical, but no rigorous proof is provided to support or to invalidate this assumption. An ad hoc remedy consisting of determining *starter* direction cosines other than −1.0 to cure the dip was proposed and the direction cosine is allowed to take values of magnitudes larger than 1.0 [5]. No attempts were made, so far, to address the closure of the set of discrete ordinates equations.

Our efforts in this endeavor are directed toward developing a set of discrete ordinates that makes the set of equations complete. The analytic solution for these equations is necessarily an approximation by way of introducing the discrete ordinates. It was pointed out that in the limit as the number of discrete ordinates becomes very large we expect that the solution becomes numerically the exact solution to the problem at hand [1]. In what follows, for the interest of avoiding repeating cumbersome expressions we will isolate the terms of the angular derivatives of eqs. (1.1) and (1.3), and focus our analysis on these derivatives. This shortcut approach is legitimized by the fact that the spatial quantities of the equations remain exact in their native form.

## 2.2 A new structure of the discrete ordinates set

We first define our set of discrete ordinates using the traditional set of weights and abscissas of GL quadrature:

$$S_{2N} \equiv \{w_n, \mu_n : w_n \in (0, 1), \mu_n \in (0, 1), n = 1, 2, \ldots, N\} \qquad (2.1)$$

Our choice of GL quadrature is merely a convenience. The user is free to adopt a quadrature of his choice provided that the abscissas are symmetric on [−1, +1]. The GL quadrature is described as the best achievable approximation for the integral of a function such as $f(\mu)$ if this function is not a polynomial [1].

We label the ordinates so that $\mu_1 < \mu_2 < \ldots < \mu_N < 1.0$ and we order them in the sequence shown in figure 2.1. By this arrangement, the inequality $\mu_n > \mu_{n-1}$ is always true for all centripetal and centrifugal quantities. The superscripts plus and minus shown in figure 2.1 are merely descriptive of the direction of flow of energy relative to the origin of the polar coordinates,

[❅] improperly described as *asymmetry* by many authors.

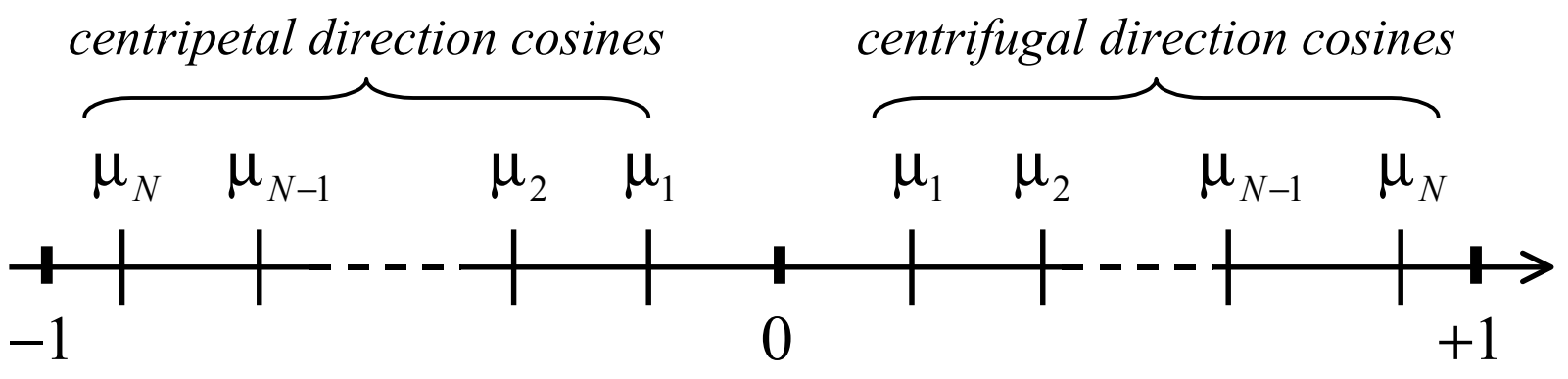

**Figure 2.1.** Layout of the angular grid for the centripetal and the centrifugal direction cosines. The boundary values ±1 are actually references for the actual boundary limits for the direction cosines. None of the direction cosines should be construed as a negative quantity.

which is the center of the sphere of figure 1. All direction cosines are positive as specified by the set $S_{2N}$ and the weights are normalized so that:

$$\sum_{n=1}^{N} w_n = 1$$

We define a *discrete ordinates derivative* on $S_{2N}$ the operator $\delta/\delta\mu_n$ where $\delta\mu_n$ is a *discrete differential entity* about the direction cosine $\mu_n$, not to be confused with the traditional finite difference $\Delta\mu = \mu_n - \mu_{n-1}$. We find the worth of $\delta\mu_n$ from the very definition of Gauss quadrature: if $f(\mu)$ is a polynomial of degree no greater than $4N-1$, then at some point $\mu_n \in [a, b] \subset (-1, +1)$, we must have exactly:

$$w_n f(\mu_n) \equiv w_n f_n = \int_a^b f(\mu)\,\mathrm{d}\mu \tag{2.2}$$

That is to say, find a rectangle of width $w_n$ about $\mu_n \in [a, b]$ such that neither $w_n$ is equal to the difference $\Delta\mu$ nor is $\mu_n$ the midpoint of the closed interval $[a, b]$, but the area of the rectangle is exactly the area under the curve bounded by $a$ and $b$. By no means $f_n$ is intended to be an average value of $f(\mu)$ on $[a, b]$ and that is demonstrated by the omission of $1/(b-a)$ as a factor of the integral of (2.2). It follows that if we wish to integrate the discrete ordinates derivative of our function $f(\mu)$ on the same domain $[a, b]$, we must have:

$$w_n \left.\frac{\mathrm{d}f}{\mathrm{d}\mu}\right|_{\mu_n} = w_n \frac{\delta f_n}{\delta\mu_n} = \int_a^b \mathrm{d}f = f_b - f_a \tag{2.3}$$

The identity (2.3) provides that, if $w_n$ is a weight of a gaussian quadrature the differential $\delta\mu_n$ can no longer be set as a finite difference $\Delta\mu_n = b - a$. In order to satisfy identity (2.3) it is imperative to set $\delta\mu_n = w_n$. This identity implies that $w_n$ is a gaussian weight but $\delta\mu_n$ is not the difference between two abscissas and hence it is not known *a priori*. It follows that the permutation $w_n = \delta\mu_n$ is not permitted for then it would imply that $\delta\mu_m$ is known *a priori* and then we define the weight from it; in GL quadrature the ordinates $\mu$, and hence $\Delta\mu$ are known, but $\delta\mu$ is not. Therefore, for gaussian discrete ordinates that satisfy identity (2.3) we must have:

$$\delta\mu_n = w_n \quad \Rightarrow \quad \delta f_n = f_n - f_{n-1} \tag{2.4}$$

The right side identity of (2.4) implies that $\mu_n$ is the upper limit $b$ of integral (2.3) and $\mu_{n\text{-}1}$ is the lower limit $a$. It follows that the angular derivative of $f$ transforms into a discrete ordinates derivative by the representation:

$$\left.\frac{\partial f}{\partial \mu}\right|_{\mu_n} \rightarrow \frac{f_n - f_{n-1}}{w_n} \tag{2.5}$$

If we extend the application of (2.3) to the angular derivative term of eq. (1.1), we will have:

$$\int_a^b \nu^2 \partial_\mu \psi \, d\mu = w_n \left[\nu^2 \partial_\mu \psi\right]_n \neq \nu_n^2 \left(w_n \left.\partial_\mu \psi\right|_n\right) \tag{2.6}$$

The inequality expressed by (2.6) is instructive about why eq. (1.1) is not conservative under quadrature: the weight $w_n$ used for the purpose of defining $\delta\mu_n$ as in identity (2.4) is not commutative with respect to the multiplication, that is, $w_n \nu_n^2 \neq \nu_n^2 w_n$. We have to think of $w_m$ as a non-commutative operator that operates on the operand nested inside the square brackets of eq. (2.6), which is the ensemble of the coefficient of the derivative and the derivative itself. In such a case we should have:

$$\int_a^b \nu^2 \partial_\mu \psi \, d\mu = w_n \left[\nu^2 \partial_\mu \psi\right]_n = w_n \frac{\nu_n^2 \left.\partial_\mu \psi\right|_n - \nu_{n-1}^2 \left.\partial_\mu \psi\right|_{n-1}}{w_n} \tag{2.7}$$

The difficulty in resolving the rightmost side of eq. (2.7) is that the angular derivatives are not discrete ordinates derivatives. They are differential calculus derivatives evaluated at the μ-values specified by the indicated indices, i.e., they are fixed numbers sitting there, neither they are known nor can we apply the transformation (2.5) on them.

In order to be able to work with eq. (1.1) with discretized angular derivative, we would have to define the set of discrete ordinates along with its appropriate group structure. Such a set structure should assign to the weight $w_n$ of eq. (2.6) a property that would make it an operator on the discrete domain equivalent to the integration by parts of differential calculus. While such a group structure is technically achievable, it is not worth the efforts. Equation (1.3) is mature enough and ready to deliver the desired product. Indeed, if we choose the function $f$ of identity (2.4) to be the angular derivative of eq. (1.3), then the definition expressed by (2.4) applies directly:

$$w_n \frac{\delta\left[\nu^2 \psi\right]_n}{\delta\mu_n} = \delta\left[\nu^2 \psi\right]_n = \nu_n^2 \psi_n - \nu_{n-1}^2 \psi_{n-1} \tag{2.8}$$

which is the proper quadrature of the angular derivative on the specified domain, $[a, b]$ at present. The quadrature is simple, direct and needs not the additional computational labor that would be needed to quadrature the angular derivative term of eq. (1.1). For this reason we will continue working formally with eqs. (1.3) and (1.2). However, in order to be consistent with our analysis we will be working actually with eqs. (1.9) and (1.10) whenever the $s$-frame is appropriate.

There is still one bump in the way of the quadrature of the angular derivative: the closure under quadrature. That is, the quadrature of the angular derivative of eq. (1.3) must match the integral of the same:

$$\sum_{n=1}^{2N}\left(w_n\frac{\delta\left[v^2\psi\right]_n}{\delta\mu_n}\right)=\int_{-1}^{1}\partial_\mu\left[v^2\psi\right]d\mu \qquad (2.9)$$

We first notice that the factor 1/2 in front of the integral is omitted, as well as in front of the summation. This omission is benign at this stage of derivations.

It is readily apparent that the integral of the right side of eq. (2.9) vanishes and, therefore, the sum of the left side must vanish as well. Expand the summation of eq. (2.9) following the indices pattern shown in figure 2.1 and make use of identity (2.8), all the terms cancel out except those affected by the index $N$. Equating this result with the vanishing integral of eq. (2.9) yields:

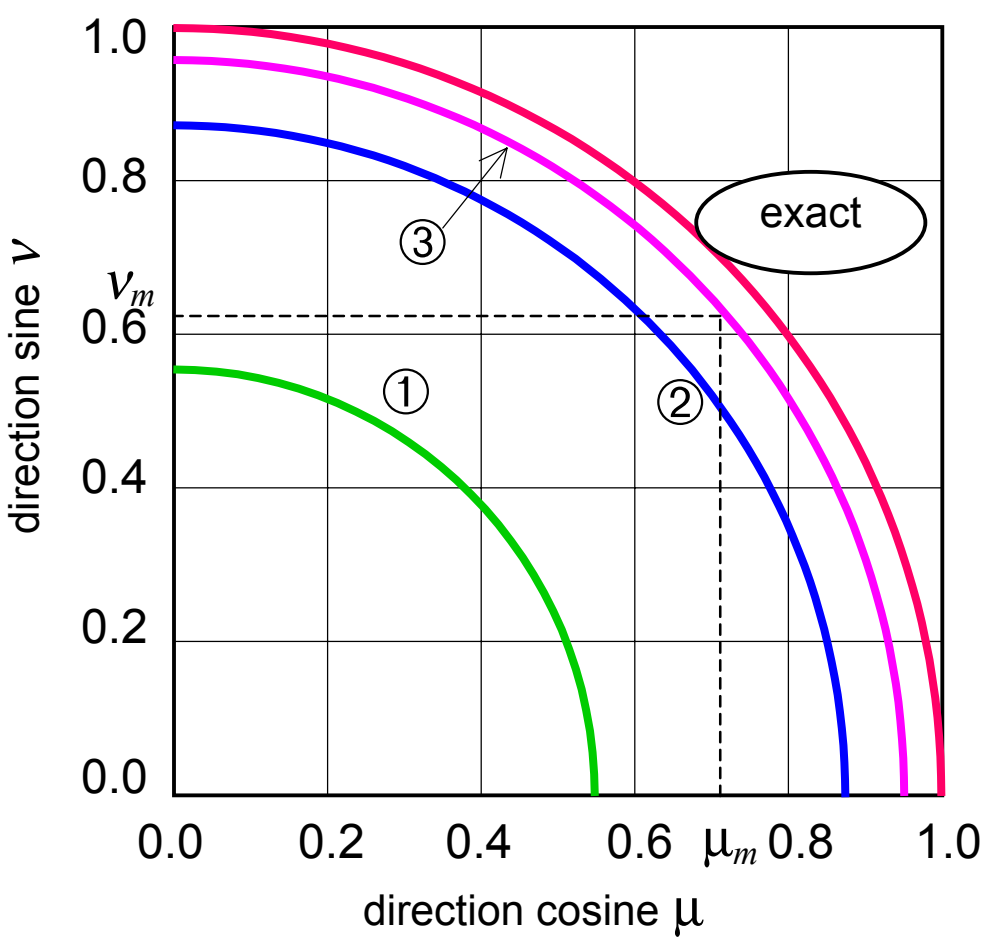

**Figure 2.2.** Graphs of DOT circles for a few values of $N$ shown in circles. The exact graph is a circle of radius 1 in $S_\infty$.

$$v_N^2\left(\psi_N^+ + \psi_N^-\right) = 0 \qquad (2.10)$$

We recognize that in the asymptotic limits as $N \to \infty$ the identity (2.10) vanishes by reason of asymptotic direction cosine $\mu_\infty = 1 \Rightarrow v_\infty = 0$. It follows that for any finite $N$ the nullity of identity (2.10) must be postulated because $\mu_N$ is never equal to 1. Also, the proposition $\mu_N = 1$ must be excluded simply because this value of direction cosine is not from the set $S_{2N}$ defined by (2.1). A direct approach to resolve this paradox is to work with un-normalized circular functions. This is tantamount to define a *discrete ordinates trigonometry* (DOT) so that the circular functions on the unit circle map onto circular functions on the DOT circle. To accomplish this we choose the radius of our trigonometric circle equal to $\mu_N$ and then discretize the circular functions on the new circle. We will have the following DOT identities:

$$v_n^2 = \mu_N^2 - \mu_n^2 \quad \Rightarrow \quad v_N = 0 \qquad (2.11)$$

Therefore, the following transformation of the sine function applies:

$$\left.\begin{matrix} \textit{Unit circle} \\ 1-\mu^2 \end{matrix}\right\} \rightarrow \left\{\begin{matrix} \textit{DOT Circle} \\ \mu_N^2-\mu_n^2 \end{matrix}\right. \qquad (2.12)$$

By this transformation the unit trigonometric circle transforms into another DOT circle. Figure 2.2 illustrates this transformation for different orders of discrete ordinates. It follows that the set of discrete ordinates defined by (2.1) along with identities (2.4) and transformation (2.12) is complete.

Another thing we took care of when we expanded the summation (2.9): the discrete ordinates derivative about $\mu_1$. In accordance with (2.4) we would have to write:

$$\delta v^2\psi\Big|_1 = v_1^2\psi_1 - v_0^2\psi_0 \qquad (2.13)$$

Considering that $v_n$ maps onto the set $S_{2N}$ by the circular relation (2.11), this mapping does not exist for $n = 0$. This is because the index $n = 0$ does not belong to the set $S_{2N}$ defined by (2.1). Therefore, if $\delta\nu\psi|_1$ is to be defined on $S_{2N}$, the term in $\nu_0$ of identity (2.13) must be dropped out notwithstanding any other considerations. With that, we will be left with only one significant term: $\delta\nu\psi|_1 = \nu_1^2\psi_1$. By doing so the integrity of the definition of the discrete ordinates derivative is preserved. The quantity $\delta\mu_1 = w_1$ is still completely specified about $\mu_1$ on $S_{2N}$.

Finally, the total intensity and the flux at any given point in the radiation field can be obtained from the specific intensities by quadrature using the set $S_{2N}$:

$$\varphi_r = \frac{1}{2}\sum_{n=1}^{N} w_n\left(\psi_{r,n}^{+} + \psi_{r,n}^{-}\right) \tag{2.14}$$

$$F_r = \frac{1}{2}\sum_{n=1}^{N} w_n\mu_n\left(\psi_{r,n}^{+} - \psi_{r,n}^{-}\right) \tag{2.15}$$

The specific intensities of eqs. (2.14) and (2.15) will be calculated from the transfer equations (1.10) discretized on the angular domain in $S_{2N}$ discrete ordinates. These solutions are discussed in Part III and Part IV.

PART III

# THE INCOMPLETE DIFFUSION MODEL

## 3.1 Prolog

The natural school of thought preaches evolution from simple to complex. In line with this natural wisdom we begin our search for the solution for the transfer equation by first solving the equation in its simplest form: the diffusion equation. This may not be a sophisticated approach by some measures, but it is insightful and instructive about our handling of the spatial variable and the angular derivative.

A number of diffusion models based on the principal theme of the asymptotic diffusion theory [7] [ 8] were developed over the years. They are expressed as a second order equation for the total intensity commonly known as Helmholtz diffusion equation [9] [10] [11]. They differ in the expression for the diffusion coefficient. For this reason we will refer to these diffusion theories as Helmholtz's diffusion. In some instances the diffusion equation is expressed as a coupled set of two first order equations for the total intensity and the flux but the set is convertible to a second order Helmholtz equation. The principal shortcoming of Helmholtz diffusion model in spherical geometry is that the solution varies as $1/r$. An exception is the case of a cold absorber sphere: the diffusion intensity varies as $1/r^2$, as it should, provided that Ficks' law prevails.

Our pursuit to developing a diffusion equation and providing its general analytic solution for monochromatic radiation is motivated by the fact that the total intensity in a sphere with central symmetry is independent of the angular derivative of the transfer equation. Our efforts in this endeavor lead to a set of first order diffusion equations, as well as one second order diffusion equation that is not a Helmholtz's, results that should not be left out without a fair address. The first order equations are particularly delightful because of their simplicity and accuracy in central symmetry problems. They are most useful for the diagnostics of the imploded pellet.

## 3.2 The first order diffusion equations

### (*a*) *Isotropic diffusion*

Let $\psi^+$ and $\psi^-$ be average values in $\mu\in[0, 1]$. By the mean value theorem there is one value of $\mu$ for which $\psi(\mu) = \psi_{avg}$, that is, the average intensity is indeed an intensity in one properly defined direction represented by $\mu_a$. We define a diffusion intensity the quantity $\phi = (\psi^+ + \psi^-)/2$ in $\mu_a$. Clearly, this direction cosine cannot be defined uniquely for $\forall r\in[0, \infty]$. A compromise is the traditional way to go. This is accomplished by choosing an average direction cosine $\bar{\mu}$ for the diffusion intensity $\phi$ that preserves the integrity of the conservation relation. A number of recipes can be used to obtain an average direction cosine ranging from intensity weighted to the simplest ones:

- The logarithmic mean value: $$\bar{\mu} = \exp\left[\int_0^1 \log\mu \, d\mu\right] = e^{-1}$$

This average direction cosine sets the diffusion intensity in a direction about 68° off the radial direction. We label such a diffusion as the *logarithmic* diffusion model.

- The $n$th root mean value: $\bar{\mu} = \sqrt[n]{\int_0^1 \mu^n d\mu} = \sqrt[n]{\frac{1}{n+1}}$

  $n = 1 \Rightarrow \bar{\mu} = 1/2$. This is the *mean value* diffusion model.

  $n = 2 \Rightarrow \bar{\mu} = 1/\sqrt{3}$. This is the *root mean square* (RMS) diffusion model. It is the angular basis of Helmholtz's diffusion and the $S_2$ discrete ordinates of Part II.

- $n \to \infty \Rightarrow \bar{\mu} = 1$. This is the *normal* diffusion model. The diffusion intensity is the same as the flux and it is defined in the radial direction, normal to the spherical surface. This model has been reported as a particular case of the *flux-limited diffusion theory* [9]. It is also the asymptotic solution of the $S_\infty$ discrete ordinates representation of the asymptotic transfer equation in the point source configuration.

These diffusion theories, and any other theories based on the average direction cosine are *isotropic*: the intensity at any point in space has the same value in any direction notwithstanding its definition in only one direction cosine.

### (*b*) *The incomplete diffusion: definition*

The principal cause that fails the isotropic diffusion theories lies in the very basic hypothesis upon which they are built: the intensity in an average direction cosine is not the same as the average intensity in the same interval of direction cosine: $\psi_{avg} \neq \psi(\mu_{avg})$. We project the diffusion theory in a different perspective. If $\mu$ is a direction cosine on [0, 1], there is always one centripetal intensity $\psi^-$ and one centrifugal intensity $\psi^+$ both of them defined in this direction cosine in azimuthal symmetry. We define our *diffusion intensity* in $\mu$ the same way a diffusion density is defined in the kinetic theory of gases, specifically: $\phi = (\psi^+ + \psi^-)/2$. Clearly, there are an infinite number of different diffusion intensities on the spectrum of direction cosines $\mu \in [0, 1]$. By this characterization this diffusion intensity is *anisotropic*. The integral sum of $\phi$ on the direction cosine $\mu \in [0, 1]$ yields the total intensity $\varphi$.

We define the discrete ordinates diffusion intensity $\phi_n$ in an identical manner on the set $S_{2N}$ and the total intensity $\varphi$ is obtained by the quadrature (2.14) on the diffusion intensity. We simplify this diffusion model further. The specific intensities are defined in only one direction at a time. Therefore, their angular derivatives must vanish. By this simplification the coupling between intensity in a given direction and the intensity in its neighboring direction is destroyed. The result is an *incomplete diffusion model* the object of our analysis. This diffusion model is valid whenever the radiation field is defined in all directions on $\mu \in [-1,+1]$ in central symmetry. In what follows, the subscript $n$ for the $n$th direction will be omitted for notational convenience except when it is expressly needed to avoid confusion. The subscript $n$ is implied whenever any one of the symbols $\psi$, $\phi$, $\mu$, $\lambda$ or $f$ is encountered; $f$ is for the diffusion flux and $\lambda$ is an opacity to be defined later.

### (*c*) *The first order equations*

In order to construct a mathematical prescription for our diffusion model we turn to the transfer equation and we attempt to clear it from its angular derivative. We cannot remove the angular derivative from eqs. (1.8), that will purge out the curvature signature from the equation. By contrast, removing the angular derivative from eqs. (1.9) does not jeopardize the curvature integrity of the equations. For this reason we define our diffusion equations from eqs. (1.9) to be:

$$\partial_r \psi^+ + \left(\frac{2}{\vartheta} + \lambda\right)\psi^+ = \lambda B \tag{3.1}$$

$$\partial_r \psi^- + \left(\frac{2}{\vartheta} - \lambda\right)\psi^- = -\lambda B \tag{3.2}$$

where $\lambda = \kappa/\mu$. These two equations are exact in the phase space ($r$, $\mu$) and need not be solved as discrete ordinates equations. However, because their exact solutions are exponential integrals in $d\mu$ that cannot be carried out in closed form, a numerical integration algorithm becomes mandatory for quantitative analysis of the solutions. Most numerical integration algorithms could accomplish the job, but the least expensive, as we indicated earlier, is the GL quadrature and that brings us back to the discrete ordinates formalism. For this reason we follow the short route and treat these two equations as if they were discrete ordinates equations. The complete treatment of the discrete ordinates equations proper is discussed in Part IV.

The diffusion equations (3.1) and (3.2) are a valid prescription for the radiative transfer problem when both of them have to be solved simultaneously. This is always the case when the medium is a black body emitter, or a cold medium exposed to impinging intensity from opposite boundaries, such as a sphere. None of them standing alone is valid. We examine the solutions of these equations in a few instructive configurations.

## 3.3 Cold medium

### (*a*) *Mantle surrounded by vacuum*

In this configuration the core is a source characterized by its radius $\varepsilon$ and an isotropic and uniformly distributed radiance $\psi_\varepsilon^+$. It is planted in a cold absorber medium of outer radius $R$ that can be finite or infinite (fig. 1.1). We assume further that the boundary of the medium is a free surface exposed to vacuum. Therefore, only centrifugal energy flow is to be considered and the intensity is described by the homogeneous part of the centrifugal equation (3.1) for $r \geq \varepsilon > 0$. The restriction $\varepsilon > 0$ is imperative by the definition of this configuration for if $\varepsilon = 0$ the problem is not defined; a mantle must have a non-zero inner radius. For this reason we do not need to apply a shift transformation to the equation. The $c$-frame is adequate to properly address this special case problem. In this case eq. (3.1) transcribes into:

$$\partial_r \psi_r^{+c} + \left(\frac{2}{r} + \lambda\right)\psi_r^{+c} = 0 \qquad R \geq r \geq \varepsilon > 0 \tag{3.3}$$

The solution of this equation standing alone is intrinsically erroneous for the reasons we indicated in sec. 3.2(*b*). The purpose of carrying out the solution is merely to generate data for comparison with the discrete ordinates equation discussed in Part IV. The solution is given as:

$$\psi_r^{+c} = \psi_\varepsilon^{+c} \exp\left[-\int_\varepsilon^r \left(\frac{2}{x}+\lambda\right)\mathrm{d}x\right] = \psi_\varepsilon^{+c}\frac{\varepsilon^2}{r^2}e^{-\lambda(r-\varepsilon)} \tag{3.4}$$

The total intensity is obtained by using the quadrature (2.14):

$$\varphi_r^c = \frac{\psi_\varepsilon^{+c}}{2}\frac{\varepsilon^2}{r^2}\sum_{n=1}^N w_n e^{-\lambda_n(r-\varepsilon)} \tag{3.5}$$

The solution (3.5) is often sought in terms of the strength of the source: $q_\varepsilon = 4\pi\varepsilon^2\psi_\varepsilon^{+c}$. With this expression for $q_\varepsilon$ eq. (3.5) simplifies to:

$$\varphi_r^c = \frac{q_\varepsilon}{8\pi}\frac{1}{r^2}\sum_{n=1}^N w_n e^{-\lambda_n(r-\varepsilon)} \tag{3.6}$$

The expression for the flux is readily obtained:

$$F_r^c = \frac{q_\varepsilon}{8\pi}\frac{1}{r^2}\sum_{n=1}^N w_n\mu_n e^{-\lambda_n(r-\varepsilon)} \tag{3.7}$$

Numerical experiments with the expression (3.5) have shown that the summation on the discrete ordinates saturates fairly quickly. A graphical sample is shown in figure 3.1 in a mantle of outer radius $R$ = 11, with a core of radius $\varepsilon$ = 1 and isotropic radiance normalized to 1. The graphs generated with $S_{10}$ and up to $S_{8000}$ are undiscernable on this graph. The explanation for this behavior lies in the coefficient $\lambda_n$: it becomes very large with small $\mu_n$ that are obtained with large $N$, thus causing the corresponding exponential to die out quickly, not enough to contribute graphically to the summation over $N$. The graph labeled $N$ = 1 in figure 3.1 is representative of the intensity in $S_2$ discrete ordinates, a RMS diffusion intensity.

### (*b*) *The point source problem*

The traditional treatment of the point source problem yields a mythical singular intensity at the origin. Traditionally, the solution is systematically constraint to the power of the source: $q$ is used for the boundary condition. The proper boundary condition must be the boundary value of the function $\psi$ of the differential equation, i.e., the radiance of the source is the natural boundary condition. The power of the source is *assistant* boundary condition. It can be used in *lieu* of the natural boundary condition for practical purposes if it is properly related to the natural boundary condition as illustrated with eq. (3.6), but not if it is derived from the traditional condition:

$$\lim_{r\to 0}(4\pi r^2 F) = q \tag{3.8}$$

where $F$ is the flux and $q$ is the strength of the source. The right side of expression (3.8) is deceptive: it is missing the limit condition on the $q$ term. It should read:

$$\lim_{r\to 0}(4\pi r^2 F) = \lim_{r\to 0} q \tag{3.9}$$

To address the validity of this condition we adopt the empty space postulate:

*Empty space does not generate energy*

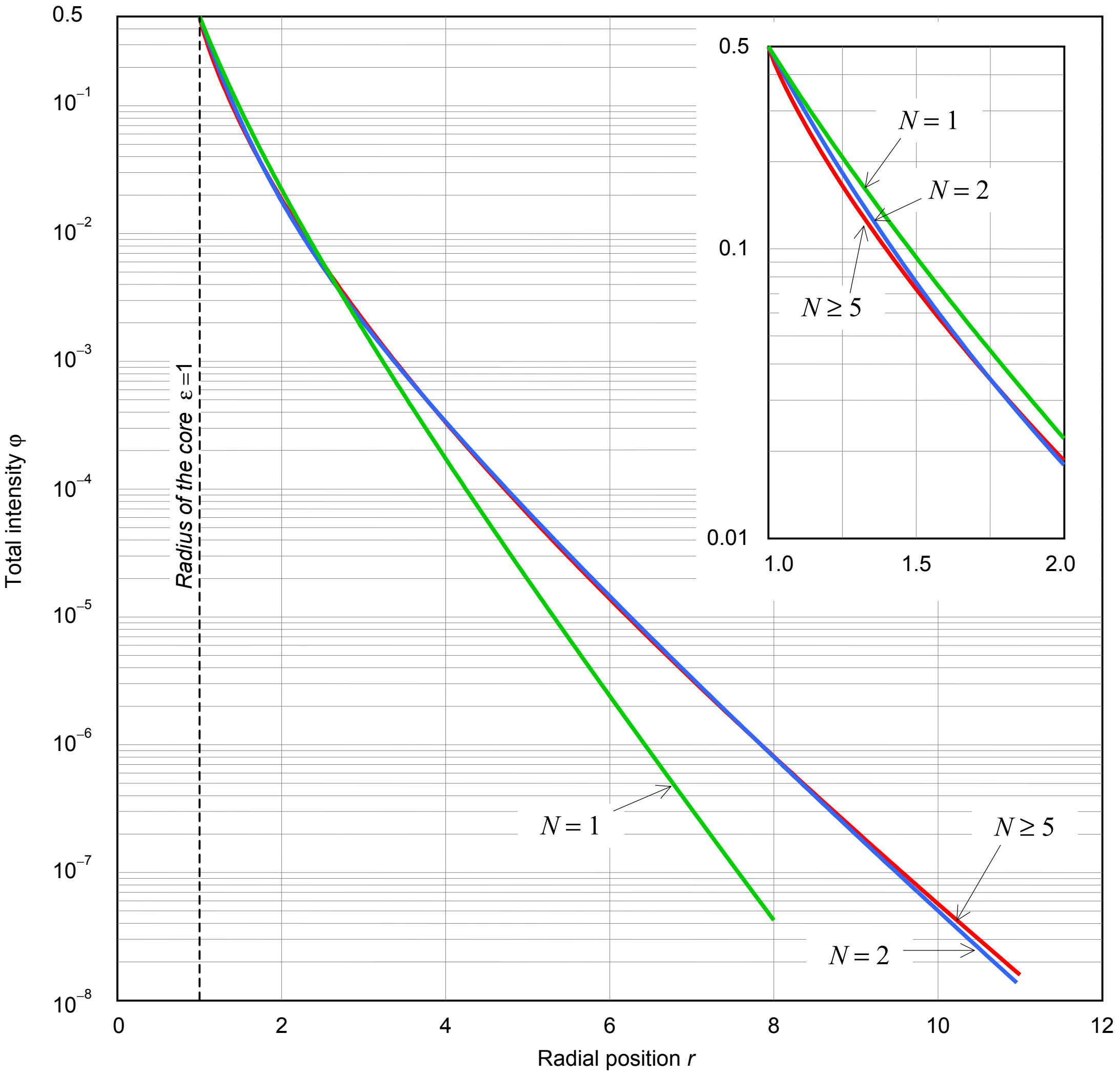

**Figure 3.1**. Graphs of the total intensity in a cold mantle of opacity κ = 1. The graphs are generated using the incomplete diffusion model. The label *N* indicates the order of discrete ordinates.

As $r \to 0$ the source is reduced to a geometric point, an empty element ∅ of space. By the empty space postulate such a source does not emit radiation. Unfortunately, the traditional practice is to maintain a defiant non-zero value for $q$ at $r \to 0$. By this practice the intensity of the point source problem is made singular at the origin by design. The remedy to this situation is to abandon the practice of using the source as a principal boundary condition, whether by way of the limit expressed by (3.8), or by the injection of a delta function source.

We approach the problem by rendering the mathematical image of the problem to represent the physical configuration. We represent the source of radiation by a sphere[✧] of arbitrary radius ε, however large or small it may be[✥]. The characterization *point source* is a token for a pair of

✧ The choice of a sphere is dictated by the geometry of the problem at hand: *central symmetry*. For a point source problem there is no practical difference between a spherical source and a source of arbitrary shape.
✥ This is a direct denial of the validity of the delta function source.

source-detector configuration where the detector is placed at a distance from the surface of the source much larger than the characteristic radius of the source. An equivalent statement for this condition is that the source angular diameter measured at the detector position must be very small. That is, eq. (3.5) applies to the point source problem provided it is properly constrained to the asymptotic condition:

$$r >> \varepsilon \tag{3.10}$$

This asymptotic condition inhibits the limit $r \rightarrow \varepsilon$ and *a fortiori* $r \rightarrow 0$. Such inhibition annuls the legitimacy of the limit condition (3.8) and the delta function source.

An alternative expression for the point source problem is to consider the case of a spherical source of radius $\varepsilon \rightarrow \xi > 0$ and $\xi$ is an arbitrarily small number. In this limiting configuration it is reasonable to assume that all the energy generated within the $\xi$-sphere reaches the surface. If $\eta$ is the energy emission rate per unit volume, then the radiance of the $\xi$-sphere becomes:

$$\psi_\varepsilon^{+c} = \eta \frac{4\pi\varepsilon^3}{3} \frac{1}{4\pi\varepsilon^2} = \frac{\eta\varepsilon}{3} \tag{3.11}$$

Insert eq. (3.11) into eq. (3.4) to obtain:

$$\psi_r^{+c} = \frac{\eta\varepsilon}{3} \frac{\varepsilon^2}{r^2} e^{-\lambda(r-\varepsilon)} \qquad r \geq \varepsilon \rightarrow \xi \tag{3.12}$$

A point in case is that if we bypass the restriction $r >> \varepsilon$, then setting $r = \varepsilon$ and letting $\varepsilon \rightarrow 0$, that makes the source to fade away and the intensity is lost in an empty space. In all accounts we emphasize that if the diameter of the $\xi$-sphere is larger than a tiny fraction of a mean free path of the material of the source, the expression (3.11) is no longer valid. The radiance $\psi_r^+$ has to be determined from other considerations. However, whenever the asymptotic limit $r >> \varepsilon$ is satisfied, the expression (3.7) is exact. Clearly, in this limit $\varepsilon$ can be dropped from the exponential and the opacity $\lambda \rightarrow \kappa$. The resulting expression becomes congruent with the expression of the flux predicted by the flux-limited diffusion theory [9], provided that the strength of the source is defined as $Q = \eta\varepsilon^3/3$. It follows that the flux-limited diffusion is valid far away from the surface of the source and far away from the boundary of a finite mantle.

### (*c*) ***Solid homogeneous sphere***

#### *Definition of the problem*

We assume that the sphere of radius $R$ is a cold homogeneous medium characterized by its opacity $\lambda$ and immersed in a uniform isotropic radiation field of intensity $\varphi_{medium} = \left(\psi_R^- + \psi_R^+\right)/2$ normalized to 1. Consequently the intensity impinging at the surface of the sphere is $\psi_R^-$, which is the natural boundary condition for this problem. In this configuration the radiation field in the sphere is a competing contribution from both the centripetal and the centrifugal intensities: a centripetal intensity at one side of the sphere becomes centrifugal intensity at the opposite side. Therefore, the homogeneous parts of both of the equations (3.1) and (3.2) will have to be solved simultaneously:

$$\partial_r \psi_r^{+c} + \left(\frac{2}{\vartheta} + \lambda\right) \psi_r^{+c} = 0 \tag{3.13}$$

$$\partial_r \psi_r^{-c} + \left(\frac{2}{\vartheta} - \lambda\right)\psi_r^{-c} = 0 \tag{3.14}$$

These two equations are manifestly uncoupled. In reality they are coupled by the boundary conditions $\psi_R^-$. This coupling is not transparent through the equations but it will unfold in the course of the development of the solutions.

The first thing we notice about eqs. (3.13)-(3.14) is that their derivatives are not equal at the center of the sphere for $\forall\lambda > 0$. Also, we notice that the derivative of the centrifugal intensity is always negative indicating that this intensity function is monotonically decreasing with *r*. By contrast, the derivative of eq. (3.14) vanishes at a saddle radius $r_s$ determined from:

$$\frac{2}{\vartheta_s} = \lambda \quad \Rightarrow \quad r_s = \frac{2}{\lambda} - R \ \geq\ 0 \tag{3.15}$$

It is clear from the expression (3.15) that if $R = 2/\lambda \Rightarrow r_s = 0$, the intensity function has a saddle point at the center of the sphere. In this case the centripetal intensity function is monotonically increasing with *r*. If $R = 1/\lambda \Rightarrow r_s = R$, the saddle point is at the surface of the sphere and the intensity functions is monotonically decreasing with *r*. If $1/\lambda < R < 2/\lambda$ there is a saddle point within the sphere. In that case the centripetal intensity function is monotonically decreasing with $r < r_s$ and monotonically increasing with $r > r_s$. In the former case the intensity increases as the radiation progresses toward the center and saturates to a finite value at the origin. In all accounts, if $R = 0$ there is no problem to solve to begin with: the transfer problem fades away to naught, despite that the two equations (3.13) and (3.14) survive the script of the *c*-frame.

### *The solution.*

With this information at hand we proceed in solving our transfer equations (3.13) and (3.14) in a cold homogeneous solid sphere. The mathematical solution of (3.13) with boundary condition at the center of the sphere is given by:

$$\psi_r^{+c} \ =\ \exp\left(-\int_0^r \frac{2}{R+x}\,\mathrm{d}x \ -\ \int_0^r \lambda\,\mathrm{d}x\right)\psi_0^c \qquad\qquad R \geq r \geq\ 0$$

Which can be cast in the form:

$$\psi_r^{+c} \ =\ \frac{R^2}{\vartheta^2}\, e^{-\lambda r}\, \psi_0^c \tag{3.16}$$

Similarly, we obtain the mathematical solution for the centripetal equation (3.14):

$$\psi_r^{-c} \ =\ \frac{R^2}{\vartheta^2}\, e^{\lambda r}\, \psi_0^c \tag{3.17}$$

The intensity at $r = 0$, represented by $\psi_0^c$, is identically the same in either centripetal or centrifugal direction. The value of the intensity at the center is obtained by evaluating eq. (3.17) at *R*:

$$\psi_0 \ =\ 4e^{-\lambda R}\,\psi_R^- \tag{3.18}$$

Insert eq. (3.18) into eqs. (3.16) and (3.17) to obtain the complete solution for the centripetal and the centrifugal intensities for $r \in\ [0, R]$:

$$\psi_r^{-c} = \frac{4R^2}{\vartheta^2} e^{-\lambda(R-r)} \psi_R^- \tag{3.19}$$

$$\psi_r^{+c} = \frac{4R^2}{\vartheta^2} e^{-\lambda(R+r)} \psi_R^- \tag{3.20}$$

The outflow intensity at the surface of the sphere is obtained from eq. (3.20):

$$\psi_R^{+c} = e^{-2\lambda R} \psi_R^- = \frac{\psi_0^c}{4} e^{-\lambda R} \tag{3.21}$$

It is apparent from eq. (3.18) that the intensity at the center of a hollow sphere is four times larger than the intensity at the surface. Another property is worth mentioning. Although the sphere is a pure absorber, hardly a reflector, it has an albedo that can be extracted from eq. (3.21):

$$\alpha = e^{-2\lambda R} \tag{3.22}$$

Finally, we compute the diffusion intensity $\phi_r = \left(\psi_r^+ + \psi_r^-\right)/2$ and the flux $f_r = \mu\left(\psi_r^+ - \psi_r^-\right)/2$ from eqs. (3.19) and (3.20):

$$\phi_r^c = \psi_R^- e^{-\lambda R} \frac{4R^2}{\vartheta^2} \cosh \lambda r \qquad R \geq r \geq 0 \tag{3.23}$$

$$f_r^c = -\mu \psi_R^- e^{-\lambda R} \frac{4R^2}{\vartheta^2} \sinh \lambda r \tag{3.24}$$

The total intensity $\varphi$ is obtained by applying the quadrature (2.14) to eq. (3.23):

$$\varphi_r^c = \psi_R^- \frac{4R^2}{\vartheta^2} \sum_{n=1}^{N} w_n e^{-\lambda_n R} \cosh \lambda_n r \tag{3.25}$$

The factor $1/2$ in front of the summation of the quadrature (2.14) is contained in the definition of the hyperbolic cosine of eq. (3.25). Similarly, the flux is obtained by applying the quadrature (2.15) to eq. (3.24):

$$F_r^c = -\psi_R^- \frac{4R^2}{\vartheta^2} \sum_{n=1}^{N} w_n \mu_n e^{-\lambda_n R} \sinh \lambda_n r \tag{3.26}$$

The differential conservation relation (1.11) and the integral conservation relation (1.13) are completely satisfied with either pair of equations (3.23)-(3.24), or (3.25)-(3.26).

Graphical representations of the specific intensities are generated using eqs. (3.19) and (3.20) in a sphere of radius $R = 1$, and for an incident isotropic intensity $\psi_R^- = 1$. Various values of $\lambda$ are considered. Figure 3.2 shows the profile of the centripetal intensity $\psi_r^{-c}$. Saddle points are noticeable on the curves for $1.0 \leq \lambda \leq 2.0$. For $\lambda = 1.0$ the saddle point is at the surface of the sphere and for $\lambda = 2.0$ it is at the center of the sphere. For $\lambda < 1.0$, the centripetal intensity is monotonically decreasing with $r$; it is monotonically increasing with $r$ for $\lambda \geq 2.0$.

Figure 3.3 shows the profile for the centrifugal intensity $\psi_r^{+c}$ for the same values of $\lambda$ used to generate the graphs of figure 3.2. Nothing exceptionally remarkable is shown on these graphs, the centrifugal intensity is a monotonically decreasing function with $r$. The values of the intensities at $r = 0$ are the same as the values of the corresponding intensities of figure 3.2.

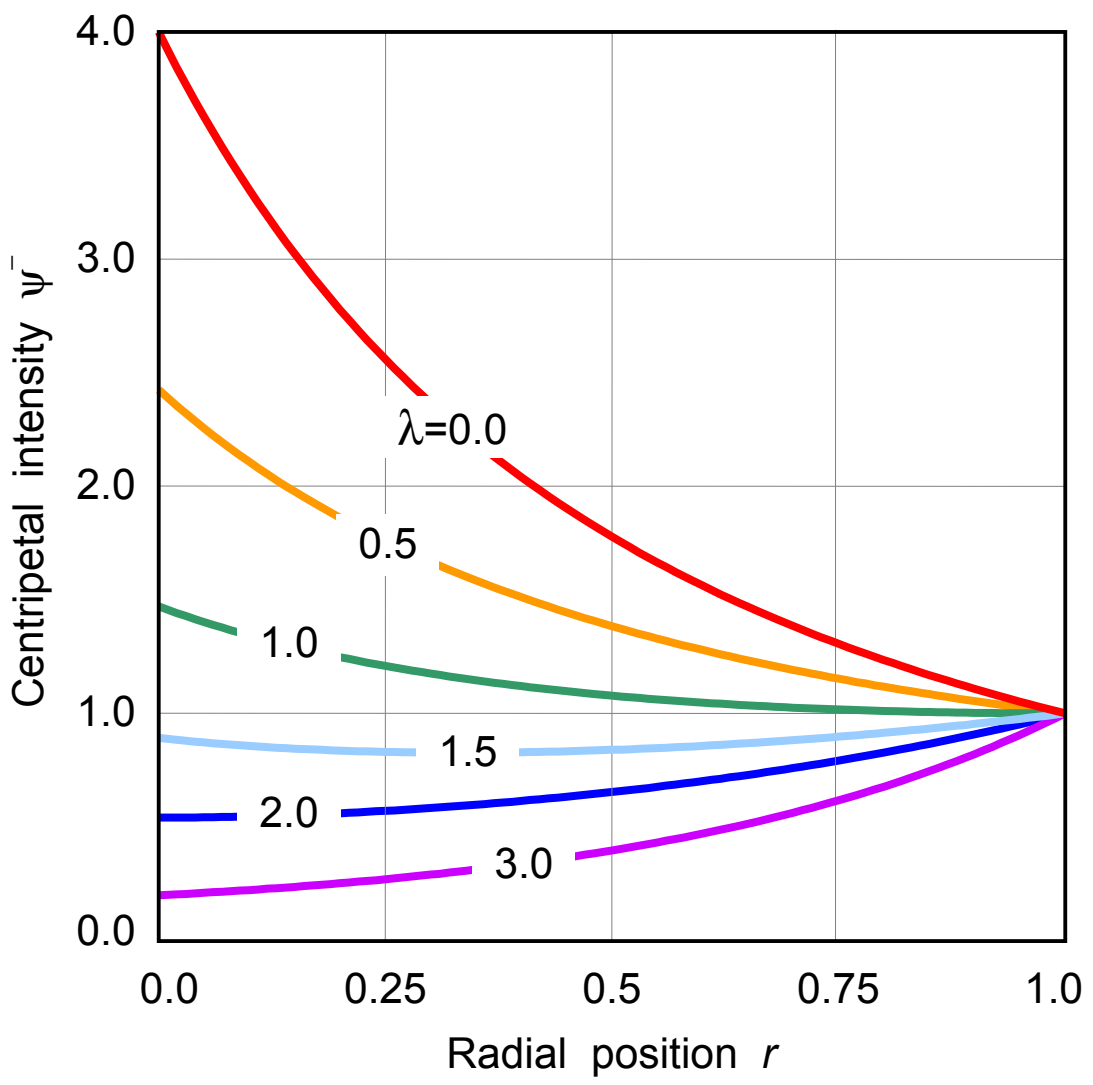

**Figure 3.2.** Grpahs of the centripetal intensity in a cold sphere of unit radius for various values of λ shown on the graph.

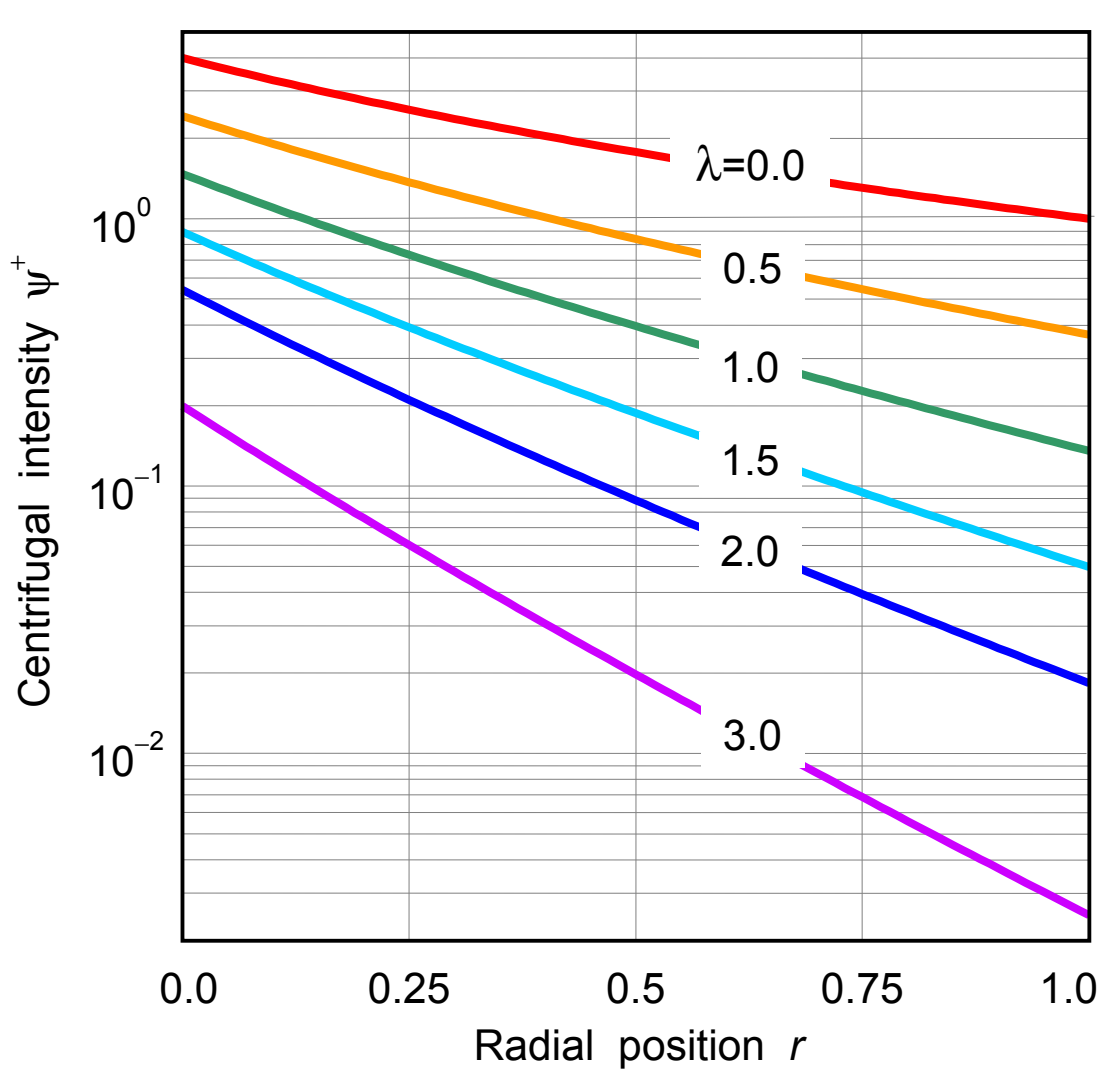

**Figure 3.3.** Graphs of the centrifugal intensity in a cold sphere of unit radius for various values of λ shown on the graph.

The profiles of the total intensity $\varphi_r^c$ of eq. (3.25) and the flux of eq. (3.26) are also produced with different orders of discrete ordinates. Figure 3.4 shows the graphs of the total intensity in a strongly absorbing sphere of $\kappa = 4$. Fluxes and total intensities are also shown in figures 3.5 and 3.6. It is apparent from these graphs that the saddle point is persistent in the total intensity although it is not possible to determine its position analytically considering the $N$ superposition of centripetal and centrifugal intensities. Another thing we learn from these graphs is that the total intensity saturates graphically with only a few abscissas from the set $S_{2N}$. The order of discrete ordinates is shown on the figures.

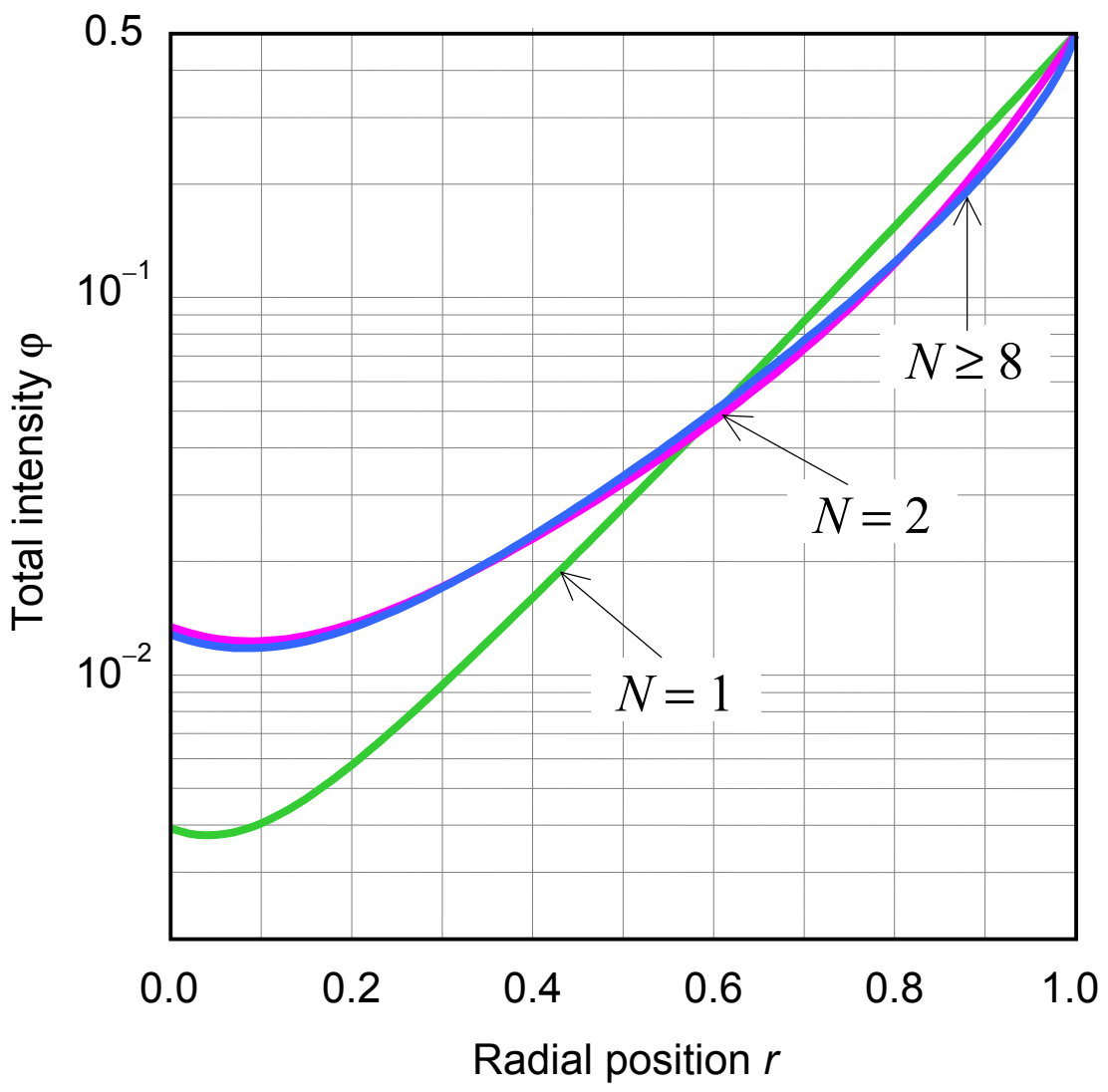

**Figure 3.4**. Graphs of the total intensity in a cold sphere of radius $R = 1$ and opacity $\kappa = 4$, immersed in a uniform isotropic radiation field.

## 3.4 Medium under LTE conditions

### (*a*) *Solid homogeneous sphere*

Under a local thermodynamic equilibrium the medium is an isotropic emitter of radiation. Therefore, the equations (3.1) and (3.2) have to be solved with the inhomogeneous parts included. The homogeneous solutions are:

$$\psi_H^+ = \frac{e^{-\lambda r}}{\vartheta^2} C^+ \; ; \qquad \psi_H^- = \frac{e^{\lambda r}}{\vartheta^2} C^- \tag{3.27}$$

The subscript $H$ stands for homogeneous. These solutions suggest that the particular solutions too would vary as the reciprocal of $\vartheta^2$. Therefore, we elect a particular solution to be a quadratic polynomial in $1/\vartheta$ such as:

$$\psi_P = a + \frac{b}{\vartheta} + \frac{c}{\vartheta^2} \tag{3.28}$$

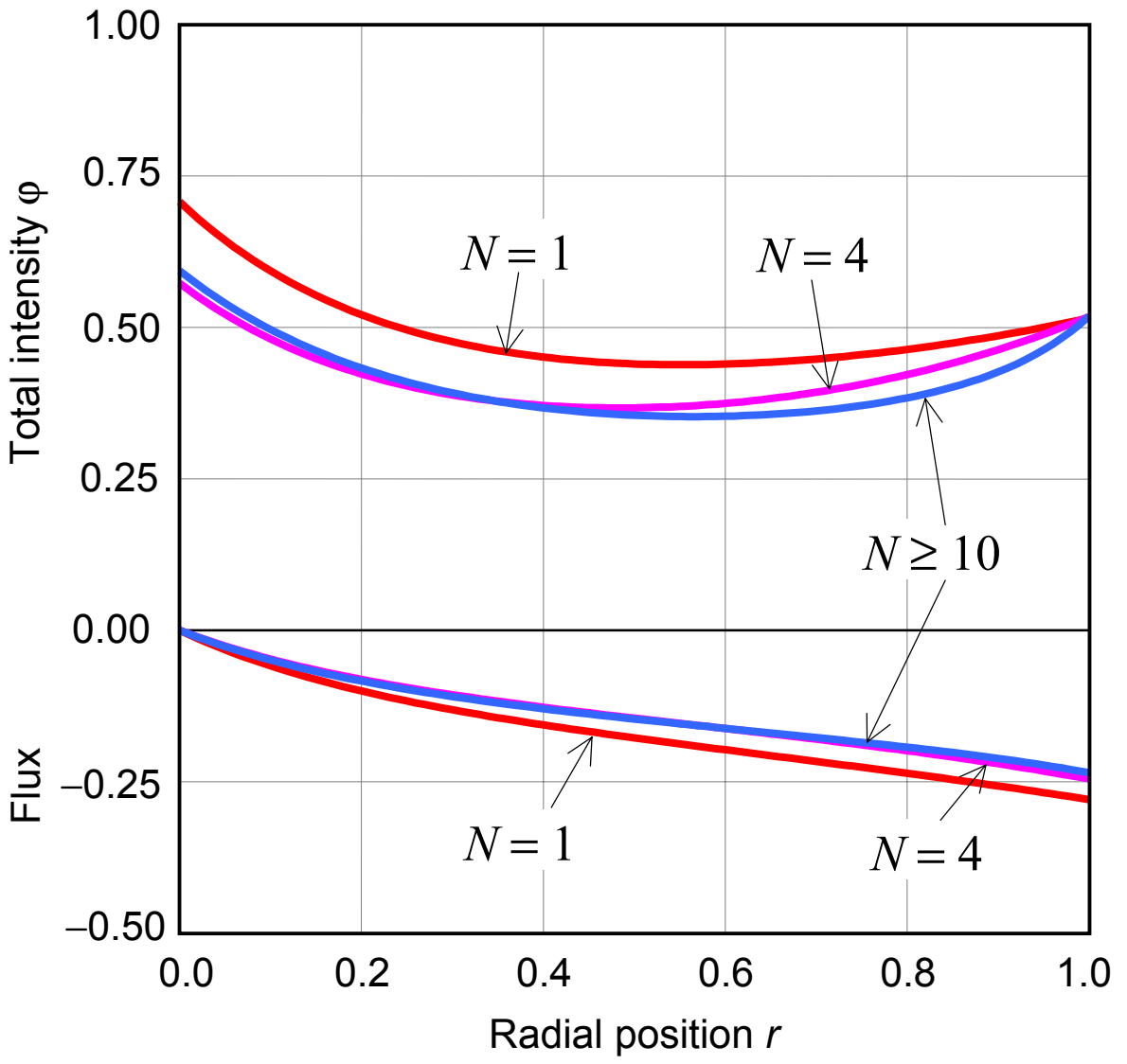

**Figure 3.5**. Graphs of the total intensity and the flux in a cold sphere of radius $R = 1$ and opacity $\kappa = 1$, immersed in a uniform isotropic radiation field. The labels $N$ are the number of discrete ordinates used to generate the graphs.

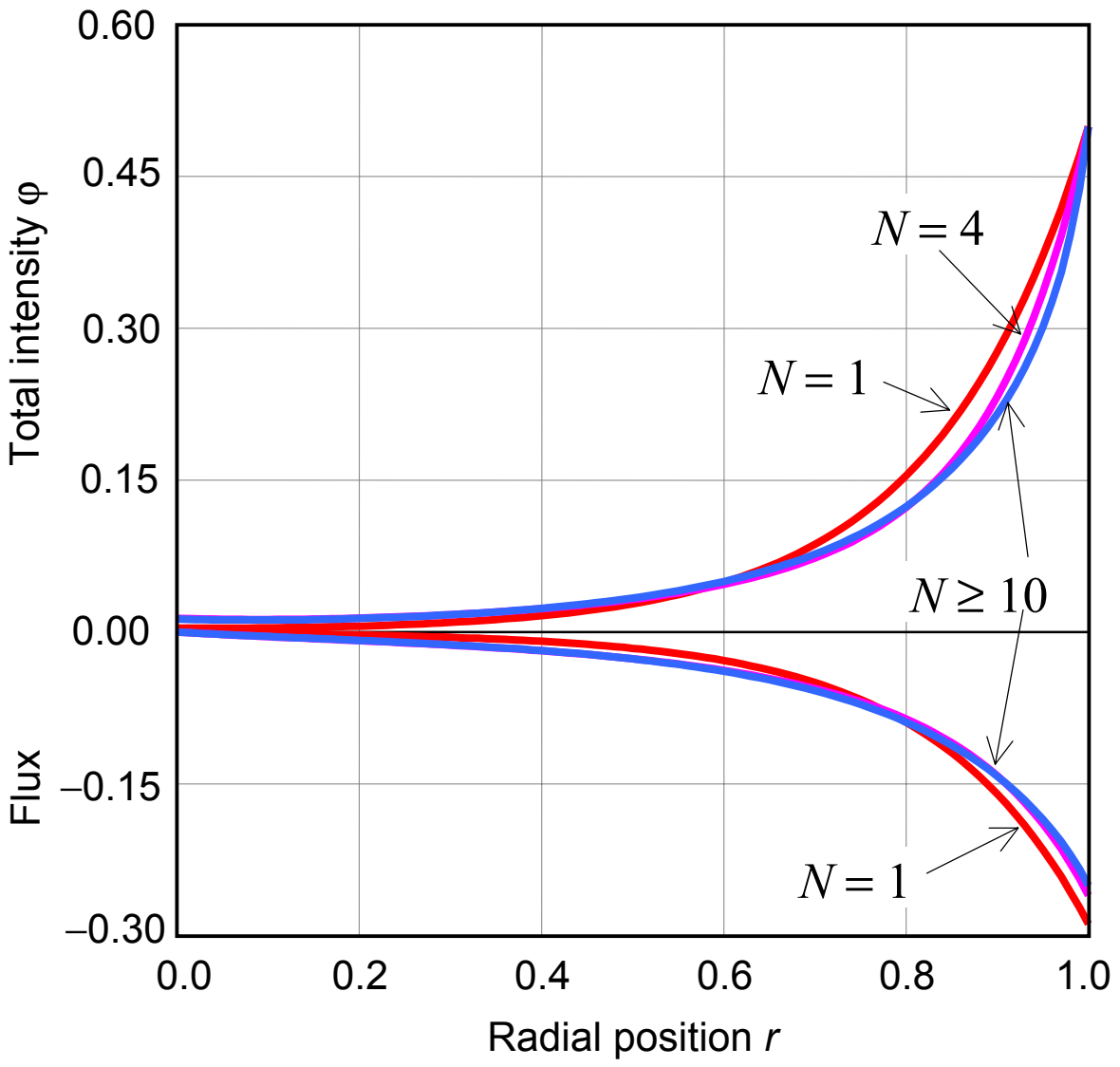

**Figure 3.6**. Graphs of the total intensity and the flux in a cold sphere of radius $R = 1$ and opacity $\kappa = 4$, immersed in a uniform isotropic radiation field. The labels $N$ are the number of discrete ordinates used to generate the graphs.

The constants $a$, $b$ and $c$ can be determined by inserting eq. (3.28) into eqs. (3.1) and (3.2). The following values are to obtain:

$$a = B\,, \qquad b = \pm\, 2B/\lambda\,, \qquad c = 2B/\lambda^2$$

The positive $b$ goes for the centripetal intensity and the negative $b$ for centrifugal intensities. We cast the particular solutions in the form:

$$\psi_P^+ = \frac{B}{\vartheta^2}\left[\vartheta^2 - \frac{2}{\lambda}\vartheta + \frac{2}{\lambda^2}\right] = \frac{B}{\vartheta^2}Q_r^+ \tag{3.29}$$

$$\psi_P^- = \frac{B}{\vartheta^2}\left[\vartheta^2 + \frac{2}{\lambda}\vartheta + \frac{2}{\lambda^2}\right] = \frac{B}{\vartheta^2}Q_r^- \tag{3.30}$$

From the properties of quadratic polynomials we determine the followings:

$$Q_r^+ \text{ and } Q_r^- > 0\,, \quad \forall r \ge 0$$

$$\partial_r Q_r^- > 0\,, \quad \forall r \ge 0$$

$$\partial_r Q_r^+ > 0\,, \quad \forall r > 0 : R > 1/\lambda$$

$$\partial_r Q_r^+ < 0\,, \quad \forall r > 0 : R < 1/\lambda$$

$$\exists \min\left(Q_r^+\right) \quad \text{at} \quad r = 1/\lambda - R \ge 0$$

These properties guarantee that the sum of the homogeneous solutions (3.27) and the particular solutions (3.29) and (3.30) is always positive:

$$\psi_r^+ = \frac{e^{-\lambda r}}{\vartheta^2}C^+ + \frac{B}{\vartheta^2}Q_r^+ \tag{3.31}$$

$$\psi_r^- = \frac{e^{\lambda r}}{\vartheta^2}C^- + \frac{B}{\vartheta^2}Q_r^- \tag{3.32}$$

One property of the $Q_r$ functions is that they are singular in vacuum. That's not of a concern to us because this singularity is not physical: in vacuum, $B = 0$ in accordance with our empty space postulate (sec. 3.3*b*). Therefore, the $Q$ terms need not be accounted for in vacuum space.

Evaluate eqs. (3.31) and (3.32) at $r = 0$ and solve for the constant of integrations $C^+$ and $C^-$. That completes the general mathematical solutions for eqs. (3.1) and (3.2):

$$\psi_r^+ = \frac{R^2}{\vartheta^2}e^{-\lambda r}\psi_0^h + \frac{B}{\vartheta^2}\left[Q_r^+ - Q_0^+e^{-\lambda r}\right] \tag{3.33}$$

$$\psi_r^- = \frac{R^2}{\vartheta^2}e^{\lambda r}\psi_0^h + \frac{B}{\vartheta^2}\left[Q_r^- - Q_0^-e^{\lambda r}\right] \tag{3.34}$$

The value of the intensity $\psi_0^h$ at the center of the sphere depends on the external boundary value $\psi_R^-$ at the outer surface of the mantle and on black body emission. Therefore, all we need is to determine the value of $\psi_0^h$ by evaluating eq. (3.34) at $R$, then solve for $\psi_0^h$ to obtain:

$$\psi_0^h = 4e^{-\lambda R}\psi_R^- + \frac{B}{R^2}\left[Q_0^- - Q_R^-e^{-\lambda R}\right] \tag{3.35}$$

Insertion of eq. (3.35) into eqs. (3.33) and (3.34) yields the complete solutions for the centrifugal and the centripetal equations in the sphere:

$$\psi_r^+ = \psi_r^{+c} + \psi_r^{+h} \qquad R \geq r \geq 0 \tag{3.36}$$

$$\psi_r^- = \psi_r^{-c} + \psi_r^{-h} \tag{3.37}$$

where $\psi_r^c$ are the specific intensities of a cold sphere given by eqs. (3.19) and (3.20), and $\psi_r^h$ are the intensities in a *hot* sphere defined in $r \in [0, R]$:

$$\psi_r^{+h} = \frac{B}{\vartheta^2}\left[Q_r^+ - \left(Q_0^+ - Q_0^-\right)e^{-\lambda r} - Q_R^-e^{-\lambda\vartheta}\right] \tag{3.38}$$

$$\psi_r^{-h} = \frac{B}{\vartheta^2}\left[Q_r^- - Q_R^-e^{-2\lambda R}e^{\lambda\vartheta}\right] \tag{3.39}$$

One thing we need to be concerned about is whether either of eqs. (3.38) and (3.39) can ever be negative. Indeed, they are always positive for $\forall R > 0$ and $\forall\lambda > 0$. The case of $R = 0$ is to be excluded for then we have no sphere to wonder about. The particular case of an infinite medium is rigorously modeled by eqs. (3.38) and (3.39): set $R \to \infty$ to obtain $\psi_r^{+h} = \psi_r^{-h} = B$ as it should be.

The diffusion intensity and the flux can be calculated from eqs. (3.36) and (3.37):

$$\phi_r = \frac{1}{2}\left[\left(\psi_r^{+c} + \psi_r^{-c}\right) + \left(\psi_r^{+h} + \psi_r^{-h}\right)\right] = \phi_r^c + \phi_r^h \tag{3.40}$$

$$f_r = \frac{\mu}{2}\left[\left(\psi_r^{+c} - \psi_r^{-c}\right) + \left(\psi_r^{+h} - \psi_r^{-h}\right)\right] = f_r^c + f_r^h \tag{3.41}$$

The diffusion intensity $\phi_r^c$ and diffusion flux $f_r^c$ of a cold sphere are given by eqs. (3.23)-(3.24). The diffusion intensity $\phi_r^h$ and the flux $f_r^h$ in a hot sphere are obtained from eqs. (3.38)-(3.39) for $r \in [0, R]$:

$$\phi_r^h = \frac{B}{\vartheta^2}\left(\frac{Q_r^+ + Q_r^-}{2} - \frac{Q_0^+ - Q_0^-}{2} e^{-\lambda r} - Q_R^- e^{-\lambda R} \cosh \lambda r\right) \tag{3.42}$$

$$f_r^h = \frac{\mu B}{\vartheta^2}\left(\frac{Q_r^+ - Q_r^-}{2} - \frac{Q_0^+ - Q_0^-}{2} e^{-\lambda r} + Q_R^- e^{-\lambda R} \sinh \lambda r\right) \tag{3.43}$$

The eqs. (3.42) and (3.43) verify the differential conservation relation (1.11) and the integral conservation relation (1.13).

The first thing we learn from eqs. (3.42) and (3.43) is that the flux vanishes at the center of the sphere while the diffusion intensity $\phi$ has a maximum at the center. The value of the maximum depends on the radius of the sphere and the opacity $\lambda$: for each pair of values $(R,\lambda)$ there is a maximum of diffusion intensity $\phi^h$. The maximum of all of these maxima can be determined from eq. (3.42) by setting $r = 0$. The resulting equation is transcendental in $\xi \equiv R\lambda$, the optical radius of the sphere. We found numerically that the maximum of the maxima of the diffusion intensities at the center of the spheres is $\phi_{max}^h = 1.63400B$, and the minimum of the minima of the intensities is $\phi_{min}^h = 0.34735B$ at the surface of all the spheres. These extrema intensities occur in all hot spheres of fixed optical radius $\xi = 2.38837$; these numbers are accurate to the fifth decimal place.

Also, the equations (3.42)-(3.43) satisfy the condition in the asymptotic region which transcribes into:

$$R \to \infty \;\Rightarrow\; \begin{cases} \phi_r = B \\ f_r = 0 \end{cases}$$

The total intensity $\varphi^h$ is obtained by applying the quadrature (2.14) to eq. (3.42):

$$\varphi_r^h = \sum_{n=1}^{N} w_n \phi_{n,r}^h \tag{3.44}$$

Hence the total intensity in a sphere with uniform and isotropic internal emission and exposed to external isotropic radiation field becomes the sum:

$$\varphi_r = \varphi_r^c + \varphi_r^h \tag{3.45}$$

Similarly, the flux is obtained from eq. (3.43):

$$F_r^h = \sum_{n=1}^{N} w_n f_{n,r}^h \tag{3.46}$$

and if the sphere is exposed to external radiation field the flux becomes:

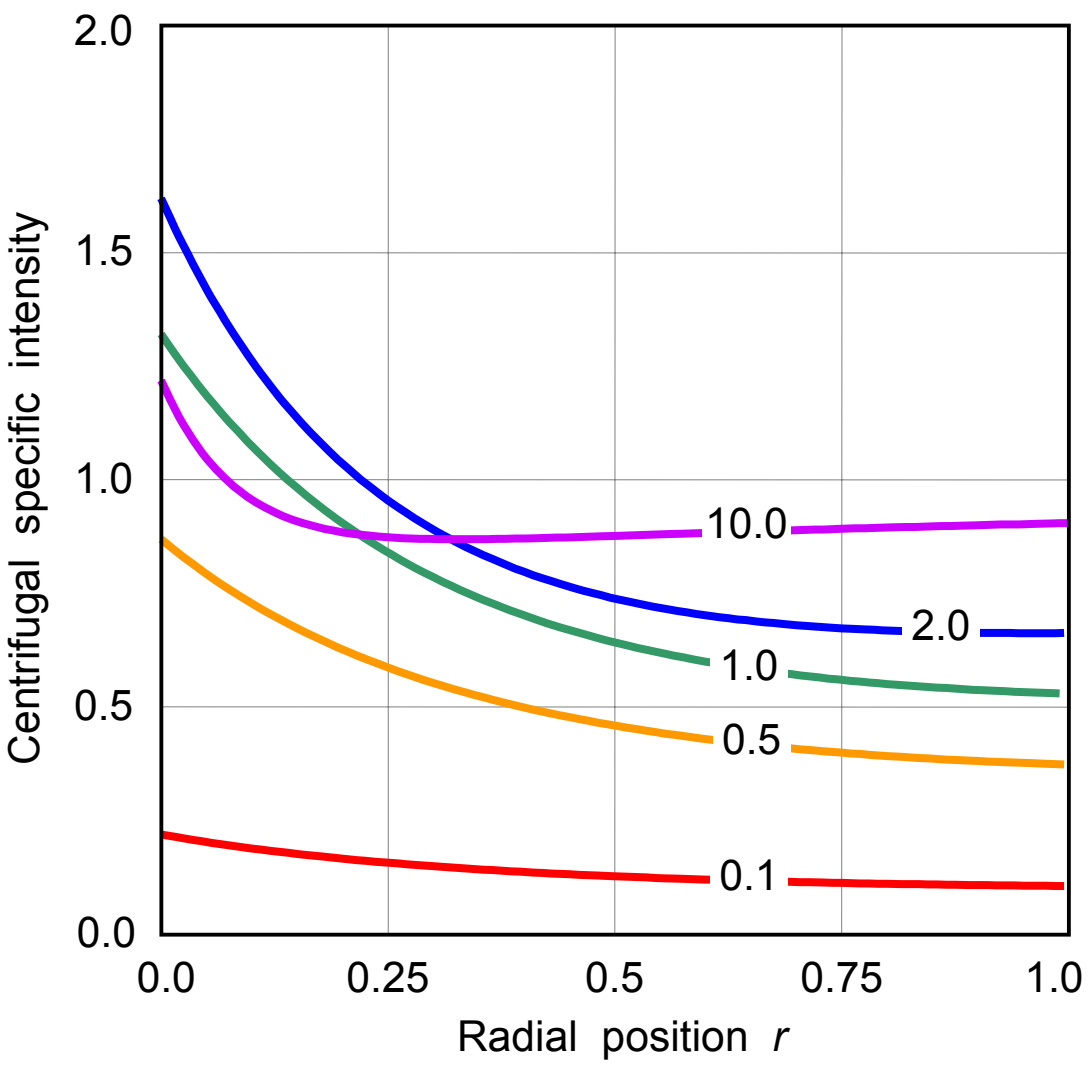

**Figure 3.7.** Graphs of the centrifugal intensities in a hot sphere and for various values of the opacity λ shown as numbers on the curves; the black body constant $B$=1. The sphere is surrounded by vacuum.

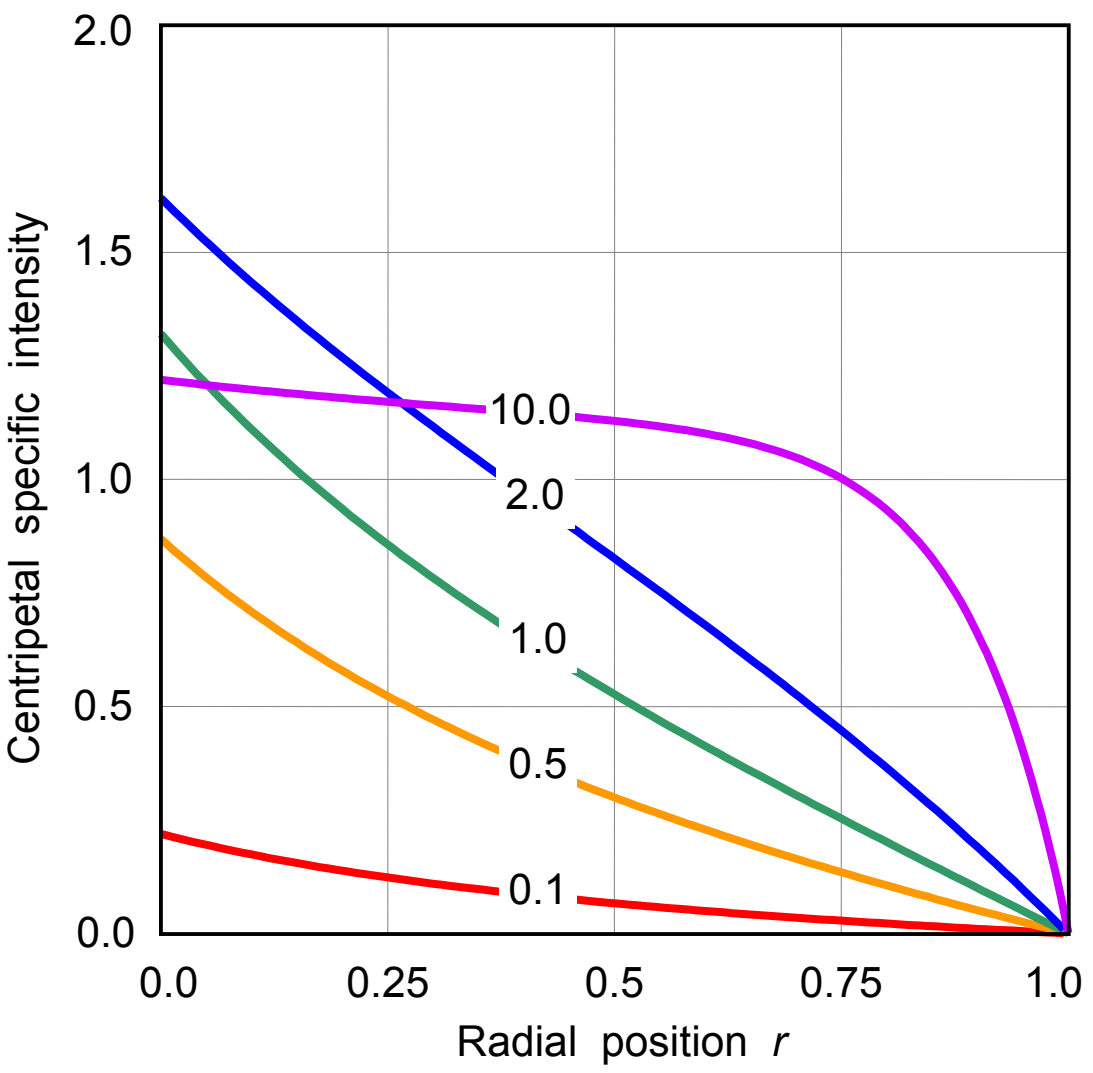

**Figure 3.8.** Graphs of the centripetal intensities in a hot sphere and for various values of the opacity λ shown as numbers on the curves; the black body constant $B$=1. The sphere is surrounded by vacuum..

$$F_r = F_r^c + F_r^h \tag{3.47}$$

The cold intensity $\varphi^c$ and flux $F^c$ are given by eqs. (3.25) and (3.26), respectively.

Graphical representations of the specific intensities $\psi^{\pm}$ are shown in figures 3.7 and 3.8 for various values of λ. These graphs are generated using eqs. (3.38) and (3.39). Graphs for the diffusion intensities are shown in figure 3.9 for various values of λ. Figure 3.10 shows the variations of the diffusion intensity $\phi^h$ as a function of $r$ in different spheres of constant optical radius ξ. These graphs are generated using eq. (3.42) by maintaining $\lambda R = \xi$.

Another set of graphs for the total intensity $\varphi^h$ are generated using eq. (3.44) and a normalized black body constant $B$ = 1. These graphs are informative about the shape of the intensity profile as well as about the effectiveness of the proposed incomplete diffusion equations in spheres with different opacities. Figure 3.11 shows that a very large number of discrete ordinates is needed to saturate the intensity in a sphere of a very small absorption coefficient, in this particular case the opacity $\kappa = 10^{-5}$. This is the smallest value of the absorption coefficient with which eq. (3.42) is numerically stable with 16 significant digits arithmetics. Smaller values of the absorption coefficient produce distorted intensity profile with any $N \geq 1$. Figure 3.12 shows the effects of the numerical instabilities on the intensity profile for $\kappa = 10^{-6}$. The graph becomes smoother with increasing $N$ up to 7000 but it is still too rugged to have any usefulness other than the trend. Increasing $N$ further causes the roughness of the graph to get worse.

The numerical instabilities with very small κ are explained by eq. (3.42). The $Q$ functions are singular for κ = 0. However, with κ = 0 the black body emission $B$ vanishes and there will be no radiation field resulting from absorption and emission. Consequently, eq. (3.42) is no longer defined in this medium. On the other hand, starting from pure mathematical considerations, the bracketed quantity of eq. (3.42) vanishes in the limit as $\kappa \to 0$. Unfortunately, the analytical limit does not always transcribe into a smooth limit on digital machines a process that explains the numerical instabilities in finite arithmetics.

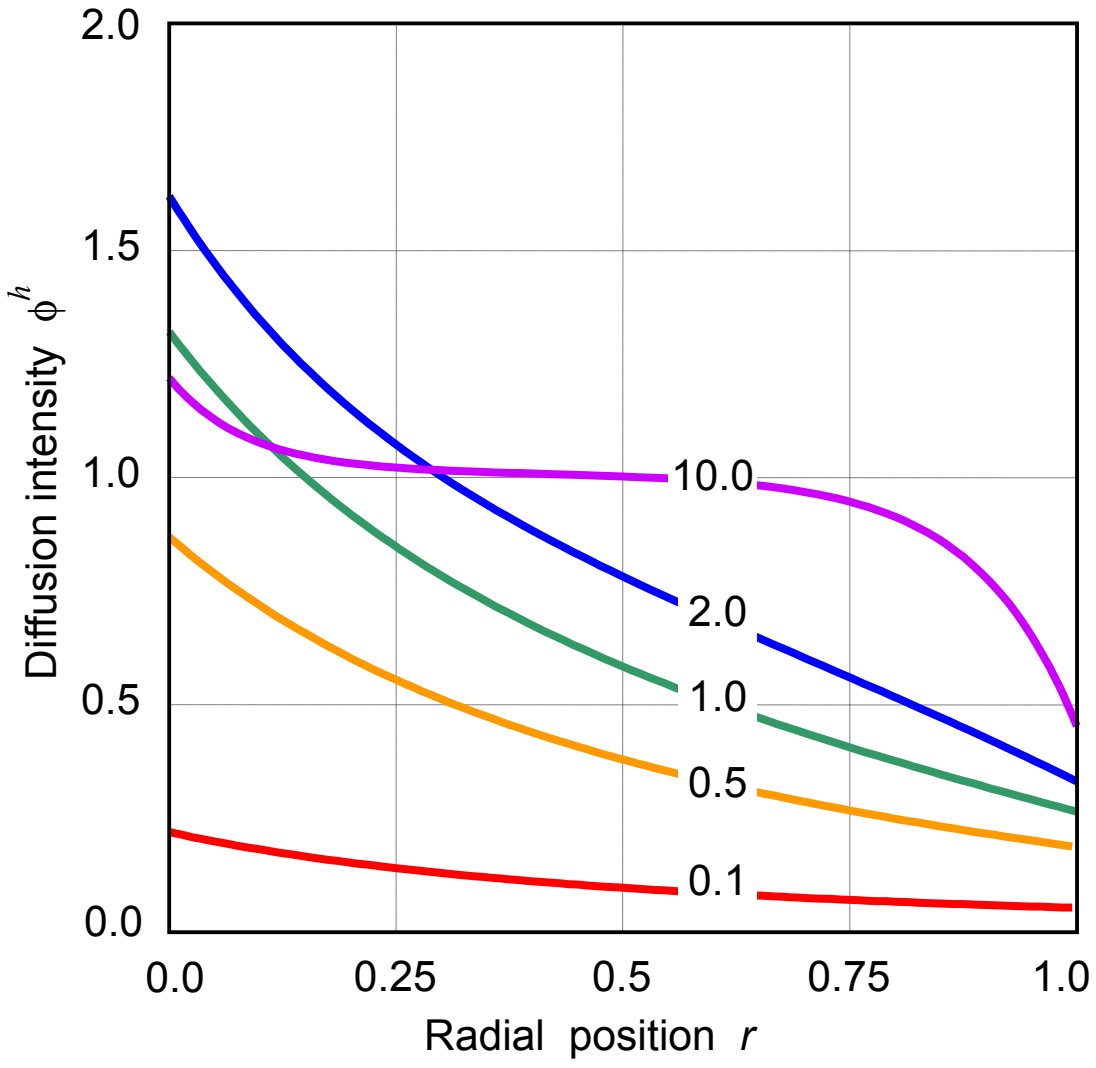

**Figure 3.9.** Graphs of the diffusion intensities in a hot sphere and for various values of the opacity λ shown as numbers on the curves; the black body constant $B$=1. The sphere is surrounded by vacuum.

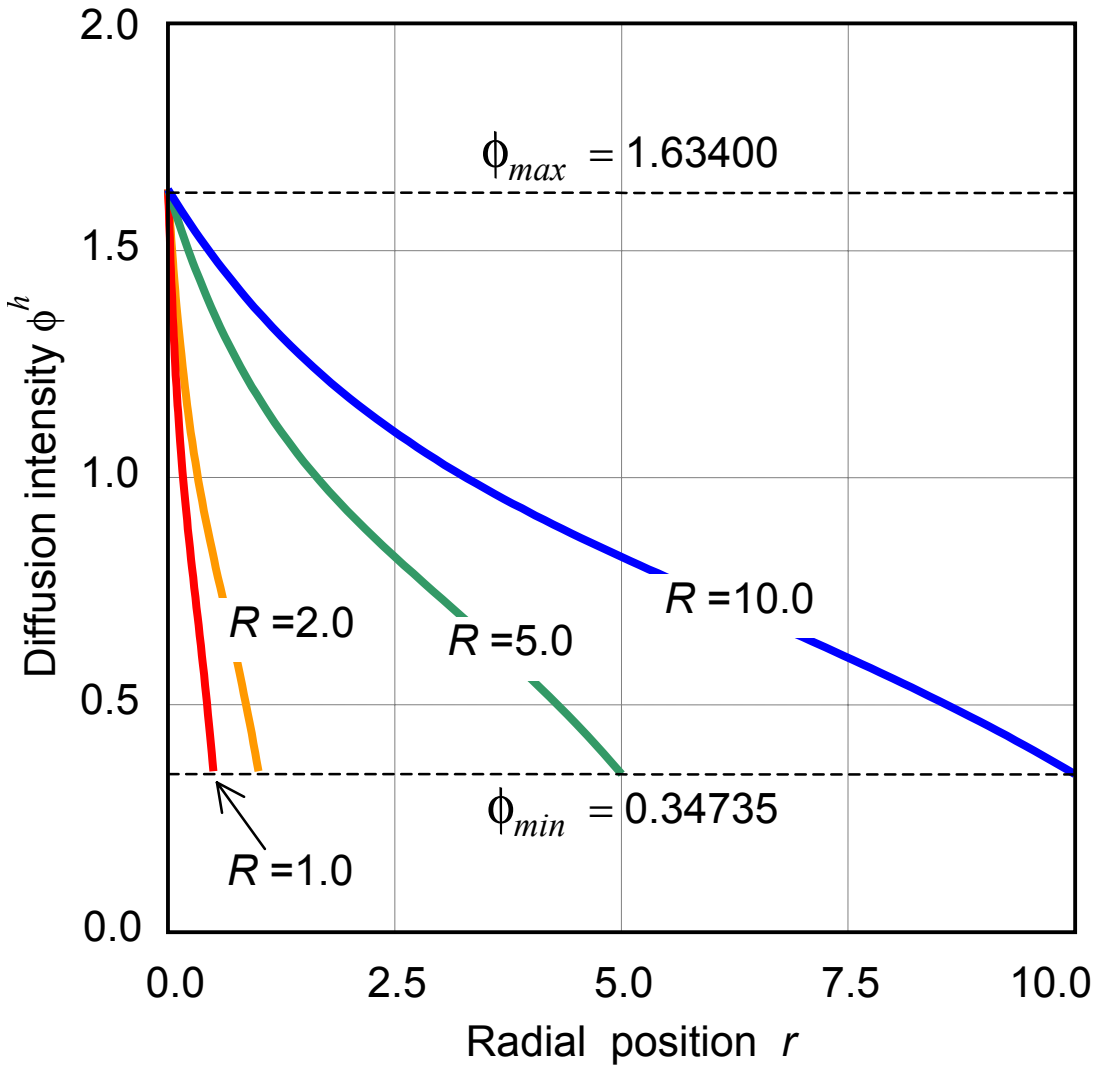

**Figure 3.10.** Graphs of the diffusion intensities in a hot sphere of constant optical radius ξ = 2.38837 and for various values of the radius $R$. This optical thickness produces maximum intensity at the center of the sphere. The black body constant $B$=1. The sphere is surrounded by vacuum.

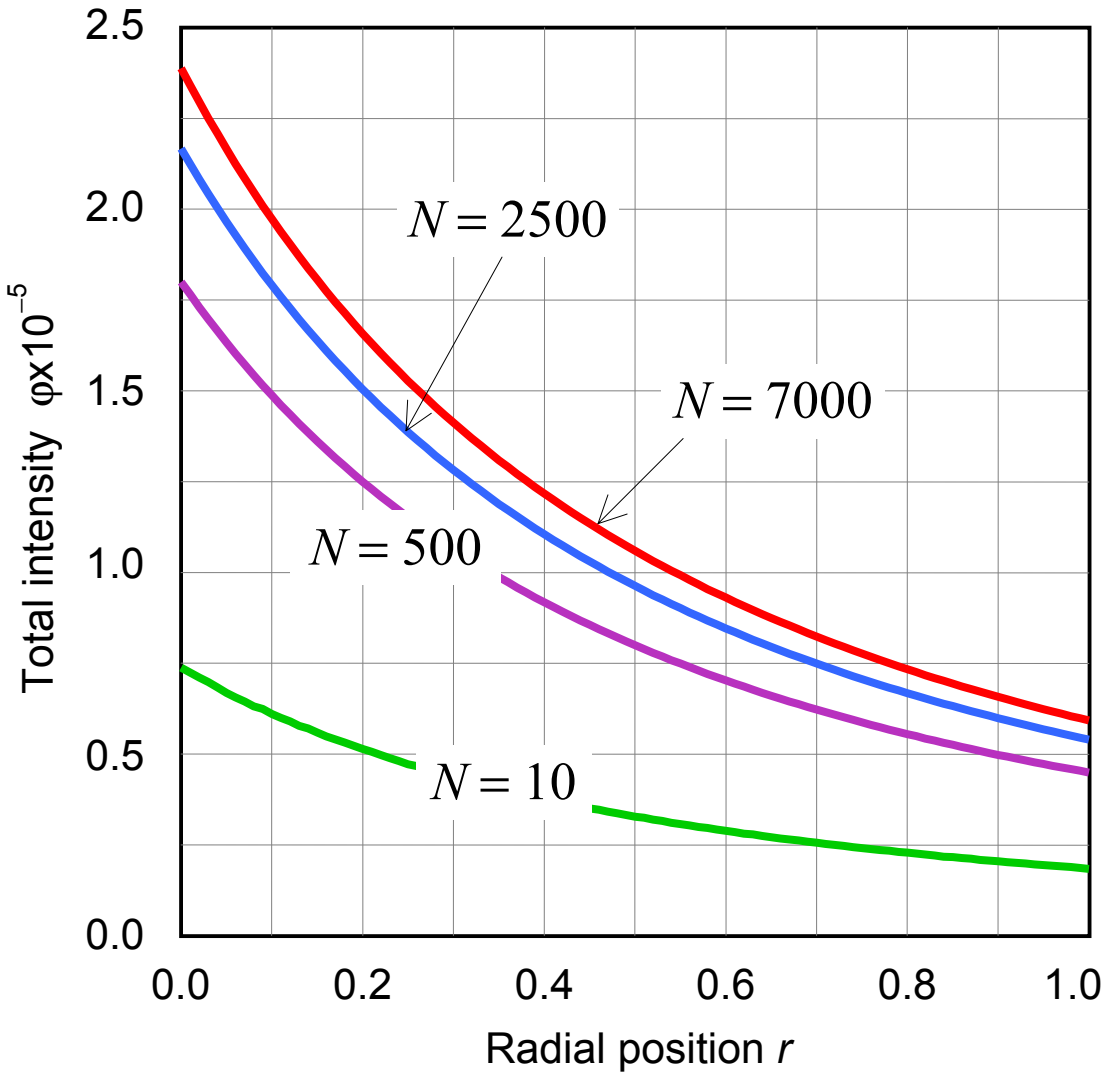

**Figure 3.11**. Graphs of the total intensity in a hot sphere of normalized black body constant $B$ = 1 and opacity $\kappa = 10^{-5}$ generated in $S_{2N}$ discrete ordinates.

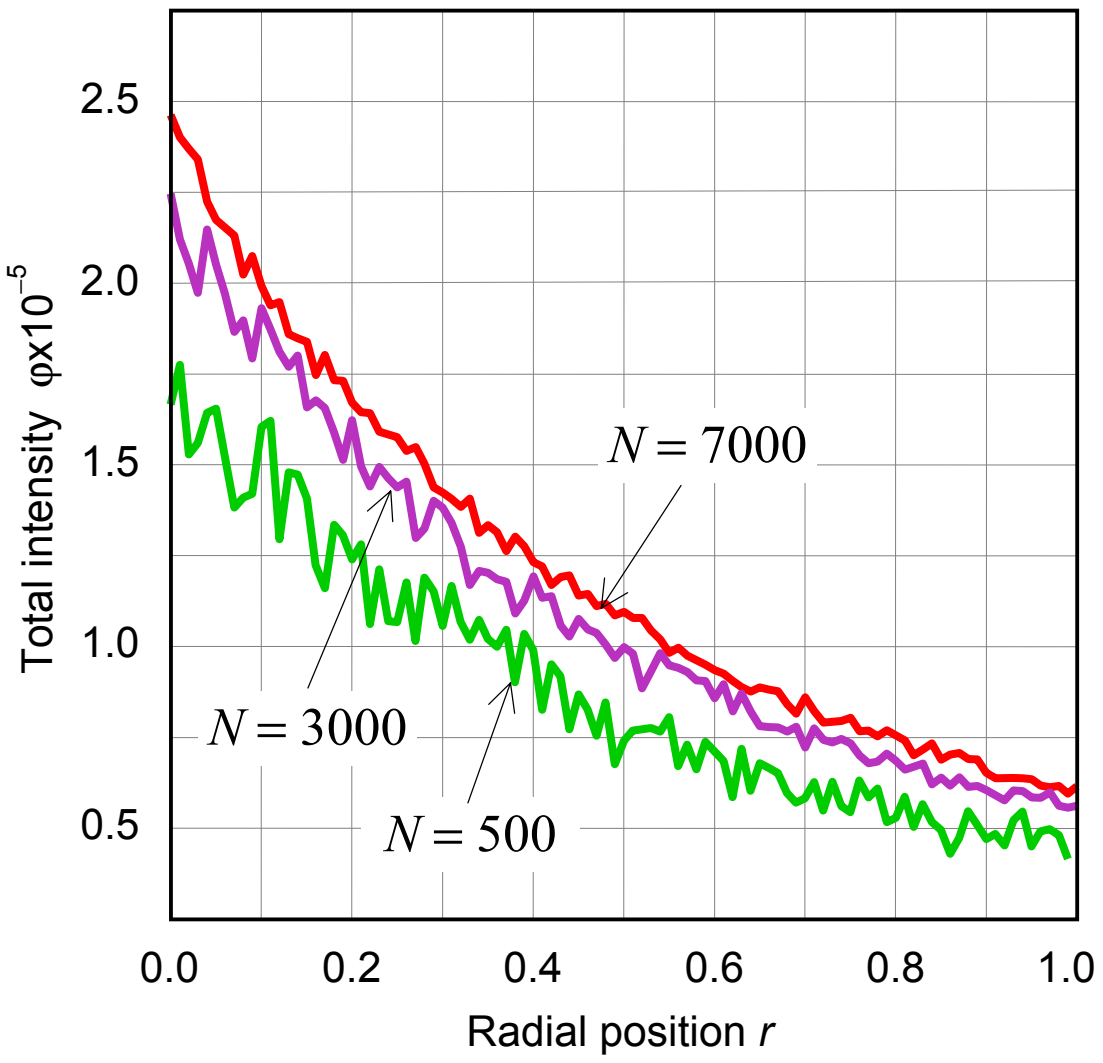

**Figure 3.12**. Graphs of the total intensity in a hot sphere of normalized black body constant $B$ = 1 and opacity $\kappa = 10^{-6}$ generated in $S_{2N}$ discrete ordinates.

Figures 3.13-3.14 show the total intensity profiles in spheres of moderate to large absorption coefficients. In these cases the intensity saturates with only a few discrete ordinates. The diffusion intensity $\phi^h$ is nearly the same as the total intensity $\varphi^h$ in a sphere of a very large absorption coefficient, such as $\kappa = 20$. This case is illustrated in figure 3.14: the intensity is flat in strongly

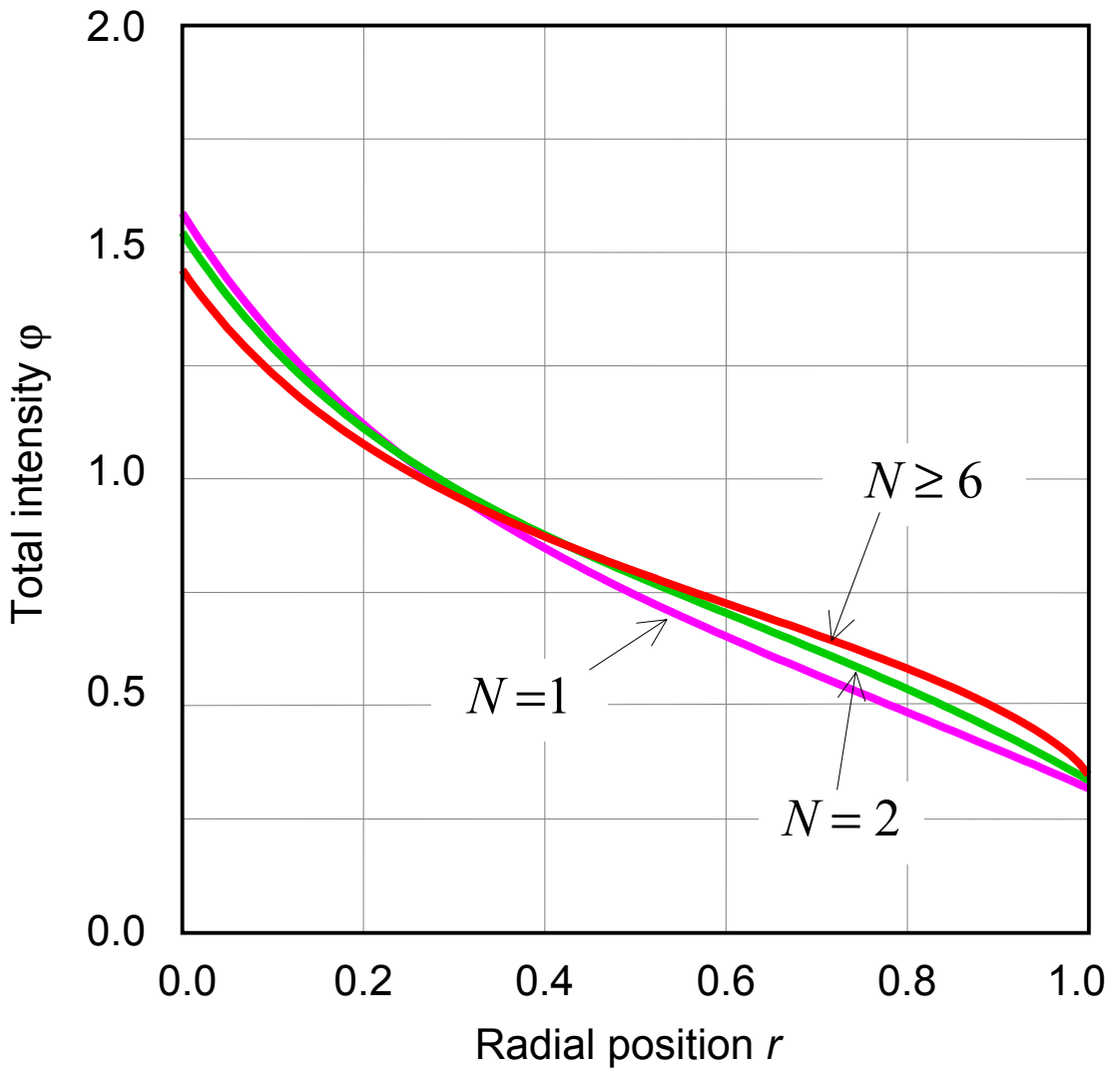

**Figure 3.13**. Graphs of the total intensity in a hot sphere of opacity κ = 1 and a black body constant $B = 1$. The intensities are calculated in $S_{2N}$ discrete ordinates.

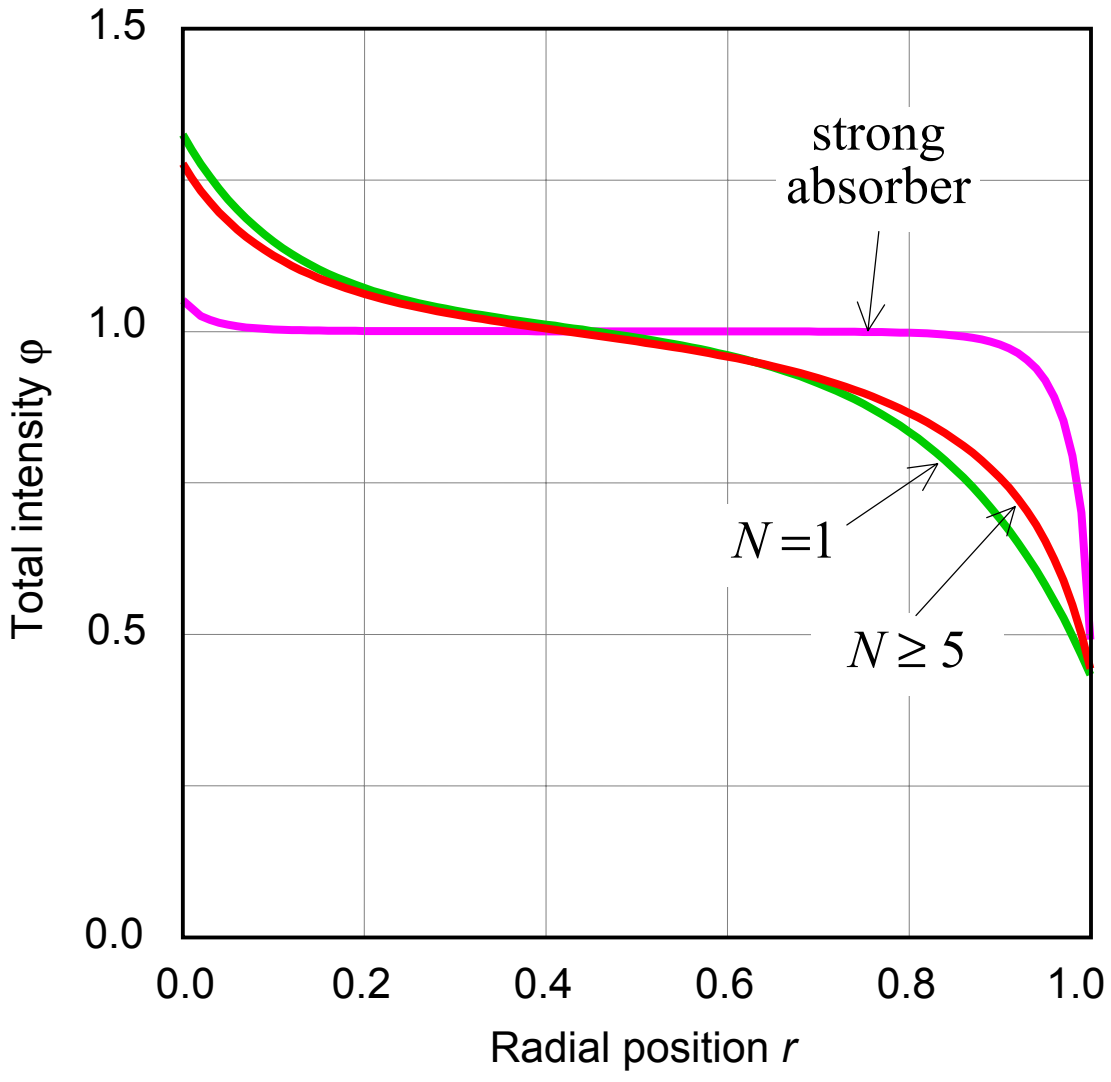

**Figure 3.14**. Graphs of the total intensity in a hot sphere of opacity κ = 4 and a black body constant $B = 1$. The intensities are calculated in $S_{2N}$ discrete ordinates; the strong absorber in $S_2$, κ = 20.

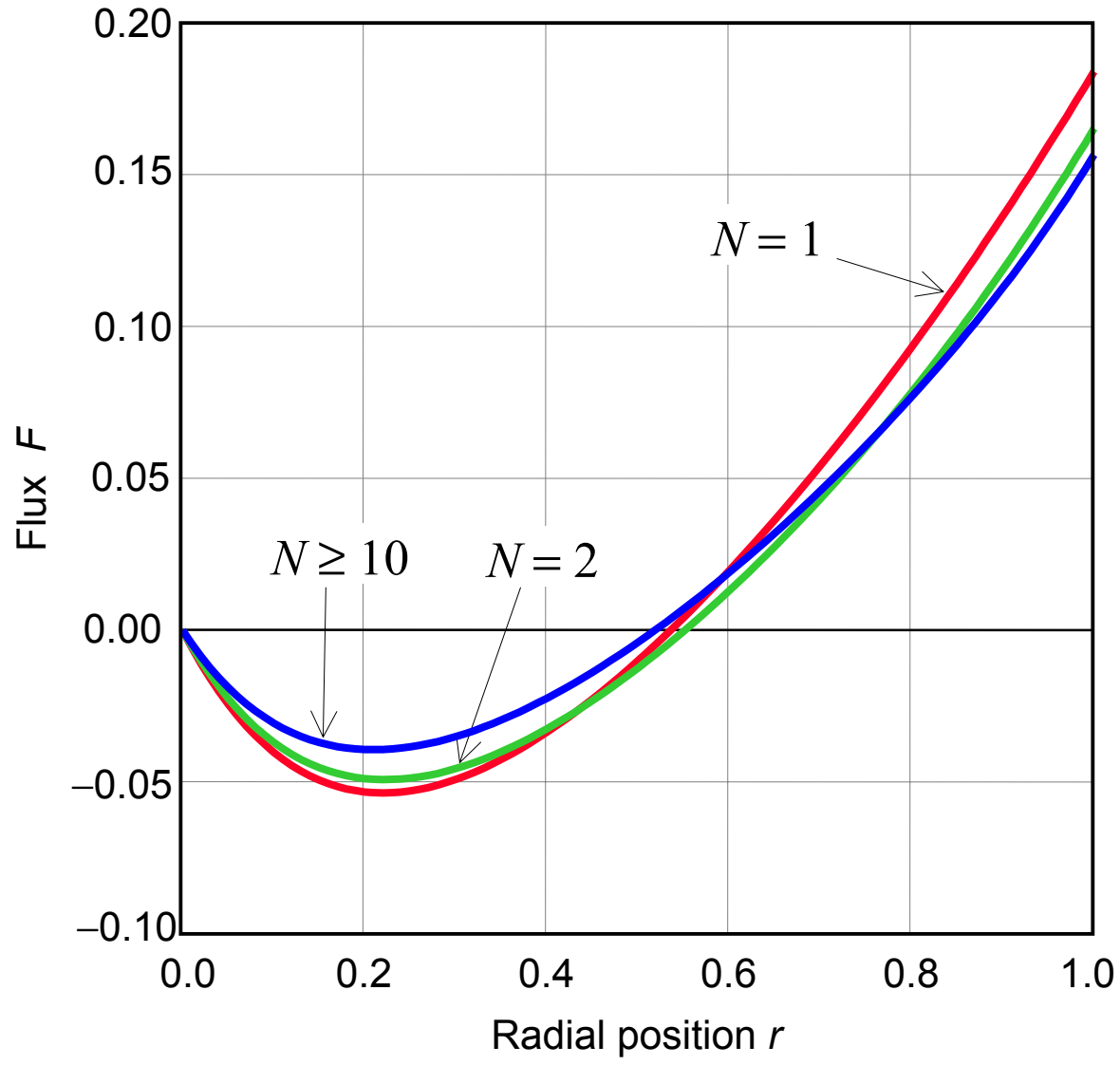

**Figure 3.15**. Graphs of the flux in a hot sphere of radius $R = 1$ and opacity κ = 1; the free surface of the sphere is exposed to vacuum. The labels $N$ are the number of discrete ordinates used to generate the

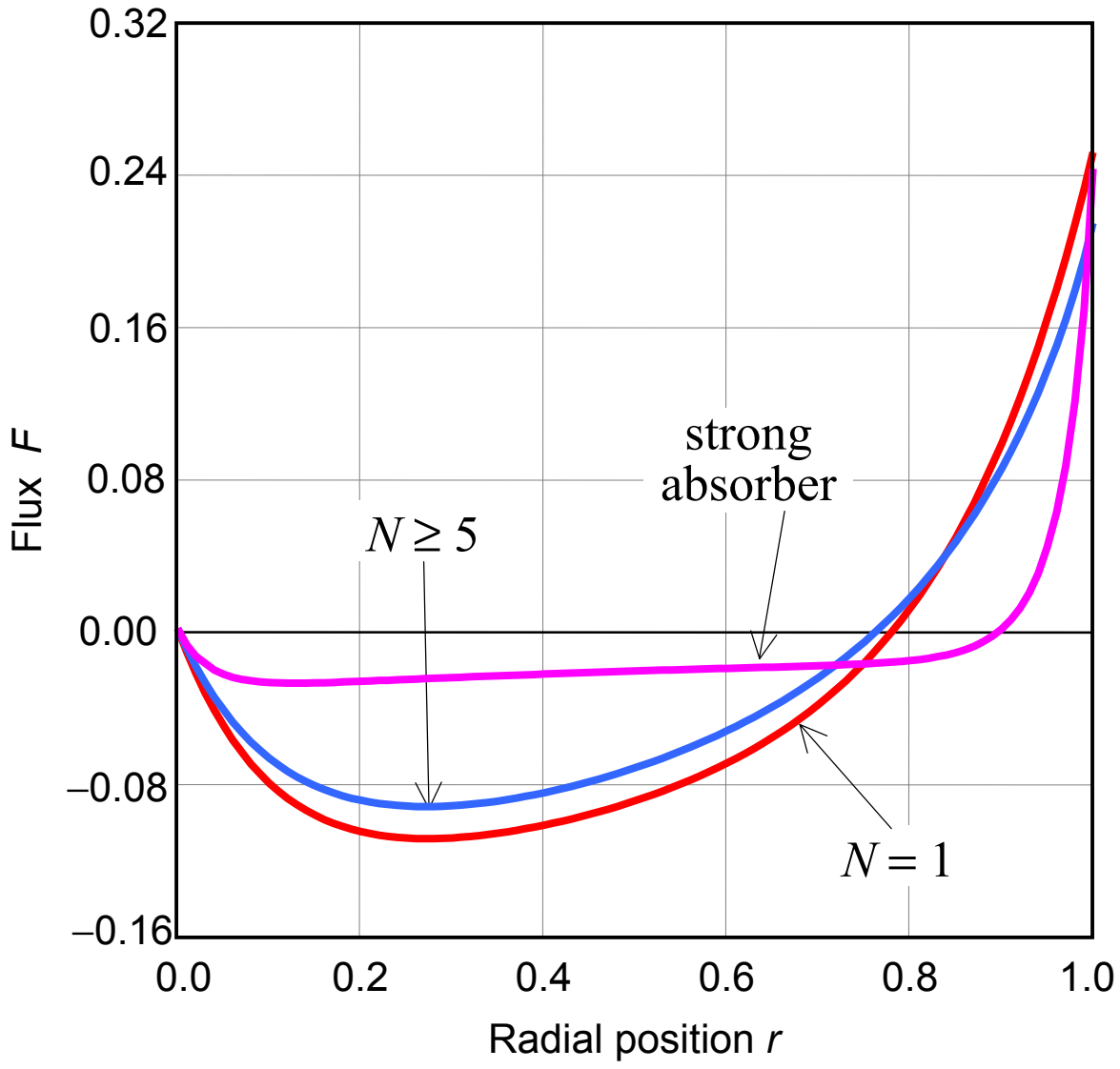

**Figure 3.16**. Graphs of the flux in a hot sphere of radius $R = 1$ and opacity κ = 4; the free surface of the sphere is exposed to vacuum. The labels $N$ are the number of discrete ordinates used to generate the absorptive spheres. This graph is generated with $N = 1$, which is the diffusion intensity $\phi^h$. The graph generated with $N = 2$ is pretty much the same and it is not any different for larger $N$.

It is apparent from these figures that the intensity at the center is always prodding upward even in a very strongly absorbing medium such as the one depicted by figure 3.14. This spear tip phenomenon can be explained analytically by the derivative of the diffusion intensity given by eq. (3.42). At the center the derivative of this intensity is proportional to $-1/\xi^2$, always negative, where $\xi$ is the optical radius of the sphere. Therefore, the derivative of the total intensity at the center is always negative except in an optically infinite medium for $\xi \to \infty$. Physically, this means that there is a ball of radius $r_b$, determined from the point of inflection of the intensity function, within which the energy attenuation rate is smaller than the convergence rate defined from $-1/r^2$, which becomes large for small $r$. The net result is an accumulation of energy at the center that culminates with spear tip intensity.

Also, graphs of the fluxes in the spheres were generated and are shown in figures 3.15 and 3.16. The graph of the flux labeled $N = 2$ is not shown in figure 3.15. It is barely discernable from the graph for $N = 5$.

### (*b*) *Multi-layered sphere*

In this configuration we have a core to be labeled as mantle 1, wrapped by a concentric mantle 2 itself wrapped by mantle 3, ... etc. (fig. 3.17). Each mantle is a homogeneous isothermal medium characterized by its emission density $\lambda_m B_m$ and bounded by its inner radius $R_{m-1}$ and outer radius $R_m$; $m = 1, 2, ..., M$. A mantle may be divided to any number of shells as needed by the user. The intensity at point $r$ in a mantle is labeled as $\psi_{m,r}$. At the boundaries of the mantle the index for the independent variable $r$ is replaced by the index of the boundary radius as illustrated in figure 3.18. This nomenclature of indices applies to the shift function as well.

The transfer equations (3.1) and (3.2) defined on $m$th mantle read:

$$\partial_r \psi^+_{m,r} + \left( \frac{2}{\vartheta_{m,r}} + \lambda_m \right) \psi^+_{m,r} = \lambda_m B_m \qquad r \in [R_{m-1}, R_m] \tag{3.48}$$

$$\partial_r \psi^-_{m,r} + \left( \frac{2}{\vartheta_{m,r}} - \lambda_m \right) \psi^-_{m,r} = -\lambda_m B_m \tag{3.49}$$

The first thing we learn from these two equations is that the derivatives are continuous within a mantle and on its boundary from the interior, but they are discontinuous at the interface boundary itself between two inhomogeneous mantles. This is because the emission density $\lambda B$ has one value on the left side of the interface, another value on the right. It is a simple exercise to demonstrate that the second derivatives are also discontinuous at the interface and at the center of the sphere as well. The one-sided continuity of the derivative does not preclude the continuity of the intensity at the boundary. On physical ground the intensities must be continuous everywhere in the sphere. These properties are carried on to the total intensities and the flux.

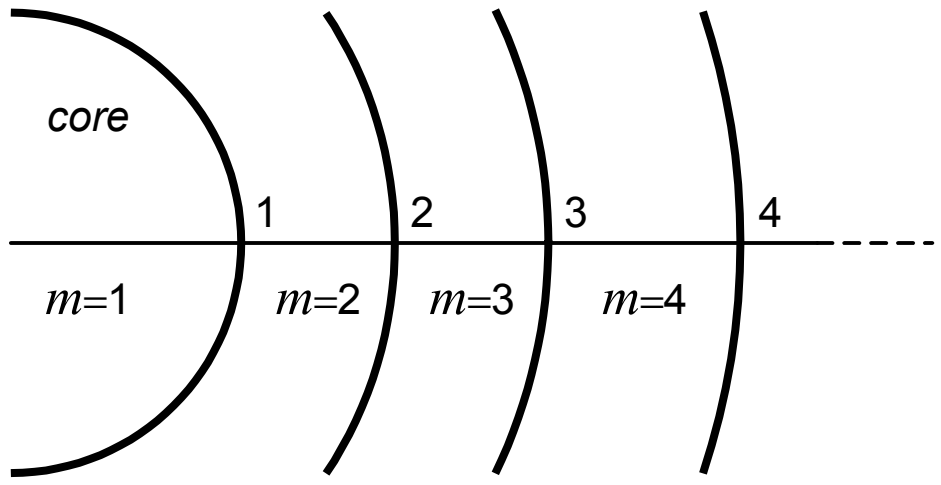

**Figure 3.17.** Disposition of the mantles and the labeling of the radii

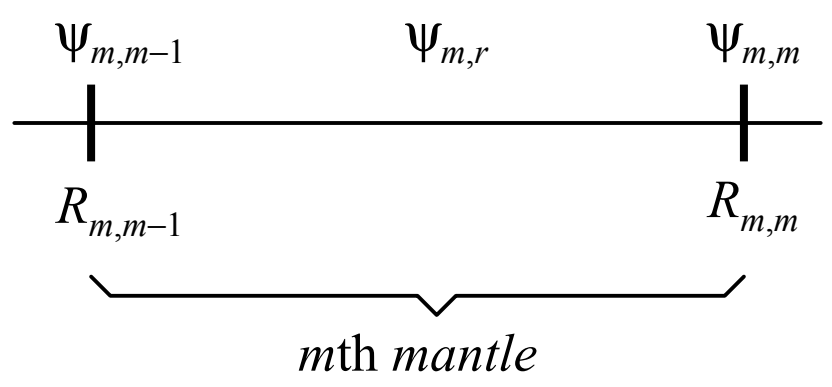

**Figure 3.18.** Labeling of the intensities in a mantle and on its boundaries

The general solutions for eqs. (3.48) and (3.49) may be cast in the form:

$$\psi^-_{m,r} = \frac{\vartheta^2_{m,m}}{\vartheta^2_{m,r}} e^{\lambda_m(r-R_{m-1})}\psi^-_{m,m-1} + \frac{B_m}{\vartheta^2_{m,r}}\left[Q^-_{m,r} - Q^-_{m,m}e^{\lambda_m(r-R_{m-1})}\right] \tag{3.50}$$

$$\psi^+_{m,r} = \frac{\vartheta^2_{m,m-1}}{\vartheta^2_{m,r}} e^{-\lambda_m(r-R_{m-1})}\psi^+_{m,m-1} + \frac{B_m}{\vartheta^2_{m,r}}\left[Q^+_{m,r} - Q^+_{m,m-1}e^{-\lambda_m(r-R_{m-1})}\right] \tag{3.51}$$

The solutions as functions of $r$ are most useful to calculate the energy absorbed in the volume of interest. Most often it is convenient to do computations with the end-points form of the solutions. In order to convert eqs. (3.50) and (3.51) to end-points forms we have to keep in mind that the intensities at all points must be expressed in the same shifted frame of reference. It is most convenient in the present case to shift the frame to the radius of the outermost boundary of the sphere so that $\vartheta_{M,r} \equiv R_M + r$ applies to all mantles for $\forall r \in [0, R_M]$. The end-points solutions are most conveniently expressed by the following recursion relations:

*Centripetal intensities*, for $m = M, M{-}1, ..., 2, 1$:

$$\psi^-_{m-1} = \frac{\vartheta^2_{M,m}}{\vartheta^2_{M,m-1}} e^{-\lambda_m\Delta_m}\psi^-_m + \frac{B_m}{\vartheta^2_{M,m-1}}\left[Q^-_{M,m-1} - Q^-_{M,m}e^{-\lambda_m\Delta_m}\right] \tag{3.52}$$

*Centrifugal intensities*, for $m = 1, 2, ..., M{-}1, M$:

$$\psi^+_m = \frac{\vartheta^2_{M,m-1}}{\vartheta^2_{M,m}} e^{-\lambda_m\Delta_m}\psi^+_{m-1} + \frac{B_m}{\vartheta^2_{M,m}}\left[Q^+_{M,m} - Q^+_{M,m-1}e^{-\lambda_m\Delta_m}\right] \tag{3.53}$$

where $\Delta_m \equiv R_m - R_{m-1}$ and the $Q$ functions are defined by eqs (3.29) and (3.30). The indices of the geometric quantities, as well as the intensities are the ordinals of the radial boundaries as illustrated in figure 3.18, the indices of the opacity $\lambda$ and the black body constant are the ordinals of the mantles. The boundary index of the center of the sphere is 0. The asymptotic condition can be verified directly by setting $R_M \to \infty \Rightarrow \psi_{M,r} = B, \forall r < \infty$.

It is convenient for computational purposes to rewrite eqs. (3.52) and (3.53) in terms of the ART properties❖. Define the following operators for the $m$th mantle:

*Centripetal radiance* at the concave surface:

$$\mathsf{R}^-_{m,m-1} = \frac{B_m}{\vartheta^2_{M,m-1}}\left[Q^-_{M,m-1} - Q^-_{M,m}e^{-\lambda_m\Delta_m}\right] \tag{3.54}$$

*Centrifugal radiance* at the convex surface:

$$\mathsf{R}^+_{m,m} = \frac{B_m}{\vartheta^2_{M,m}}\left[Q^+_{M,m} - Q^+_{M,m-1}e^{-\lambda_m\Delta_m}\right] \tag{3.55}$$

*Centripetal transmittance*, from the boundary at $R_m$ to the boundary at $R_{m-1}$:

❖ Albedo, Radiance and Transmittance [2]

$$\mathsf{T}_{m-1}^{m} = \frac{\vartheta_{M,m}^{2}}{\vartheta_{M,m-1}^{2}} e^{-\lambda_m \Delta_m} \tag{3.56}$$

*Centrifugal transmittance*, from the boundary at $R_{m-1}$ to the boundary at $R_m$:

$$\mathsf{T}_{m}^{m-1} = \frac{\vartheta_{M,m-1}^{2}}{\vartheta_{M,m}^{2}} e^{-\lambda_m \Delta_m} \tag{3.57}$$

The subscript $n$ for the $n$th direction is omitted for national convenience. Rewrite equations (3.52) and (3.53) in terms of these operators, keeping in mind that $R_0 = 0$ and $\psi_{1,0} = \psi_0$ is independent of direction, and $\psi_{M,M}^{-}$ is boundary condition:

$$\psi_{m-1}^{-} = \mathsf{T}_{m-1}^{m} \psi_{m}^{-} + \mathsf{R}_{m,m-1}^{-} \qquad m = M, M-1, ..., 2, 1 \tag{3.58}$$

$$\psi_{m}^{+} = \mathsf{T}_{m}^{m-1} \psi_{m-1}^{+} + \mathsf{R}_{m,m}^{+} \qquad m = 1, 2, ..., M-1, M \tag{3.59}$$

The rule of contraction of tensors applies, but there is no implied summation on the repeated index in these equations: for every $m$ there is only one $m - 1$.

The equations (3.58) and (3.59) are the complete algorithm for the calculations of the specific intensities at all the interfaces within the sphere as well as at the center and at the outer surface. Clearly, since the intensity $\psi_{M,M}^{-}$ is the trigger boundary condition, whether it has a zero or a non-zero value, the calculation should start with eq. (3.58). After the completion of the sweep in decreasing $m$, eq. (3.59) completes the solution in the increasing $m$ sweep. Once all the specific intensities are calculated at all interfaces the diffusion intensities and fluxes are calculated in the specified $n$th direction:

$$\phi_{n,m} = \left(\psi_{n,m}^{+} + \psi_{n,m}^{-}\right)/2 \tag{3.60}$$

$$f_{n,m} = \mu_n \left(\psi_{n,m}^{+} - \psi_{n,m}^{-}\right)/2 \tag{3.61}$$

The total intensity and the flux at each interface is obtained by applying the quadrature (2.14):

$$\varphi_m = \sum_{n=1}^{N} w_n \phi_{n,m} \tag{3.62}$$

$$F_m = \sum_{n=1}^{N} w_n f_{n,m} \tag{3.63}$$

The advantage of the (3.58)-(3.59) algorithm is that $M$ needs not be limited to the number of material in the heterogeneous sphere. The algorithm is valid in homogeneous spheres as well. In this case $M$ may be set so that only intensities at selected spherical surfaces within the sphere are calculated.

The proposed algorithm has another use: it determines the boundary values to be used in equations (3.50)-(3.51). Hence, the solutions of the transfer equation in a heterogeneous sphere are completely defined as functions of $r$. The calculation of the fluxes at the interfaces is straightforward and the energy absorbed in a mantle can be obtained by integrating the total intensities from eqs. (3.50) and (3.51) in each mantle on $[R_{m-1}, R_m]$.

**Table I**. Geometric and physical data for the two heterogeneous hot spheres used to generate the graphs for the total intensities and the fluxes.

| Material and thermal properties | | Radial thickness of the mantles | | |
|---|---|---|---|---|
| | | 0.0 – 0.4 | 0.4 – 0.6 | 0.6 – 1.0 |
| Sphere 1 | $B$ | 1 | 4 | 10 |
| | $\kappa$ | 1 | 2 | 3 |
| Sphere 2 | $B$ | 10 | 4 | 1 |
| | $\kappa$ | 3 | 2 | 1 |

Graphical illustrations are produced in three heterogeneous spheres made up of a core wrapped with two concentric mantles. The core and the mantles have different opacities $\kappa$ and characterized by different black body constants shown in Table I. The free surface of the outer mantel is exposed to vacuum. The graph with label *exposed* is generated for a sphere its outer surface exposed to an isotropic and uniform radiation field setting a boundary condition to the problem $\psi_R^- = 10$. Graphs for the total intensities $\varphi$ are generated with eqs. (3.58) and (3.59) in $S_{2N}$ discrete ordinates. Figure 3.19 shows the intensities in a sphere with radially increasing temperature gradient and figure 3.20 shows the graphs of the intensities with radially decreasing temperature gradient. The external energy input is the same for both of the spheres.

## 3.5 A second order diffusion equation

A second order diffusion equation valid for LTE conditions can be derived by first adding eqs. (3.1) and (3.2) to obtain a relation between the diffusion intensity and the flux:

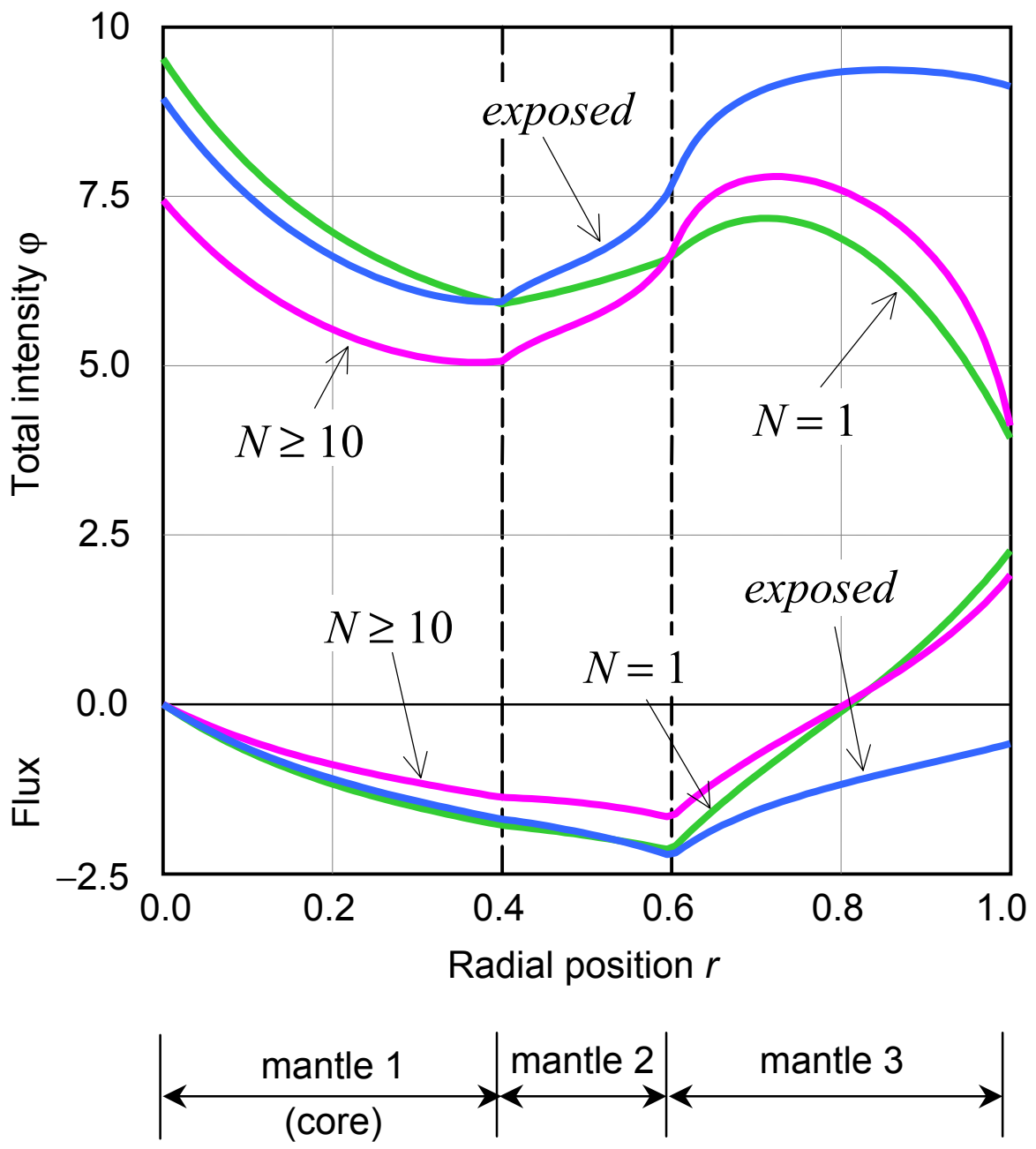

**Figure 3.19**. Graphs of the total intensity and the flux in a hot non-homogeneous sphere 1 of Table I. The label $N$ is for the discrete ordinates. The graph with label *exposed* is generated with $N = 10$.

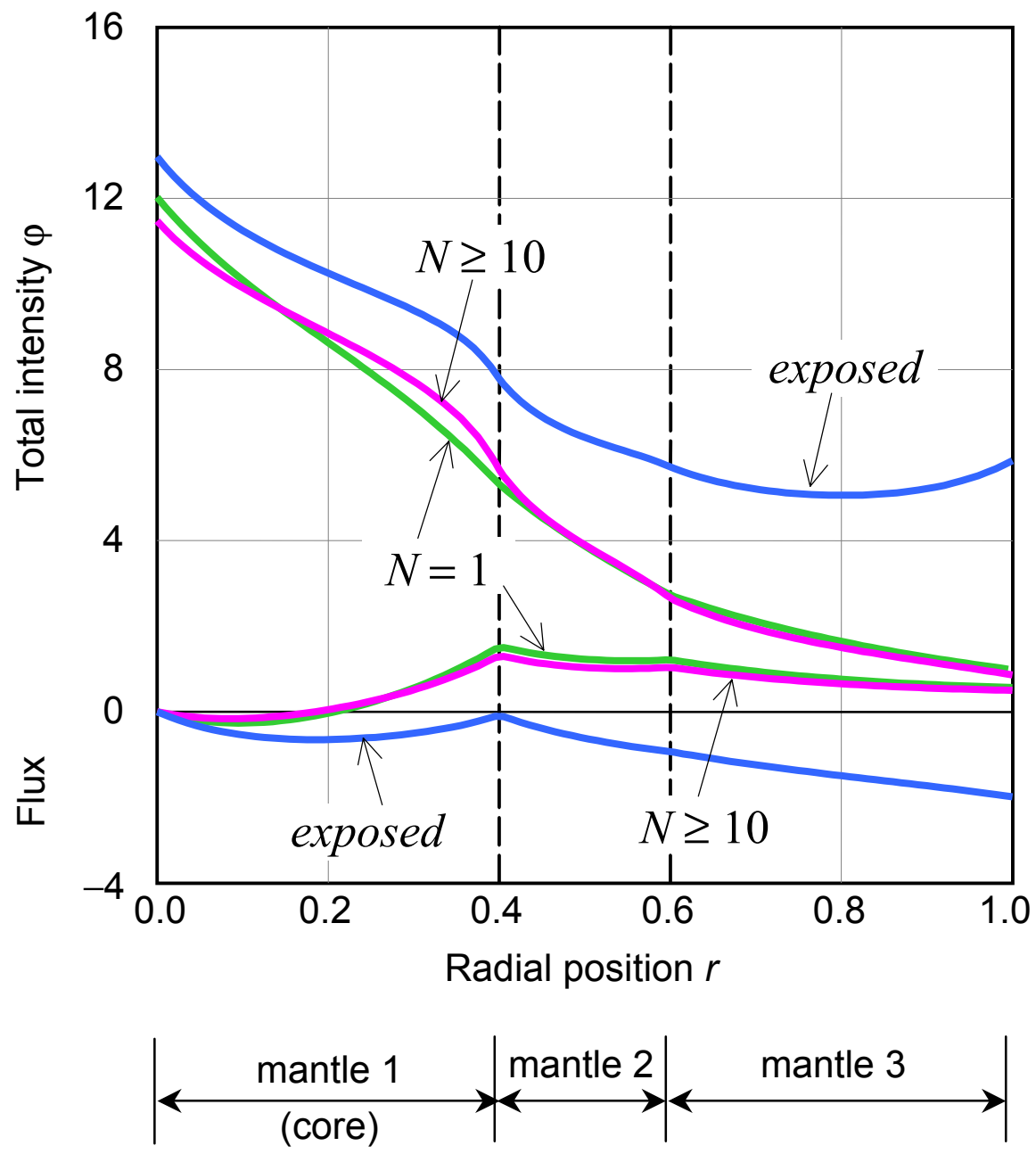

**Figure 3.20**. Graphs of the total intensity and the flux in a hot non-homogeneous sphere 2 of Table I. The label $N$ is for the discrete ordinates. The graph with label *exposed* is generated with $N = 10$.

$$\partial_r \phi + \frac{2}{\vartheta}\phi + \frac{\lambda}{\mu} f_\mu = 0 \tag{3.64}$$

Multiply eq. (3.64) by $\vartheta^2$, differentiate once with respect to $r$ and eliminate the flux term using the conservation relation (1.11) to obtain:

$$\partial_r^2 \phi + \frac{4}{\vartheta}\partial_r \phi + \left[\frac{2}{\vartheta^2} - \lambda^2\right]\phi = -\lambda^2 B \tag{3.65}$$

This is another RMS diffusion equation. It is not reconcilable with Helmholtz diffusion equation in two accounts: the coefficient $2/\vartheta^2$ of the intensity $\phi$, and the factor 4 of the first derivative. This result is expected. We derived eq. (3.65) from a transfer equation where we have set the angular derivative equal to zero. If the angular derivative were approximated with the standard finite difference, then we would obtain the Helmholtz equation exactly in $\varphi$. That is, Helmholtz equation is integrated over $\mu \in [-1,+1]$, but eq. (3.65) is still awaiting integration over $\mu \in [0,1]$. If we integrate eq. (3.65) over the direction cosine we obtain an integro-differential equation:

$$\partial_r^2 \varphi + \frac{4}{\vartheta}\partial_r \varphi + \frac{2}{\vartheta^2}\varphi - \kappa^2 \int_{\zeta\to 0}^{1} \phi \mu^{-2}\, d\mu = -\kappa^2 B \int_{\zeta\to 0}^{1} \mu^{-2}\, d\mu \tag{3.66}$$

We realize that this is not a pleasant integral, but there is no advantage to integrate the differential equation over the direction cosine. We did not do this practice in the previous sections. To avoid the impasse of $\mu \to 0$, we solve eq. (3.65) for the diffusion intensity $\phi$ then we integrate the solution on the angular domain to obtain the total intensity $\varphi$.

The general solution of eq. (3.65) can be obtained directly without any hint of the solution obtained from the set of first order transfer equations. We begin our search for the general solution by first noticing that in empty space, i.e., $\lambda = 0$, eq. (3.65) is Euler second order equation. The solutions of this equation are $\vartheta^{-2}$ and $\vartheta^{-1}$. On the other hand, the attenuation in $\lambda$ in a pure cold absorber is going to be $e^{-\lambda r}$ and $e^{+\lambda r}$. Therefore, the solution of the homogeneous part of eq. (3.65) is a linear combination of the four possible solutions: $e^{\pm\lambda r}/\vartheta^2$ and $e^{\pm\lambda r}/\vartheta$. The last two solutions in $1/\vartheta$ do not satisfy the equation, thus we retain only the solutions in $1/\vartheta^2$. We seek a particular solution quadratic in $1/\vartheta$ and we determine its coefficients as we did in sec. 3.4(a). The general solution for our second order diffusion equation may be cast in the form:

$$\phi = \frac{e^{-\lambda r}}{\vartheta^2} C_1 + \frac{e^{\lambda r}}{\vartheta^2} C_2 + \frac{B}{\vartheta^2}\left[\vartheta^2 + \frac{2}{\lambda^2}\right] \tag{3.67}$$

To determine the constants of integration $C_1$ and $C_2$ we first apply Ficks' law to eq. (3.67) and evaluate the whole at the center of the sphere where $F_{\mu,0} = 0$ to obtain:

$$C_1 = C_2 + 2BR/\lambda$$

This identity reduces eq. (3.67) to:

$$\phi = \frac{\cosh \lambda r}{\vartheta^2} C_2 + \frac{B}{\vartheta^2}\left[\frac{2R}{\lambda} e^{-\lambda r} + \vartheta^2 + \frac{2}{\lambda^2}\right] \qquad R \ge r \ge 0 \tag{3.68}$$

Insert eq. (3.68) into the conservation relation (1.11) and integrate over the volume of the sphere in the shifted frame of reference:

$$4R^2 F_{\mu,R} + \kappa \int_0^R \vartheta^2 \phi \, \mathrm{d}r = \kappa B \int_0^R \vartheta^2 \, \mathrm{d}r \tag{3.69}$$

where the flux $F_{\mu,R}$ is obtained from Ficks' law evaluated at $R$. The solution of this equation gives the expression for the constant $C_2$ as a function of $R$ and $\lambda$. We do not see at this point the advantage of working with the second order diffusion equation. What future developments will add to the problem is yet to be revealed.

PART IV

# THE DISCRETE ORDINATES TRANSFER EQUATIONS

## 4.1 The set of coupled equations

The transfer equation is now ready for discretization with the proposed discrete ordinates formalism. Whether we are working in the *c*-frame or the *s*-frame, depending on the configuration of the problem, the discrete ordinates are the same in either frame. The shifted equations in the *n*th direction on $S_{2N}$ may be written in the form:

$$\frac{\mu_n}{\vartheta^2}\,\partial_r\left(\vartheta^2\psi_n^+\right) + \frac{1}{\vartheta}\,\frac{\delta\left[\nu^2\,\psi^+\right]_n}{w_n} + \kappa\psi_n^+ = \kappa B \tag{4.1}$$

$$-\frac{\mu_n}{\vartheta^2}\,\partial_r\left(\vartheta^2\psi_n^-\right) - \frac{1}{\vartheta}\,\frac{\delta\left[\nu^2\,\psi^-\right]_n}{w_n} + \kappa\psi_n^- = \kappa B \tag{4.2}$$

Define the parameters:

$$\beta_n^n = \frac{\nu_n^2}{w_n\mu_n};\quad \beta_n^{n-1} = \frac{\nu_{n-1}^2}{w_n\mu_n};\quad \lambda_n = \frac{\kappa}{\mu_n} \tag{4.3}$$

These parameters have the following properties:

$\beta_n^k > 0,\quad \forall n,\, k \neq N$

$\beta_n^{n-1} > \beta_n^n\ ,\ \forall n > 1$

$\beta_N^N = 0,\quad \forall N$

$N \to \infty \;\Rightarrow\; \beta_1^1 \to \infty$

$\beta_1^0$ not defined, fill its place by a zero

smaller $n$ $\Rightarrow$ larger $\beta$ terms

The second property comes from the fact that $\mu_n > \mu_{n-1} \Rightarrow \nu_{n-1} > \nu_n$. Make use of identities (4.3) and rearrange eq. (4.1) to read:

$$\partial_r\psi_n^+ + \left[\frac{2+\beta_n^n}{\vartheta} + \lambda_n\right]\psi_n - \frac{1}{\vartheta}\beta_n^{n-1}\psi_{n-1}^+ = \lambda_n B \tag{4.4}$$

Also, eq. (4.2) rearranged reads:

$$\partial_r\psi_n^- + \left[\frac{2+\beta_n^n}{\vartheta} - \lambda_n\right]\psi_n^- - \frac{1}{\vartheta}\beta_n^{n-1}\psi_{n-1}^- = -\lambda_n B \tag{4.5}$$

It is readily apparent that for $N = 1$ these two equations are $S_2$ discrete ordinates equations, a set of RMS diffusion equations that can be represented by (3.1) and (3.2) of the incomplete diffusion model.

The two equations (4.4) and (4.5) differ from those of the sec. 3.2 by the presence of nonzero $\beta$ terms. This is enough of a complication to make a quantitative analysis of the homogeneous parts of eqs. (4.4) and (4.5) intractable. A qualitative analysis is still instructive however limited it may be. On physical ground we must have:

$$\psi_n^+ > \psi_{n-1}^+ \Rightarrow \beta_n^n \psi_n^+ > \beta_n^{n-1} \psi_{n-1}^+$$

Therefore, the derivative of the centrifugal intensity, eq. (4.4), is always negative on [0, $R$]. This property guarantees that the centrifugal intensity decreases monotonically with increasing $r$. The same conclusion applies on the domain [$\varepsilon$, $R$] whether $R$ is finite or infinite.

By contrast, for some set of directions there exist values for $\beta$ for which the homogeneous part of the centripetal equation (4.5) vanishes at some radial position, say $r_n$, that is different for different directions. When such a value of $r_n > 0$ exists the intensity has a saddle point. Unfortunately, this radial position cannot be determined before the solution is carried out. This is because the $n$th intensity at $r_n$ is coupled with the intensities from the solutions in all the other directions in a cascade relationship. How this coupling relationship will determine the saddle point, if there is one, is not clear at this point. By induction from our analysis in Part III we expect that saddle points in high order discrete ordinates are bound to exist for at least a few directions clustered around $N$ where $\beta_n^n < 2$.

## 4.2 Matrix representation of the transfer equations

We develop the equations in the $s$-frame since in most configurations both centripetal and centrifugal intensities compete in setting the radiation field. When the problem under consideration does not require shift, the symbol $\vartheta$ of the shifted equations can be safely replaced by $r$, no other changes are needed.

### (*a*) *The centrifugal equation*

Define the matrix elements[*]:

$$\mathrm{B}_n^n = \left(2+\beta_n^n\right); \quad \mathrm{B}_n^{n-1} = \beta_n^{n-1}; \quad \Lambda_n^n = \lambda_n \tag{4.6}$$

and rewrite eq. (4.4) in matrix form:

$$\partial_r \begin{bmatrix} \psi_1^+ \\ \psi_2^+ \\ \psi_3^+ \\ \cdots \\ \psi_N^+ \end{bmatrix} = - \begin{bmatrix} \frac{1}{\vartheta}\mathrm{B}_1^1 + \Lambda_1^1 & 0 & 0 & \cdots & 0 \\ -\frac{1}{\vartheta}\mathrm{B}_2^1 & \frac{1}{\vartheta}\mathrm{B}_2^2 + \Lambda_2^2 & 0 & \cdots & 0 \\ 0 & -\frac{1}{\vartheta}\mathrm{B}_3^2 & \frac{1}{\vartheta}\mathrm{B}_3^3 + \Lambda_3^3 & \cdots & 0 \\ \cdots & \cdots & \cdots & \cdots & \cdots \\ 0 & 0 & 0 & \cdots & \frac{1}{\vartheta}\mathrm{B}_N^N + \Lambda_N^N \end{bmatrix} \begin{bmatrix} \psi_1^+ \\ \psi_2^+ \\ \psi_3^+ \\ \cdots \\ \psi_N^+ \end{bmatrix} + B \begin{bmatrix} \lambda_1 \\ \lambda_2 \\ \lambda_3 \\ \cdots \\ \lambda_N \end{bmatrix} \tag{4.7}$$

---

[*] There should be no confusion between the beta matrix element $\mathrm{B}_n^k$ and the black body constant $B$. The latter is in italic type font and not indexed.

This equation may be written in condensed form as:

$$\partial_r \mathbf{\Psi}^+ = -\left(\frac{1}{\vartheta}\mathbf{B} + \mathbf{\Lambda}\right)\mathbf{\Psi}^+ + B\boldsymbol{\lambda} = -\tilde{\mathbf{H}}\mathbf{\Psi}^+ + B\boldsymbol{\lambda} \tag{4.8}$$

This is the $S_{2N}$ matrix equation on [0, $R$] for the centrifugal intensity. The mathematical boundary condition at the lower end of the domain of definition, i.e., at the center of the sphere, is $\mathbf{\Psi}_0^+$. Although this boundary condition is not known *a priori*, it will be resolved from the solution of the centripetal equation. When the problem at hand is defined on [ε, $R$], the natural boundary condition will have to be $\mathbf{\Psi}_\varepsilon^+$, which has to be known *a priori*.

### (*b*) *The centripetal equation*

Make use of the identities (4.6) and rewrite eq. (4.5) in matrix form:

$$\partial_r \begin{bmatrix} \psi_1^- \\ \psi_2^- \\ \psi_3^- \\ \cdots \\ \psi_N^- \end{bmatrix} = -\begin{bmatrix} \frac{1}{\vartheta}\mathrm{B}_1^1 - \Lambda_1^1 & 0 & 0 & \cdots & 0 \\ -\frac{1}{\vartheta}\mathrm{B}_2^1 & \frac{1}{\vartheta}\mathrm{B}_2^2 - \Lambda_2^2 & 0 & \cdots & 0 \\ 0 & -\frac{1}{\vartheta}\mathrm{B}_3^2 & \frac{1}{\vartheta}\mathrm{B}_3^3 - \Lambda_3^3 & \cdots & 0 \\ \cdots & \cdots & \cdots & \cdots & \cdots \\ 0 & 0 & 0 & \cdots & \frac{1}{\vartheta}\mathrm{B}_N^N - \Lambda_N^N \end{bmatrix} \begin{bmatrix} \psi_1^- \\ \psi_2^- \\ \psi_3^- \\ \cdots \\ \psi_N^- \end{bmatrix} - B \begin{bmatrix} \lambda_1 \\ \lambda_2 \\ \lambda_3 \\ \cdots \\ \lambda_N \end{bmatrix} \tag{4.9}$$

or, using matrix notation:

$$\partial_r \mathbf{\Psi}^- = -\left(\frac{1}{\vartheta}\mathbf{B} - \mathbf{\Lambda}\right)\mathbf{\Psi}^- - B\boldsymbol{\lambda} = -\mathbf{H}\mathbf{\Psi}^- - B\boldsymbol{\lambda} \tag{4.10}$$

This is the $S_{2N}$ matrix equation on [0, $R$] for the centripetal intensity. The equation (4.10) is an initial value problem in space. Therefore, the mathematical boundary condition must be $\mathbf{\Psi}_0^-$ the boundary value at the lower limit of the domain of definition. This boundary condition is not known *a priori* but it will be resolved in terms of the natural boundary condition $\mathbf{\Psi}_R^-$ by inverting the mathematical solution.

### (*c*) *The transfer matrix equation*

The two equations (4.8) and (4.10) merged together yield the matrix equation of the radiative transfer equation defined on [0, $R$]:

$$\partial_r \begin{bmatrix} \mathbf{\Psi}^- \\ \mathbf{\Psi}^+ \end{bmatrix} = \begin{bmatrix} -\mathbf{H} & \mathbf{0} \\ \mathbf{0} & -\tilde{\mathbf{H}} \end{bmatrix} \begin{bmatrix} \mathbf{\Psi}^- \\ \mathbf{\Psi}^+ \end{bmatrix} + B \begin{bmatrix} -\boldsymbol{\lambda} \\ +\boldsymbol{\lambda} \end{bmatrix} \tag{4.11}$$

or, in compact form:

$$\partial_r \mathbf{\Psi} = \mathbf{G}\mathbf{\Psi} + B\boldsymbol{\lambda} \tag{4.12}$$

and the boundary condition is the vector:

$$\mathbf{\Psi}_0 = \left[\mathbf{\Psi}_0^- \ \ \mathbf{\Psi}_0^+\right]^T \tag{4.13}$$

At the center of the sphere we must have $\mathbf{\Psi}_0^- = \mathbf{\Psi}_0^+$. We kept the superscripts plus and minus in place merely for notational convenience. They have no bearing on the solution. The formal solution for eq. (4.12) may be written in the form:

$$\mathbf{\Psi} = \mathcal{G}(r)\,\mathbf{\Psi}_0 + B\int_0^r \mathcal{G}(r-x)\,\boldsymbol{\lambda}\,\mathrm{d}x \tag{4.14}$$

The lower limit of the integral of this equation is set to 0 consistently with the subscript of the intensity vector. The lower limit of the integral and the subscript of the boundary intensity vector should be ε if the problem is defined on [ε, $R$].

The function $\mathcal{G}(r)$ is the fundamental matrix of eq. (4.12). It follows that the solution for the transfer equation in the proposed configuration reduces to determining the fundamental matrix $\mathcal{G}$. However, eq. (4.11) is instructive to the effect that the centripetal and the centrifugal equations are not coupled when they are expressed in terms of the boundary condition $\mathbf{\Psi}_0$. Therefore, it is less cumbersome to solve the centripetal and the centrifugal equations individually in terms of $\mathbf{\Psi}_0$ and then resolve the solution in terms of the natural boundary condition by the appropriate inversion.

## 4.3 The cold medium

### (*a*) *Mantle surrounded by vacuum*

The problem of sec. 3.3(*a*) is revisited but now with the angular derivative in place. The solution of the homogeneous part of equation (4.8) yields the intensity vector $\mathbf{\Psi}^+$ on the domain [ε, $R$] in the $c$-frame frame of reference:

$$\mathbf{\Psi}_r^+ = e^{\tilde{\Gamma}(r)}\,\mathbf{\Psi}_\varepsilon^+ \qquad r \geq \varepsilon \tag{4.15}$$

where $\tilde{\mathbf{\Gamma}}(r)$ is the integral of the matrix coefficient $\tilde{\mathbf{H}}$ of eq. (4.8):

$$\tilde{\mathbf{\Gamma}}(r) = \mathbf{B}\ln\rho - \mathbf{\Lambda}(r-\varepsilon)\ ; \quad \rho = \varepsilon/r \tag{4.16}$$

Its elements are:

$$\tilde{\Gamma}_n^n(r) = \mathrm{B}_n^n \ln\rho - \lambda_n(r-\varepsilon) \tag{4.17}$$

$$\tilde{\Gamma}_n^{n-1}(r) = -\mathrm{B}_n^{n-1}\ln\rho \tag{4.18}$$

and the matrix $\tilde{\mathbf{\Gamma}}$ has the same pattern as the coefficient matrix of eq. (4.7):

$$\tilde{\mathbf{\Gamma}} = \begin{bmatrix} \tilde{\Gamma}_1^1 & 0 & 0 & \cdots & 0 \\ \tilde{\Gamma}_2^1 & \tilde{\Gamma}_2^2 & 0 & \cdots & 0 \\ 0 & \tilde{\Gamma}_3^2 & \tilde{\Gamma}_3^3 & \cdots & 0 \\ \cdots & \cdots & \cdots & \cdots & \cdots \\ 0 & 0 & 0 & \cdots & \tilde{\Gamma}_N^N \end{bmatrix} \tag{4.19}$$

It is a simple mathematical exercise to demonstrate that the eigenvalues of matrix $\tilde{\mathbf{\Gamma}}$ are the diagonal elements expressed as identity (4.17), and the eigenvectors are obtained by the recursion relations:

$$\left.\begin{array}{lll} \widetilde{V}_n^n = 1 & & \forall n \\ \widetilde{V}_n^k = 0 & & \text{for } k > n \\ \widetilde{V}_n^k = \dfrac{\widetilde{\Gamma}_n^{n-1}}{\widetilde{\Gamma}_k^k - \widetilde{\Gamma}_n^n}\widetilde{V}_{n-1}^k & & \text{for } n = 2,3,\ldots,N;\ k < n \end{array}\right\} \tag{4.20}$$

There is no summation implied in the third identity (4.20). It follows that the eigentable $\widetilde{\mathbf{V}}$ built with the eigenvectors (4.20), and its inverse $\widetilde{\mathbf{V}}^{-1}$, are lower triangular matrices that diagonalize the exponential matrix of eq. (4.15). These matrices are functions of the independent variable *r*. The computational representation of the solution expressed by (4.15) may be written in the form:

$$\mathbf{\Psi}_r^+ = \widetilde{\mathbf{V}}\left[e^{\widetilde{\Gamma}_n^n}\right]\widetilde{\mathbf{V}}^{-1}\mathbf{\Psi}_\varepsilon^+ \tag{4.21}$$

Insert the value for $\widetilde{\Gamma}_n^n$ of (4.17) into this equation and rearrange to obtain:

$$\mathbf{\Psi}_r^+ = \rho^2\widetilde{\mathbf{V}}\left[\rho^{\beta_n^n}e^{-\lambda_n(r-\varepsilon)}\right]\widetilde{\mathbf{V}}^{-1}\mathbf{\Psi}_\varepsilon^+ \tag{4.22}$$

Define the lower triangular spherical attenuation operator:

$$\widetilde{\mathbf{S}}_r \equiv \widetilde{\mathbf{V}}\left[\rho^{\beta_n^n}e^{-\lambda_n(r-\varepsilon)}\right]\widetilde{\mathbf{V}}^{-1} \qquad r \in [\varepsilon, \infty] \tag{4.23}$$

then:

$$\psi_r^+ = \rho^2\,\widetilde{\mathbf{S}}_r\mathbf{\Psi}_\varepsilon^+ \tag{4.24}$$

and:

$$\varphi_r = \frac{1}{2}\sum_{n=1}^{N} w_n\psi_{r,n}^+ \tag{4.25}$$

Some useful properties of the attenuation operator:

$$r \to \infty \quad \Rightarrow \quad \widetilde{\mathbf{S}}_\infty \to \mathbf{0} \quad \text{(null matrix)}$$

$$r = \varepsilon \quad \Rightarrow \quad \widetilde{\mathbf{S}}_\varepsilon = \mathbf{I} \quad \text{(identity matrix)}$$

Additionally, for arbitrarily large *N* and $r >> \varepsilon \Leftrightarrow \rho << 1$, all the β's are positive except $\beta_N^N = 0$. In this asymptotic limit $\widetilde{\mathbf{S}}_r \to e^{-\kappa r}$ since all exponentials in $\kappa/\mu$ vanish and do not contribute to the quadrature. In this limit the spherical operator $\widetilde{\mathbf{S}}_r$ yields the point kernel of the point source problem in $S_\infty$ regardless of how large or small ε may be.

The total intensity calculated from the expression (4.25) is shown graphically in figure 4.1 for various values of *N*. The sample sphere is a cold homogeneous mantel surrounded by vacuum and embedding a source of isotropic radiance normalized to 1. It appears that the intensity saturates graphically with relatively small number of discrete ordinates. The effect of the omission of the angular derivative from the transfer equation is demonstrated graphically in figure 4.2. The incomplete diffusion underestimates the intensity and the flux in the cold mantle configuration. This is expected since the total intensity in this problem is obtained on one-half of the domain of the angular cosine: the angular derivative does not vanish upon integration over the direction cosine on the interval [0, 1]. This conclusion is not valid if we have scattering. In that case eq. (4.8) standing alone is not sufficient to describe the transfer problem.

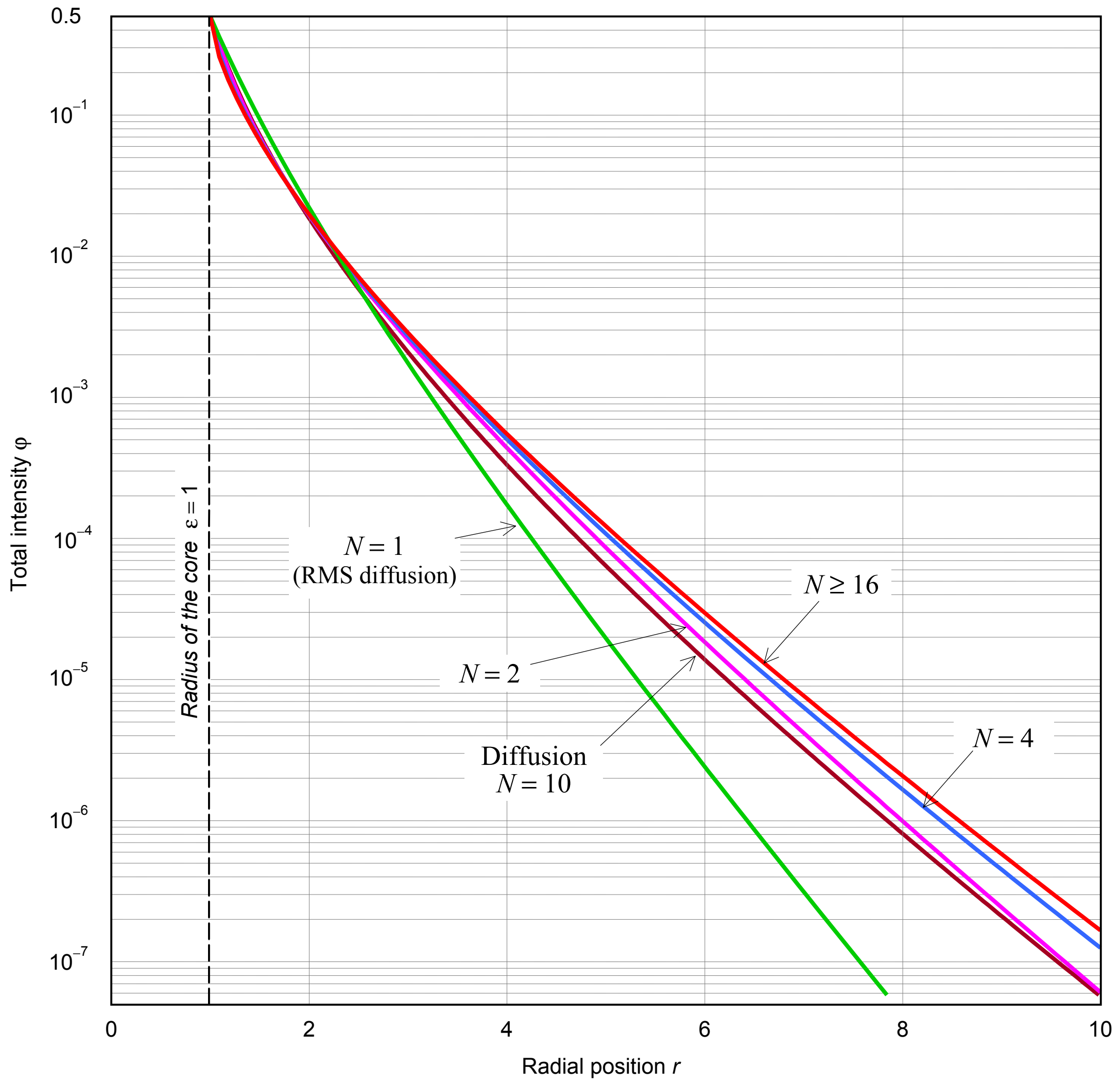

**Figure 4.1**. Graphs of the total intensity in a cold medium embedding a core of isotropic radiance normalized to 1. The opacity of the medium is $\kappa = 1$. The graphs are generated in $S_{2N}$ discrete ordinates. A graph for the diffusion intensity is shown for comparison.

### (*b*) *Multi-layered medium*

The quantitative comparison, illustrated by figures 4.1 and 4.2, between the intensities and the fluxes obtained with the incomplete diffusion model and the discrete ordinates model are conclusive in favor of the discrete ordinates model: we cannot drop the angular derivative from the transfer equation and expect correct predictions for the intensity and the flux in this cold mantle configuration. Therefore, the expressions developed in sec. 3.4(*b*) using the incomplete diffusion model are not valid for the cold multi-layered medium hosting a core and its free surface is exposed to vacuum. For this problem, the discrete ordinates solution is expressly needed. To obtain this solution, in the *c*-frame, we follow the nomenclature described in sec. 3.4(*b*) and shown in figures 3.17 and 3.18. The solution given by eq. (4.21) now reads:

$$\mathbf{\Psi}_m^+ = \widetilde{\mathbf{V}}_m \left[ e^{\widetilde{\Gamma}_{m,n}^n} \right] \widetilde{\mathbf{V}}_m^{-1} \mathbf{\Psi}_{m-1}^+ \qquad m = 2, 3, \ldots, M \tag{4.26}$$

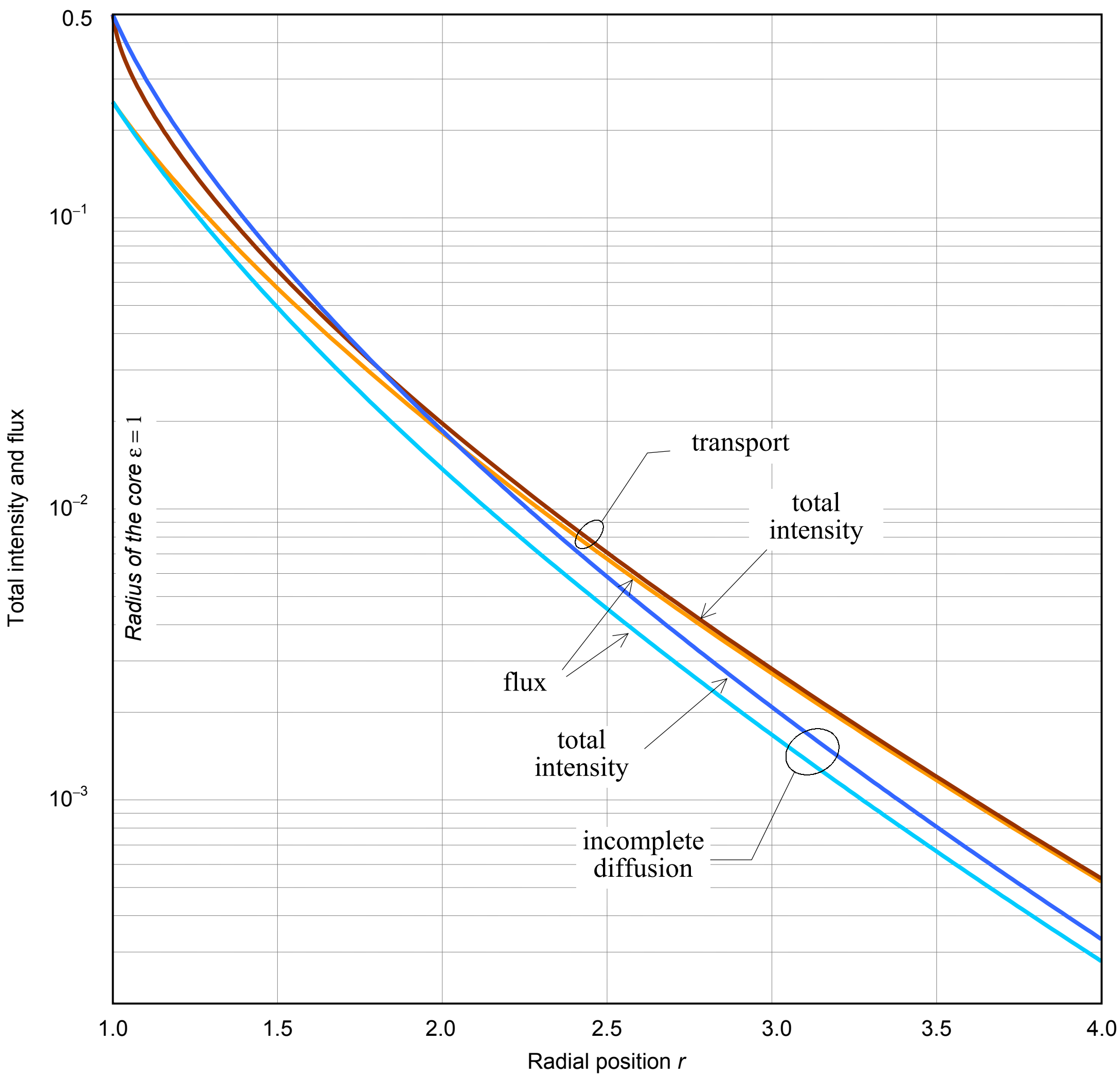

**Figure 4.2**. Graphs of the total intensity and the flux in a cold mantle generated from the solution of the incomplete diffusion equation and the transport discrete ordinates equations. The opacity of the mantle is $\kappa = 1$ and the radiance of the core is isotropic normalized to 1. All graphs are generated with $N = 10$.

where $M$ is the number of media and $m = 1$ is the index for the core*, $m = 2$ is the index for the first mantle, $m = 3$ the index for the second mantle, ... etc. In this nomenclature the radiance of the core is now $\psi_1^+$ where the subscript 1 takes the place of ε. Consequently, eq. (3.22) transcribes into the end-points solution:

$$\mathbf{\Psi}_m^+ = \rho_m^2 \widetilde{\mathbf{V}}_m^{m-1} \left[ \rho_m^{\beta_n^n} e^{-\lambda_{m,n}(R_m - R_{m-1})} \right] \left( \widetilde{\mathbf{V}}_m^{m-1} \right)^{-1} \mathbf{\Psi}_{m-1}^+ \tag{4.27}$$

where:

$$\rho_m = \frac{R_{m-1}}{R_m} \tag{4.28}$$

Define:

$$\widetilde{\mathbf{S}}_m^{m-1} = \widetilde{\mathbf{V}}_m^{m-1} \left[ \rho_m^{\beta_n^n} e^{-\lambda_{m,n}(R_m - R_{m-1})} \right] \left( \widetilde{\mathbf{V}}_m^{m-1} \right)^{-1} \tag{4.29}$$

Then, the transmittance becomes:

* This indexing pattern relates to figure 3.18. The index of the core could be 0 as well.

$$\mathbf{T}_m^{m-1} = \rho_m^2 \widetilde{\mathbf{S}}_m^{m-1} \tag{4.30}$$

and the recurrence relation follows:

$$\mathbf{\Psi}_m^+ = \mathbf{T}_m^{m-1} \mathbf{\Psi}_{m-1}^+ \qquad m = 2, 3, ..., M \tag{4.31}$$

and the total intensity is obtained using eq. (4.25); the summation in this case is on the running index $n = 1, ..., N$, the index $m$ is reserved for the material.

We experimented numerically with eq. (4.31). The graph of the total intensity in a composite mantle made of different materials is shown in figure 4.3. The core is a sphere of radius $\varepsilon = 1$ and its radiance is isotropic and normalized to 1. The opacity of the first mantle is $\kappa_1 = 0.1$, the second mantle $\kappa_2 = 2$ and the third mantle $\kappa_3 = 1$. The outer surface of the third mantle is exposed to vacuum but could be perceived as an infinite medium as well. It appears from the graphs that the total intensity saturates graphically relatively quickly with increasing order of the discrete ordinates, like all the cases we have considered in the previous sections.

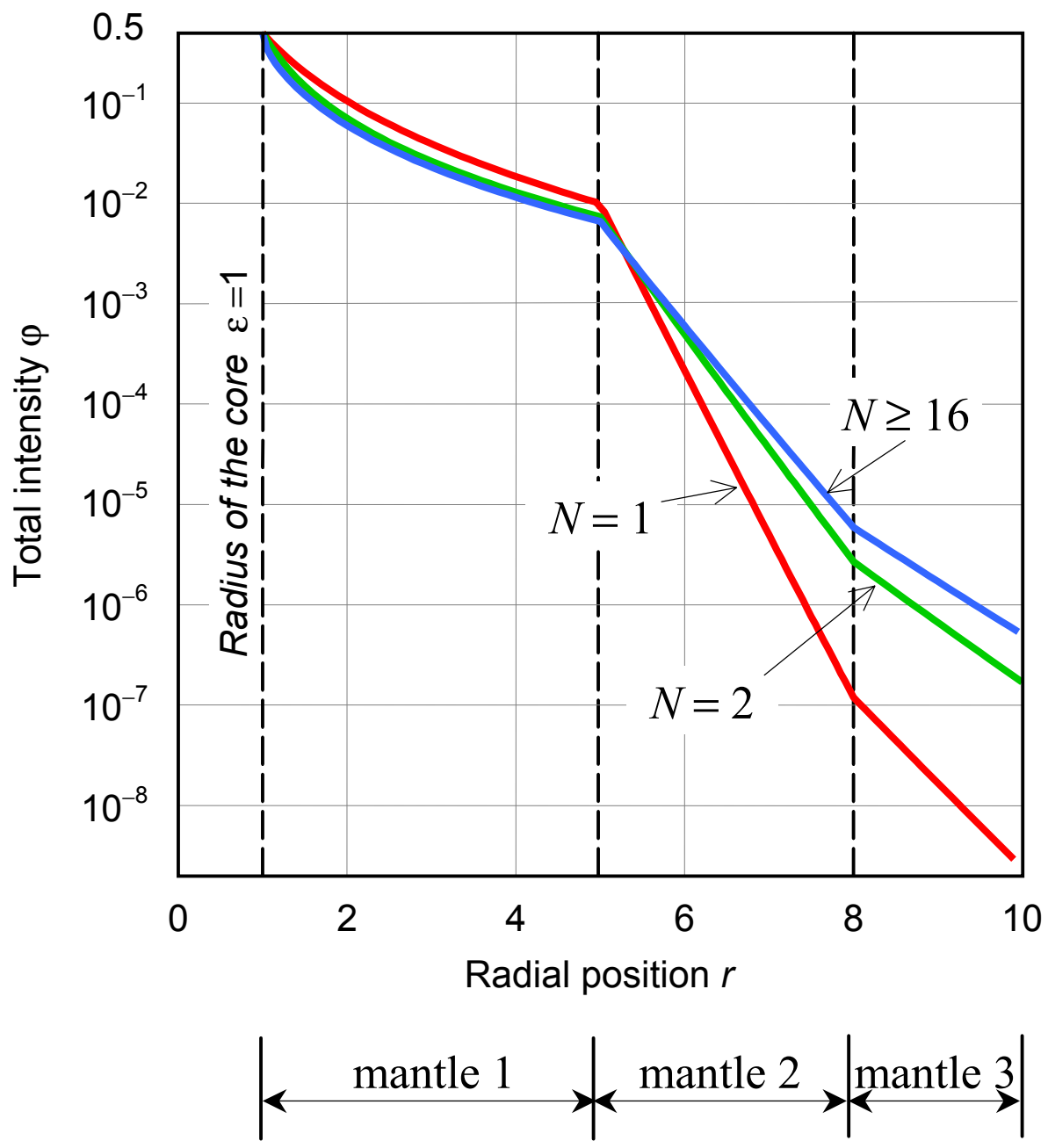

**Figure 4.3**. Graphs of the total intensity in a cold medium made of three homogeneous absorbing mantles of different opacities. The core is a source its radiance is isotropic normalized to 1.

## (*c*) *Effectiveness of the end-points solution*

The end-points solution expressed by eq. (4.31) is a powerful tool for the numerical analysis of the intensity in a cold heterogeneous as well as a homogeneous mantle. Unfortunately, the graphs shown in figure 4.4 reveal a discrepancy between the continuous in $r$ solution and the end-points solution as if one of the two solutions is more accurate than the other. We contend that two solutions are analytically equally accurate. However, if the mantle is optically thick, the end-points solution is more accurate in finite arithmetics than the continuous in $r$ solution.

The explanation of the discrepancy may be extracted from the analysis of the exponential term of the diagonal matrix of the attenuation operator **S**. At large $r$ deep in the mantle, the exponent $\lambda_n(r - \varepsilon)$ of eq. (4.22) becomes much larger than $\lambda_{n,m}(R_m - R_{m-1})$ of eq. (4.27). The former optical depth inflicts the intensity a greater attenuation than what it would have been the case with smaller exponents used recursively through the desired optical depth, which is case with the end-points solution. We emphasize that this is only true in finite arithmetics; otherwise the two forms of the solution are equally accurate.

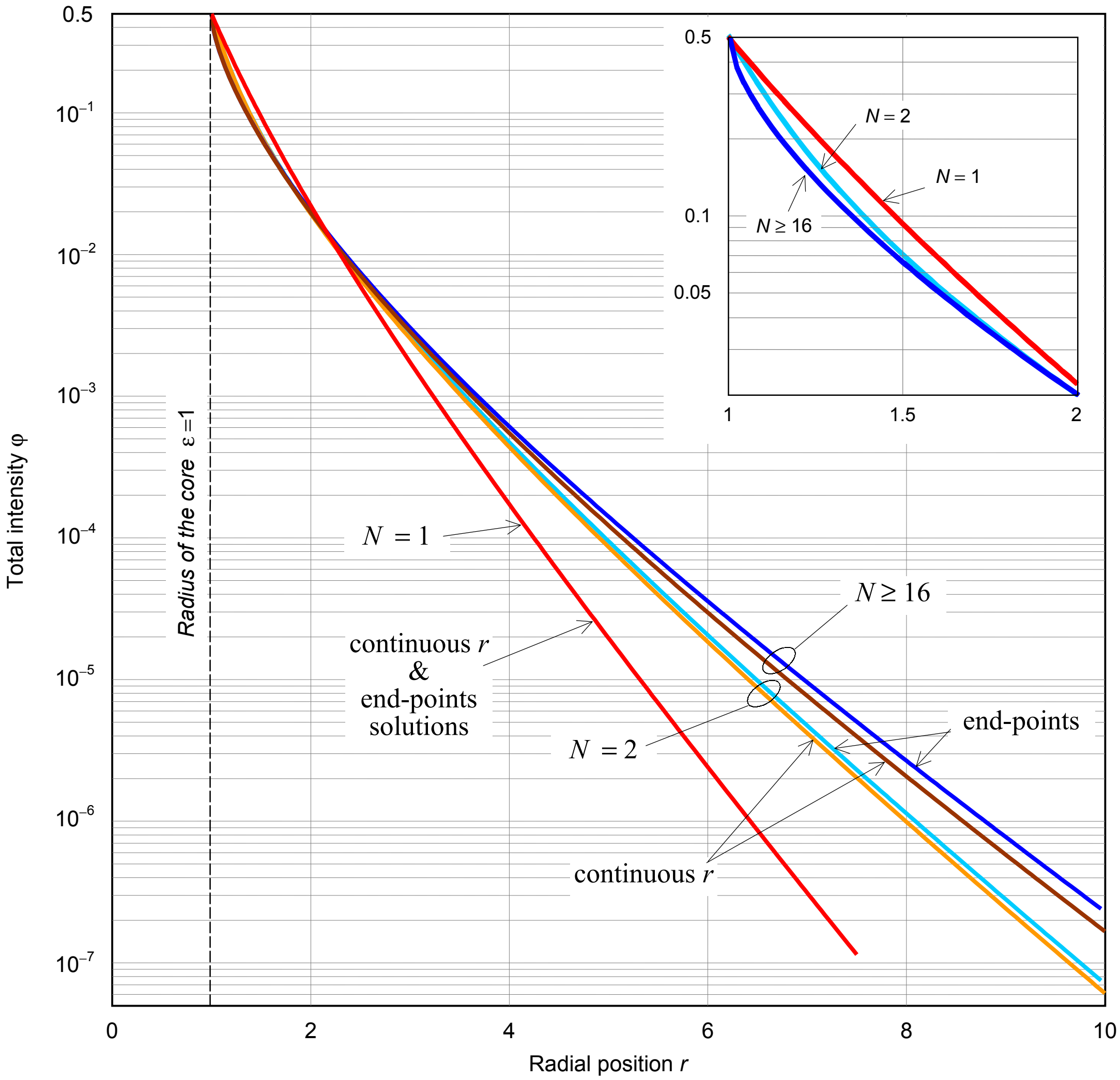

**Figure 4.4**. Graphs of the total intensity generated with the continuous *r* solution and the end-points solution of the discrete ordinates equations in a mantle of opacity κ=1; the label *N* indicates the order of the discrete ordinates. The graphs from both of the solutions are superimposed in the first 2 mfp of the mantle; the first mfp is shown in the blow-up.

In order to confirm our contention quantitatively, we experimented further by setting $R_{m-1} = \varepsilon$ and $R_m = r$ for several values of small and large *r*. The two formulations of the intensity, namely, the continuous in *r* solution and the end-points solution, reproduced exactly the same numerical values for the intensity at all values of *r*. This result is informative about the consistency and the correctness of the numerical formulation of the solution but it is not sufficient to prove our contention. To that end we compared the intensity calculated using eqs. (4.22) and (4.31) with the *path integrated intensity*. We obtained the expression of this intensity as follows.

Consider a spherical surface of radius ε characterized by its isotropic radiance $\psi_\varepsilon^+$ and embedded in a cold medium of opacity κ. A pencil of radiation emitted from a point *P* at the surface intersects with the radial axis at *Q* making an angle *θ* with the axis (fig. 4.5). The distance *s* traveled along this path is such that:

$$\varepsilon^2 = s^2 + r^2 - 2rs\mu \tag{4.32}$$

where $\mu = \cos\theta$ and $r$ is the radial position of point $Q$, which can be extended to infinity as desired. Solve eq. (4.32) for $s > 0$ and rearrange to obtain:

$$s = r\left(\mu - \sqrt{\mu^2 - \left(1 - \rho^2\right)}\right) \tag{4.33}$$

where $\rho = \varepsilon/r \le 1$ and $\rho = 0$ at $r \to \infty$. At fixed $r$, the maximum of $s(\mu)$ occurs when the radical of eq. (4.33) is equal to zero, which corresponds to a minimum of $\mu$. Hence:

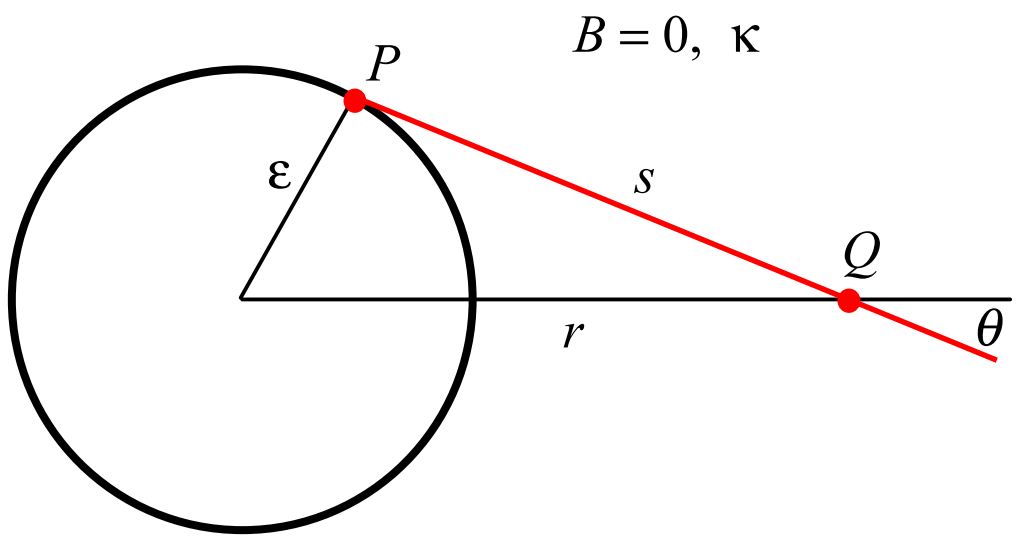

**Figure 4.5**. Geometry of the path integrated transfer problem in a cold mantle of outer radius $R$ that can be finite or infinite.

$$\mu^2 - (1 - \rho^2) \ge 0 \quad \Rightarrow \quad \min(\mu) \equiv \mu_0 = \sqrt{1 - \rho^2} \tag{4.34}$$

and the maximum of $\mu$ is 1, which corresponds to a minimum of $s$. It follows that the specific intensity at point $Q$ in a direction defined by $\mu \in [\mu_0, 1]$ is given by:

$$\psi_Q^+ = \psi_\varepsilon^+ e^{-\kappa s(\mu)} \tag{4.35}$$

The total intensity at $Q$ is obtained by integrating eq. (4.35) on $\mu$:

$$\varphi_r = \frac{\psi_\varepsilon^+}{2}\int_{\mu_0}^{1} e^{-\kappa s(\mu)} d\mu \tag{4.36}$$

and the flux:

$$F_r = \frac{\psi_\varepsilon^+}{2}\int_{\mu_0}^{1} \mu e^{-\kappa s(\mu)} d\mu \tag{4.37}$$

We evaluated these integrals numerically and compared the intensity of eq. (4.36) with the total intensity obtained using the end-points solution given by eq. (4.31). The match was perfect at all depths in the medium provided that the shells thicknesses $\lambda_{n,m}(R_m - R_{m-1})$ is not too large. By contrast, the total intensity evaluated from eq. (4.22) failed this numerical test at large depths. These observations confirm our contention about the effectiveness of the end-points solution in finite arithmetics.

### (*d*) *Solid homogeneous sphere*

The cold homogenous sphere is immersed in a uniform and isotropic radiation field. The vector $\mathbf{\Psi}_R^-$ represents the natural boundary condition for the centripetal intensity at the surface of the sphere. For the reasons discussed in sec. 3.3(c), we have to solve the homogeneous parts of the two equations (4.8) and (4.10) simultaneously:

$$\partial_r \mathbf{\Psi}^- = -\left(\frac{1}{\vartheta}\mathbf{B} - \lambda\right)\mathbf{\Psi}^- \tag{4.38}$$

$$\partial_r \mathbf{\Psi}^+ = -\left(\frac{1}{\vartheta}\mathbf{B} + \lambda\right)\mathbf{\Psi}^+ \tag{4.39}$$

Although the two equations are manifestly uncoupled, they are coupled by the intensity $\mathbf{\Psi}_0$ at the center of the sphere. This is indeed the mathematical boundary condition for both of the equa-tions. With this boundary condition the formal solutions for eqs. (4.38) and (4.39) in *s*-frame are:

$$\boldsymbol{\Psi}_r^- = \mathbf{e}^{\boldsymbol{\Gamma}(r)}\,\boldsymbol{\Psi}_0 \qquad r \in [0, R] \tag{4.40}$$

$$\boldsymbol{\Psi}_r^+ = \mathbf{e}^{\widetilde{\boldsymbol{\Gamma}}(r)}\,\boldsymbol{\Psi}_0 \qquad r \in [0, R] \tag{4.41}$$

### *Solution of the centripetal equation*

Elements of matrix $\boldsymbol{\Gamma}(r)$:

$$\Gamma_n^n = \mathrm{B}_n^n \ln\rho + \lambda_n r \qquad \rho = R/\vartheta \tag{4.42}$$

$$\Gamma_n^{n-1} = -\mathrm{B}_n^{n-1} \ln\rho$$

Eigenvalues of $\boldsymbol{\Gamma}(r)$:

$$\Gamma_n^n = \mathrm{B}_n^n \ln\rho + \lambda_n r$$

Eigentable of $\boldsymbol{\Gamma}(r)$:

$$\left.\begin{array}{ll} \mathrm{V}_n^n = 1 & \forall n \\ \mathrm{V}_n^k = 0 & \text{for } k > n \\ \mathrm{V}_n^k = \dfrac{\Gamma_n^{n-1}}{\Gamma_k^k - \Gamma_n^n}\mathrm{V}_{n-1}^k & \text{for } n = 2, 3, \ldots, N\,;\ k < n \end{array}\right\} \tag{4.43}$$

Centripetal intensity vector:

$$\boldsymbol{\Psi}_r^- = \rho^2\,\mathbf{V}_r\left[\rho^{\beta_n^n}\,e^{\lambda_n r}\right]\mathbf{V}_r^{-1}\,\boldsymbol{\Psi}_0 = \rho^2\,\mathbf{S}_r\boldsymbol{\Psi}_0 \tag{4.44}$$

Evaluate eq. (4.44) at $R$ and solve for $\boldsymbol{\Psi}_0$:

$$\boldsymbol{\Psi}_0 = 4\mathbf{V}_R\left[2^{\beta_n^n}\,e^{-\lambda_n R}\right]\mathbf{V}_R^{-1}\,\boldsymbol{\Psi}_R^- = 4\,\mathbf{S}_R^{-1}\,\boldsymbol{\Psi}_R^- \tag{4.45}$$

Insert eq. (4.45) into eq. (4.44) to obtain the complete solution for eq. (4.38):

$$\boldsymbol{\Psi}_r^- = 4\,\rho^2\,\mathbf{S}_r\,\mathbf{S}_R^{-1}\,\boldsymbol{\Psi}_R^- \tag{4.46}$$

where $\mathbf{S}_r$ is a centripetal attenuation operator, a lower triangular matrix.

### *Solution of the centrifugal equation*

Elements of matrix $\widetilde{\boldsymbol{\Gamma}}(r)$:

$$\widetilde{\Gamma}_n^n = \mathrm{B}_n^n \ln\rho - \lambda_n r \qquad \rho = R/\vartheta \tag{4.47}$$

$$\widetilde{\Gamma}_n^{n-1} = -\mathrm{B}_n^{n-1} \ln\rho$$

Eigenvalues of $\widetilde{\boldsymbol{\Gamma}}(r)$:

$$\widetilde{\Gamma}_n^n = \mathrm{B}_n^n \ln\rho - \lambda_n r$$

Eigentable of $\widetilde{\boldsymbol{\Gamma}}(r)$:

$$\left.\begin{array}{ll} \widetilde{\mathrm{V}}_n^n = 1 & \forall n \\ \widetilde{\mathrm{V}}_n^k = 0 & \text{for } k > n \\ \widetilde{\mathrm{V}}_n^k = \dfrac{\widetilde{\Gamma}_n^{n-1}}{\widetilde{\Gamma}_k^k - \widetilde{\Gamma}_n^n}\widetilde{\mathrm{V}}_{n-1}^k & \text{for } m = 2, 3, \ldots, N;\ k < n \end{array}\right\} \tag{4.48}$$

Centrifugal intensity vector:

$$\boldsymbol{\Psi}_r^+ = \rho^2\,\widetilde{\mathbf{V}}_r\left[\rho^{\beta_n^n}\,e^{-\lambda_n r}\right]\widetilde{\mathbf{V}}_r^{-1}\,\boldsymbol{\Psi}_0 = \rho^2\,\widetilde{\mathbf{S}}_r\boldsymbol{\Psi}_0 \tag{4.49}$$

Insert eq. (4.45) into eq. (4.49) to obtain the complete solution for eq. (4.39):

$$\boldsymbol{\Psi}_r^+ = 4\,\rho^2\,\widetilde{\mathbf{S}}_r\,\mathbf{S}_R^{-1}\,\boldsymbol{\Psi}_R^- \tag{4.50}$$

A property of the centrifugal attenuation operator: $\widetilde{\mathbf{S}}_0 = \mathbf{I}$. It is apparent from eq. (4.50) that despite that the sphere is a pure absorber, it has an albedo, a lower triangular matrix:

$$\boldsymbol{\alpha} = \widetilde{\mathbf{S}}_R \mathbf{S}_R^{-1} \tag{4.51}$$

The total intensity and the flux are obtained by applying the quadrature formula to the specific intensities given by eqs. (4.46) and (4.50):

$$\varphi_r = \frac{1}{2}\sum_{n=1}^{N} w_n \left(\psi_{n,r}^{+} + \psi_{n,r}^{-}\right) \tag{4.52}$$

$$F_r = \frac{1}{2}\sum_{n=1}^{N} w_n \mu_n \left(\psi_{n,r}^{+} - \psi_{n,r}^{-}\right) \tag{4.53}$$

Total intensities and fluxes in different spheres and in discrete ordinates $N \leq 48$ were calculated with eqs. (4.46) and (4.50); data obtained with $N > 48$ were plagued with numerical instabilities similar to those illustrated in figure 3.12. The graphs of these calculations superimposed perfectly with the graphs generated in the incomplete diffusion model (see Part III). This result is expected since the radiation field within the sphere is determined from all centripetal and centrifugal intensities. We have indicated in sec. 1.3 that in this case the net contribution from the angular derivative is zero.

The same congruence of the graphs was observed whether the continuous in $r$ solution is used or the end-points solution. This observation is in contrast with discrepancies observed in the cold mantle and illustrated by figure 4.4. We interpret this as follows: in the cold mantle problem we have only one **S** operator. In the sphere, we argue, we have the operator **S** and the inverse operator $\mathbf{S}^{-1}$ noticeable in eqs. (4.46) and (4.50). If the former underestimates the intensity, the latter overestimates it: one operator suppresses the numerical excursion of the other and the intensities are in order.

## 4.4 Medium under LTE conditions

We have addressed this problem in the premise of the incomplete diffusion model in sec. 3.4 and obtained the general solution complete in continuous in $r$ and in end-points form. Any further development is going to be a formal academic exercise that adds nothing to the solution we already have in central symmetry. This is because, as we have seen in the previous sections, the total intensity in such a geometric setting in an isothermal medium is independent of the angular derivative. Notwithstanding, the complete formal solution that incorporates the discrete ordinates terms of the angular derivative should be instructive.

We define our problem, as we did before, as a homogeneous isothermal sphere characterized by a black body constant $B$ and opacity $\kappa$. We assume further that the sphere is immersed in a uniform and isotropic radiation field its impinging intensity on the surface of the sphere is $\boldsymbol{\Psi}_R^{-}$. The transfer equation for this problem is described by the two equations (4.8) and (4.10). The general solution for these equations may be cast in the form:

$$\boldsymbol{\Psi}^{-} = \boldsymbol{\Psi}^{-}_{cold} + \boldsymbol{\Psi}^{-}_{hot} \tag{4.54}$$

and:

$$\boldsymbol{\Psi}^{+} = \boldsymbol{\Psi}^{+}_{cold} + \boldsymbol{\Psi}^{+}_{hot} \tag{4.55}$$

The cold centripetal and centrifugal intensities are given by eqs. (4.44) and (4.49), respectively, with the mathematical boundary condition $\boldsymbol{\Psi}_0$; the cold intensities should be set to zero if the sphere is surrounded by vacuum. The hot intensities are obtained by the standard integration of the inhomogeneous term of an ordinary first order differential equation:

$$\boldsymbol{\Psi}^{-}_{hot} = -B\int_0^r \left(\frac{\vartheta_x}{\vartheta_r}\right)^2 \mathbf{V}_{r,x}\left[\left(\frac{\vartheta_x}{\vartheta_r}\right)^{\beta_n^n} e^{\lambda_n(r-x)}\right]\mathbf{V}^{-1}_{r,x}\boldsymbol{\lambda}\,\mathrm{d}x \tag{4.56}$$

$$\boldsymbol{\Psi}^{+}_{hot} = B\int_0^r \left(\frac{\vartheta_x}{\vartheta_r}\right)^2 \widetilde{\mathbf{V}}_{r,x}\left[\left(\frac{\vartheta_x}{\vartheta_r}\right)^{\beta_n^n} e^{-\lambda_n(r-x)}\right]\widetilde{\mathbf{V}}^{-1}_{r,x}\boldsymbol{\lambda}\,\mathrm{d}x \tag{4.57}$$

where $\vartheta_x = R + x$. It is readily apparent that a closed form integration of these expressions is not possible for $N > 1$. On the other hand, numerical algorithms become sluggish for $N$ slightly larger than 1 and, for larger values of $N$, out of the box algorithms are not suitable for these integrals. This is not necessarily a loss of anything. We demonstrated earlier that the incomplete diffusion model delivers the total intensity, the flux, and eventually the energy exactly and accurately without the inclusion of the angular derivative in the transfer equation in LTE media, be it spheres or mantles in central symmetry geometry.

# REPORT DOCUMENTATION PAGE

*Form Approved*
*OMB No. 0704-0188*

| 1. REPORT DATE *(DD-MM-YYYY)* | 2. REPORT TYPE | 3. DATES COVERED *(From - To)* |
|---|---|---|
| 03-08-2001 | Final | 07/24/2000 - 07/23/2001 |

**4. TITLE AND SUBTITLE**
Radiative Transfer in Spheres I. Analytical Solutions

**5a. CONTRACT NUMBER**

**5b. GRANT NUMBER**
N00173-001-G010

**5c. PROGRAM ELEMENT NUMBER**
BAA 946 Code 6701

**6. AUTHOR(S)**
Charles H. Aboughantous

**5d. PROJECT NUMBER**

**5e. TASK NUMBER**

**5f. WORK UNIT NUMBER**

**7. PERFORMING ORGANIZATION NAME(S) AND ADDRESS(ES)**
Louisiana State University and A & M College
Office of Sponsored Programs
330 Thomas Boyd Hall
Baton Rouge, LA 70803

**8. PERFORMING ORGANIZATION REPORT NUMBER**
PA010

**9. SPONSORING/MONITORING AGENCY NAME(S) AND ADDRESS(ES)**
Office of Navel Research
Plasma Physics Division BAA 946
Washington, DC 20375

**10. SPONSOR/MONITOR'S ACRONYM(S)**
NRL

**11. SPONSORING/MONITORING AGENCY REPORT NUMBER**

**12. DISTRIBUTION AVAILABILITY STATEMENT**
Approved for public reslease

**13. SUPPLEMENTARY NOTES**

**14. ABSTRACT**

See page i of the report

**15. SUBJECT TERMS**
Radiative transfer, Discrete ordinates, Boltzmann equation

| 16. SECURITY CLASSIFICATION OF: a. REPORT | b. ABSTRACT | c. THIS PAGE | 17. LIMITATION OF ABSTRACT | 18. NUMBER OF PAGES | 19a. NAME OF RESPONSIBLE PERSON |
|---|---|---|---|---|---|
| U | U | U | UU | 53 | Charles Aboughantous |
| | | | | | **19b. TELEPONE NUMBER (*Include area code*)** |
| | | | | | 225 - 578 - 2261 |

Standard Form 298 (Rev. 8-98)
Prescribed by ANSI-Std Z39-18